\documentclass[review]{elsarticle}

\usepackage{lineno,hyperref}
\usepackage{graphics}
\usepackage{natbib}
\usepackage{subfigure}
\usepackage{amsmath}
\usepackage{amssymb}
\usepackage{empheq}
\usepackage{multirow}
\usepackage{booktabs}

\usepackage{siunitx}
\usepackage[normalem]{ulem}
\usepackage{xcolor}

\newtheorem{thm}{Theorem}
\newtheorem{clr}{Corollary}[thm]
\modulolinenumbers[5]

\journal{International Journal of Mechanical
Sciences}

\biboptions{sort&compress}

\begin{document}

\begin{frontmatter}

\title{Stable  dynamics of  micro-machined inductive contactless suspensions}


\author[KIT]{Kirill Poletkin\corref{KP}}
\cortext[KP]{Corresponding authors}
\ead{kirill.poletkint@kit.edu}

\author[IMTEK]{Zhiqiu Lu}
\author[IMTEK]{Ulrike Wallrabe}
\author[KIT]{Jan Korvink}
\author[KIT]{Vlad Badilita\corref{KP}}
\ead{vlad.badilita@kit.edu}

\address[KIT]{Institute of Microstructure Technology, Karlsruhe Institute of Technology, Germany}
\address[IMTEK]{Department of Microsystems Engineering-IMTEK, University of Freiburg, Germany}

\begin{abstract}
In this article we present a qualitative approach to study the dynamics and stability of micro-machined inductive contactless suspensions (MIS).
In the framework of this approach, the induced eddy  current into a levitated micro-object is considered as a
collection of $m$-eddy current circuits.  Assuming small displacements and the quasi-static behavior of the levitated micro-object, a generalized model of MIS is obtained and  represented as a set of six linear differential equations corresponding to six degrees of freedom in a rigid body by using the Lagrange-Maxwell formalism.  The linear model allows us to investigate the general stability properties of 
MIS as a dynamic system, and these properties are synthesized in three major theorems. In particular 
we prove that the stable levitation in the MIS without damping is impossible. Based on the approach presented herewith, we give general guidelines for 
designing MIS. Additionally, we demonstrate the successful application of this technique to  study the dynamics and stability of  symmetric and axially symmetric MIS designs, both based on 3D micro-coil technology.
\end{abstract}


\end{frontmatter}

\linenumbers

\section*{Nomenclature}
\begin{tabular}{l l l}
  $\mathbf{A}$ &    & diagonal matrix of the micro-object mass and its moments of inertia\hspace{1.5cm} \\
  $\mathbf{B}$ &    & diagonal matrix of damping coefficients\hspace{1.5cm} \\
    CS&    & contactless suspension\hspace{1.5cm} \\
  $g$ &    & acceleration of gravity $\big($m$^2$/s$\big)$\hspace{1.5cm} \\
   $i_{cj}$ &    & $j$- coil current $\big($A$\big)$\hspace{1.5cm} \\ 
   $i_{k}$ &    & $k$-eddy current $\big($A$\big)$\hspace{1.5cm} \\ 
   L-MEMS&    & levitating micro-electro-mechanical systems\hspace{1.5cm} \\
   $L$ &    & Lagrange function $\big($J$\big)$\hspace{1.5cm} \\ 
    $L_{jj}^c$ &    & self inductance of $j$-coil $\big($H$\big)$\hspace{1.5cm} \\ 
       $L_{js}^c$ &    & mutual inductance between $j$- and $s$- coils ($j\neq s$)  $\big($H$\big)$\hspace{1.5cm} \\ 
  $L_{kk}^{pm}$ &    & self inductance of $k$-eddy current circuit  $\big($H$\big)$\hspace{1.5cm} \\ 
  $M_{jk}$ &    & mutual inductance between $j$- coils and $k$-eddy current circuit  $\big($H$\big)$\hspace{1.5cm} \\
   $M$&    & mass of levitating micro-object $\big($kg$\big)$\hspace{1.5cm} \\
   MIS&    & micro-machined inductive contactless suspension\hspace{1.5cm} \\
  $m$ &    & number of eddy current circuits \\ 
  $n$ &    & number of coils \\ 
  $F_{l}$ &    & generalized force $\big($N$\big)$\hspace{1.5cm} \\ 
 PM&    & proof mass\hspace{1.5cm} \\
 $\mathbf{P}$ &    & matrix of  coefficients of the nonconservative positional forces\hspace{1.5cm} \\
 $\mathbf{\overline{q}}$ &    & vector of generalized coordinates\hspace{1.5cm} \\
  $q_{l}$ &    &lateral linear generalized coordinate $\big($m$\big)$\hspace{1.5cm} \\ 
    $q_{v}$ &    &vertical linear generalized coordinate $\big($m$\big)$\hspace{1.5cm} \\ 
       $\mathbf{R}$ &    & matrix of  coefficients of stiffness\hspace{1.5cm} \\

      $R_{kk}$ &    & resistance of $k$-eddy current circuit  $\big($$\Omega$$\big)$\hspace{1.5cm} \\
      $R_{ks}$ &    & resistance of a common circuit for $k$- and $s$- eddy current circuits  $\big($$\Omega$$\big)$\hspace{1.5cm} \\
  $T_{l}$ &    & generalized torque $\big($N m$\big)$\hspace{1.5cm} \\ 
  $T$ &    & kinetic energy $\big($J$\big)$\hspace{1.5cm} \\
  $t$ &    & time $\big($s$\big)$\hspace{1.5cm} \\
  $W_m$ &    & energy stored within electromagnetic field $\big($J$\big)$\hspace{1.5cm} \\
\end{tabular}

\subsection*{Greek}
\begin{tabular}{l l l}
   $\alpha$,  $\beta$, $\theta$ &    &angular generalized coordinates $\big($rad$\big)$\hspace{1.5cm} \\
   $\mu_l$ &    &damping coefficients \hspace{1.5cm} \\
   $\mu_0$ &    & magnetic permeability of vacuum $\big($H/m$\big)$\hspace{1.5cm} \\
  $\Pi$ &    & potential energy $\big($J$\big)$\hspace{1.5cm} \\
  $\Psi$ &    & dissipation energy $\big($J$\big)$\hspace{1.5cm} \\
   $\omega$ &    & frequency $\big($rad/s$\big)$\hspace{1.5cm} \\
  $\jmath$ &    & imaginary unit $\sqrt{-1}$\hspace{1.5cm} \\
\end{tabular}

\subsection*{Symbols}
\begin{tabular}{l l l}
$^{\ast}$ &    & imaginary part of complex variable\hspace{1.5cm} \\
$^{T}$ &    &  transpose operator \hspace{1.5cm} \\
$\bar{}$ &    &  complex variable \hspace{1.5cm} \\
\end{tabular}

\section{Introduction}

Electro-magnetic levitation
 dramatically reduces mechanical friction between various components of micro-sensors and micro-actuators in relative movement to each other and enables significant improvements in their performance. This fact has already attracted a  great interest in the MEMS research community during the past decades 
giving birth to a new generation of micro-devices: multi-inertial sensors with a high speed rotating rotor \cite{Murakoshi2003,Nakamura2005}, micro-gyroscopes \cite{Shearwood2000,Su2015}, micro-accelerators \cite{Sari2014}, frictionless micro-bearings \cite{Coombs2005,Lu2014}, hybrid suspensions \cite{Poletkin2015},\cite{Poletkin2017},  bistable switches \cite{Dieppedale2004}, linear-micro-actuators \cite{Ruffert2006}, and nano-force sensors \cite{Abadie2012}.  It is worth noting that there is no mechanical contact or attachment between a moving (sensing) element and the housing in any of the above-mentioned micro-devices. Based on this fact, all micro-devices relying on electro-magnetic levitation can be assigned the generic name levitating-MEMS (L-MEMS). 

A key element of L-MEMS
 is a contactless suspension providing the levitation and including a force field source and a micro-object (proof mass) levitated within the force field. Depending on the force field, contactless suspensions (CS) can be simply classified as electrostatic, magnetic and hybrid CS (a combination of different principles, e.g., electrostatic, static magnetic field, variable magnetic field, diamagnetic materials). The electrostatic CS has been already established as the integrated element for L-MEMS, the fabrication process being compatible with MEMS technologies. In contrast to the electrostatic CS in which stable levitation is reached by active control, in a magnetic CS the levitation of the proof mass can be achieved passively. This fact makes the latter very attractive to be employed in L-MEMS, since this advantage opens additional opportunities to improve L-MEMS performance and increase their operational capabilities  by means of hybrid CS \cite{Liu2010,Poletkin2012,Poletkin2015,Poletkin2017}.

However, the development of magnetic CS 
is still lagging behind their electrostatic counterparts. 
It is a well-known fact that magnetic levitation utilizing a static magnetic field requires a diamagnetic or a superconducting (perfect diamagnetic) proof mass. In the case of a diamagnetic proof mass, a diamagnetic material with a susceptibility much higher than 1$\times 10^{-4}$ is needed. Unfortunately, the strongest diamagnetic materials such as  
graphite and bismuth \cite{SimonHeflingerGeim2001} are not traditional materials for MEMS processing.
In order to levitate a superconducting proof mass, a cryogen-based environment is required, which itself becomes a major limit for this application.
Magnetic levitation utilizing a time-varying magnetic field and a conducting proof mass, or magnetic levitation based on electro-magnetic induction
, does not have the disadvantages mentioned above and becomes a very promising candidate for L-MEMS as an integrated element.
Recent achievements in
 the development of 3D micro-coils \cite{Kratt2010} and new magnetic materials \cite{MariappanMoazenzadehWallrabe2016} have drastically reduced the heat dissipation  in MIS \cite{PoletkinMoazenzadehMariappanEtAl2016}, which was the typical problem for  first prototypes \cite{Yates1996,Shearwood2000,Tsai2009,Lu2014a}, 
and with that, the micromachining fabrication process for MIS can be considered as fully established.

Advanced MIS applications require new designs of this type of suspension, which  should demonstrate improvements in  MIS dynamics \cite{xu2015dynamic} and at the same time provide stable levitation \cite{Poletkin2014a}. The latter  is  a key issue for designing MIS, which has been studied since the middle of the last century, when first prototypes of bulk inductive contactless suspensions employed in material processing (e.g., melting a levitated metal sphere) were demonstrated and reported in \cite{BeamsYoungMoore1946,OkressWroughtonComenetzEtAl1952}. Since that time, two main directions have been established in order to provide a condition of stable levitation in  inductive contactless suspensions. One direction is related to directly solving the Maxwell equations. For instance, a technique based on the assumptions of a quasi-stationary electromagnetic field and perfectly conducting spheroid was developed by Ciric \cite{Ciric1970} to study axially symmetric designs. However, the given theoretical formulation in this work is so complex that another direction, which can be  labelled as qualitative, was explored to avoid dealing with field equations.

In 1965, Laithwaite \cite{Laithwaite1965} developed a qualitative technique based on lines of constant phase, which can be used to predict the behaviour of different designs, including the condition of stable levitation. Laithwaite provided an overview of inductive suspension's designs which are currently realized in existing MIS prototypes. For instance, the MIS design employing levitation and stabilization coils, which was first  demonstrated in the prototype of MIS developed by Shearwood et al. \cite{Shearwood1995} in 1995, is a typical one, being used in most MIS prototypes.
 Earlier in 1952, using a qualitative approach, Okress et al. \cite{OkressWroughtonComenetzEtAl1952} proposed to replace a levitated sphere by an alternating magnetic dipole and have successfully studied  the levitating force acting on the sphere. Recently, the same idea, together with assuming the quasi-static behavior of the levitated proof mass, was employed to develop the analytical model of MIS \cite{Poletkin2013}, and then this model was extended in order to analyze axially symmetric MIS designs \cite{Poletkin2014a}. The results were successfully applied to study the dynamics and stability of MIS based on 3D wirebonded micro-coils, as well as on planar coils \cite{Lu2014}.

The approaches mentioned above provide powerful tools for designing  inductive suspensions, and in particular for designing MIS. However, in  all of them the analysis of stability is reduced to the study  of the minimum of potential energy for a conservative  system, which is not sufficient. Indeed, in addition to  potential forces, dissipative  (due to air environment) and nonconservative positional forces (due to the electrical resistance in the conductive proof mass) are acting on the levitated proof mass. According to the classical theory of stability, dissipative forces support the stable state of a system, while positional forces destroy it. Hence, it becomes necessary to determine the balance between all of these forces (potential, dissipative, and positional)  to provide a comprehensive stable levitation condition in MIS. Stability becomes especially critical for MIS operation in vacuum, which is extremely relevant, for instance, for micro-sensor applications.

In order to fill this gap, this article presents a generalized linear analytical model of MIS in which potential, dissipative and positional forces are taken into account. Considering the induced eddy current into a micro-object as a collection of $m$-eddy current circuits, and assuming small displacements and the quasi-static behavior of the levitated micro-object (proof mass), this generalized model is represented as a set of six linear differential equations by using the Lagrange-Maxwell formalism. The number of equations in this set corresponds to the six degrees of freedom (DoF) of a rigid body.
Thus,  the linear model allows us to investigate the general stability properties of an inductive contactless suspension as a dynamic system. The results of this investigation are condensed in three theorems. In particular, a theorem of unstable levitation is formulated in which we prove that the stable levitation in an inductive suspension (levitating a conductive micro-object) without damping is impossible. This theorem represents the extension of the result of the classical theorem elaborated for the case of a stable potential system having equal natural frequencies subjected to nonconservative positional forces \cite[page 202, Theorem 6.12]{Merkin2012}. Also, we prove that MIS subjected to only positional and dissipative forces is unstable and formulate this statement in the second theorem.
Then the necessary and  the sufficient practical conditions are defined to provide asymptotically stable levitation in MIS based on Metelitsyn's inequality \cite{Metelitsyn1952} and formulated in the third theorem of asymptotically stable levitation.

The presented model is the natural continuation of the qualitative approach developed in \cite{Poletkin2014a}, the crucial difference is that here the levitated micro-object is approximated by a system of magnetic dipoles instead of one single dipole. This fact increases the accuracy in the evaluation of MIS dynamical parameters (the accuracy of modelling is dependent on the number ($m$) of the eddy current circuits taken into consideration), e.g., evaluating its stiffness without employing similarity coefficients, which were used in \cite{Lu2014}. Moreover, a generalized procedure for the designing of an inductive contactless suspension is proposed. Based on this procedure, the stability of a new MIS design, proposed to be used as a linear micro-transporter, is investigated. Following the procedure, the stability map is calculated as a function of design parameters. Subsequently, the experimental study of the fabricated prototype of the developed linear micro-transporter helped to verify the calculated stability map. The result of this experimental study is in a good agreement with the modelling predictions.

\section{Qualitative Technique }
\label{sec:Qualitative Technique}

\begin{figure}[t!]
  \center
  \includegraphics[width=3.0in]{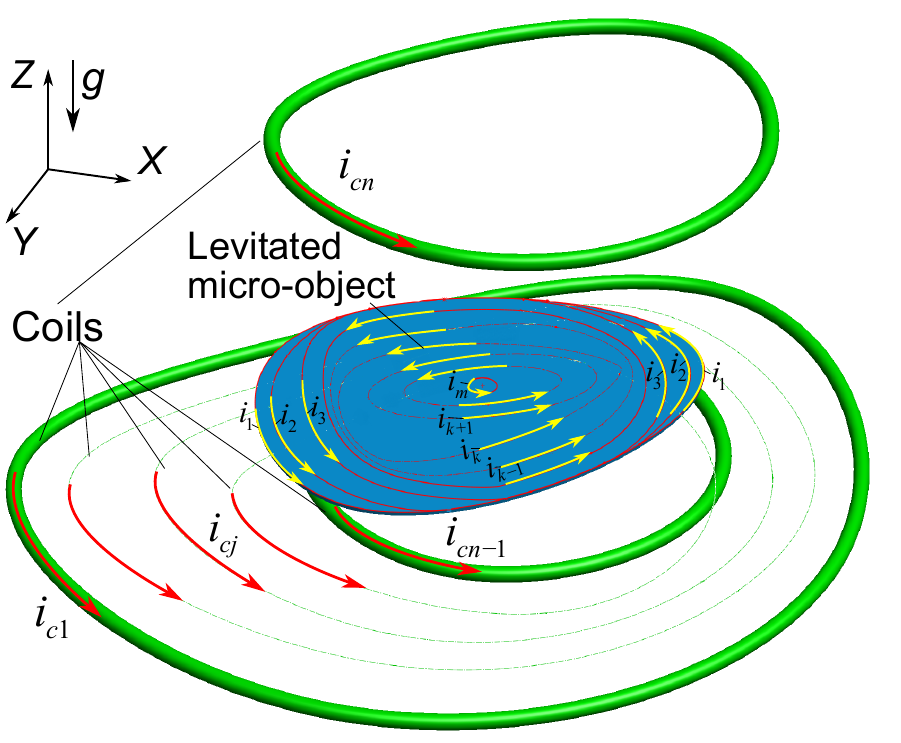}\\
  \caption{General schematic of micromachined contactless suspension with the $m$ eddy current circuits: $YXZ$ is the fixed coordinate frame; $g$ is the gravity acceleration directed along the $Z$ axis; $i_k$ is the eddy current in the $k$-th eddy current circuit; $i_{cj}$ is the current in the $j$-coil.
   }\label{fig:scheme}
\end{figure}

Let us consider the schematic of an inductive contactless suspension shown in Fig.~\ref{fig:scheme}, which consists of a system of $n$ coils, and a levitated micro-object. Each $j$-th coil is fed by its own alternating current denoted by $i_{cj}$ and generates a time-variable magnetic field in space. In turn, the alternating magnetic flux passing through the conducting  micro-object induces an eddy current. The eddy current is continuously distributed within the micro-object, however this distribution is not homogeneous. This fact helps us to selectively choose the $m$ eddy current circuits having the representative behaviour of the entire eddy current distribution, as shown in Fig.~\ref{fig:scheme}. As seen from Fig.~\ref{fig:scheme}, $i_k$ is the eddy current in the $k$-th eddy current circuit. The interaction between the currents in the coils and the eddy current produces the repulsive force levitating the micro-object at an equilibrium position, which can be characterized with respect to the fixed coordinate frame $YXZ$. Considering this micro-object as a rigid body, its behaviour relative to the equilibrium position can be characterized in general by six generalized coordinates corresponding to three linear and three angular displacements, which can be denoted by $q_l,\;l=1...6$. Let us define that coordinates,  $q_l$, with indexes  $l=1,2,3$ are the generalized linear coordinates, while indexes  $l=4,5,6$ correspond to the generalized angular coordinates.

Adapting the generalized coordinates and the assumptions introduced
above, the MIS model can be written by using the Lagrange -
Maxwell equations as follows
\begin{equation} \label{eq:Lagrange_Maxwell}
\left\{\begin{array}{l} {\displaystyle{\frac{d}{dt} \left(\frac{\partial L}{\partial i_{k} } \right)+\frac{\partial \Psi }{\partial i_{k} } =0; \; k=1,\ldots,m};} \\
{\displaystyle{\frac{d}{dt} \left(\frac{\partial L}{\partial \dot{q_l}} \right)-\frac{\partial L}{\partial q_l} +\frac{\partial \Psi }{\partial \dot{q_l}} =F_{l} ; \;l=1,2,3;}}\\
{\displaystyle{\frac{d}{dt} \left(\frac{\partial L}{\partial \dot{q_l}} \right)-\frac{\partial L}{\partial q_l} +\frac{\partial \Psi }{\partial \dot{q_l}} =T_{l} ; \;l=4,5,6,}} \\
 \end{array}\right.  \end{equation}
where $L=T-\Pi +W_{m}$ is the Lagrange function for the micro-object-coil system; $T=T(\dot{q_1},\ldots,\dot{q_6})$ is the kinetic energy of the system;  $\Pi=\Pi (q_1,\ldots,q_6)$ is the potential energy of the system; $W_{m}=W_{m}(q_1,\ldots,q_6,i_{c1},\ldots,i_{cn},i_{1},\ldots,i_{m})$ is the energy stored in the electromagnetic field; $\Psi=\Psi (\dot{q_1},\ldots,\dot{q_6},i_{1},\ldots,i_{m})$ is the dissipation function; $F_{l}$ ($l=1,2,3$) and $T_{r}$ ($l=4,5,6$) are the generalized forces and torques, respectively, acting on the micro-object relative to the appropriate generalized coordinates.

The kinetic energy is
\begin{equation}
\label{eq:kinetic} {\displaystyle T=\frac{1}{2}\sum_{l=1}^3
M\dot{q_l}^{2} +\frac{1}{2}\sum_{l=4}^6 J_l\dot{q_l}^{2}},
\end{equation}
where $M$ is the mass of the micro-object; $J_l$ is its moment of inertia in terms of the appropriate generalized angular coordinates.

The linear generalized coordinates, $q_l,\;l=1,2,3$, are defined in the orthogonal coordinate frame. Hence, for the further simplification of analysis, it can be  assumed that the generalized coordinate $q_3$ is directed along the gravity acceleration $g$, as shown in Fig.~\ref{fig:scheme}. Then, the potential energy can be defined simply as follows
\begin{equation} \label{eq:potential} \Pi =Mgq_3. \end{equation} 

The dissipation function is
\begin{equation} \label{eq:dissipation}
{\displaystyle
\Psi =\frac{1}{2}\sum_{r=1}^6 \mu_r\dot{q_r}^{2}+\frac{1}{2}\sum_{k=1}^m R_ki_k^{2} \pm \sum_{k=1}^m\sum_{s=1,s\neq k}^{m}R_{ks}i_ki_s ,
}\end{equation}
where $\mu_r$ is the damping coefficient corresponding to the appropriate generalized coordinates; $R_k$ is the electrical resistance for the $k$-th eddy current circuit within the micro-object;  $R_{ks}$ is the resistance of a common circuit for $k$-th and $s$-th eddy current circuits (for example, this case is shown in Fig.~\ref{fig:scheme} for eddy currents $i_1$, $i_2$, and $i_3$). For generality, it is assumed that the $k$-th eddy current may share a common path with the $s$-th eddy current. The plus-minus sign corresponds to eddy currents having the same or opposite direction on the common circuit. The energy stored within the electromagnetic field can be written as
\begin{equation} \label{eq:field} {\displaystyle W_{m} =\frac{1}{2}\sum_{j=1}^n\sum_{s=1}^{n}L_{js}^{c}i_{cj}i_{cs}+
\frac{1}{2}\sum_{k=1}^m\sum_{s=1}^{m}L_{ks}^{o}i_{k}i_{s}+
\frac{1}{2}\sum_{k=1}^m\sum_{j=1}^nM_{kj}i_{k}i_{cj}} ,
\end{equation}
where $L_{jj}^{c}$ is the self inductance of the $j$-coil; $L_{js}^{c}$, $j\neq s$ is the mutual inductance between $j$- and $s$-coils; $L_{kk}^{o}=L_{kk}^{o}(q_1,\ldots,q_6)$ is the self inductance of the $k$-eddy current circuit; $L_{ks}^{o}=L_{ks}^{o}(q_1,\ldots,q_6)$, $k\neq s$ is the mutual inductance between $k$-  and $s$-eddy current circuits; $M_{kj}=M_{kj}(q_1,\ldots,q_6)$ is the mutual inductance between  the $k$-eddy current circuit and  the $j$-coil.

We now show that the induced eddy currents $i_k$ can be expressed in terms of coil currents $i_{cj}$ under a particular condition discussed below. Assuming that for each coil, the current $i_{cj}$ is a periodic signal with an amplitude of $I_{cj}$ at the same frequency $\omega$, we can write
\begin{equation}\label{eq:coil current}
    i_{cj}=I_{cj}e^{\jmath\omega t},
\end{equation}
where $\jmath=\sqrt{-1}$. Taking the derivative of the Lagrange function with respect to the eddy current $i_k$, we have:
\begin{equation}\label{eq: current derivative of W}
    {\displaystyle \frac{\partial L}{\partial i_k}=\frac{\partial W_{m}}{\partial i_k} =\sum_{s=1}^{m}L_{ks}^{o}i_{s}+
\sum_{j=1}^nM_{kj}i_{cj}} ,
\end{equation}
or
\begin{equation}\label{eq: or current derivative of W}
    {\displaystyle \frac{\partial L}{\partial i_k}=\frac{\partial W_{m}}{\partial i_k} = L_{kk}^{o}i_{k}+\sum_{s=1,\; s\neq k}^{m}L_{ks}^{o}i_{s}+
\sum_{j=1}^nM_{kj}i_{cj}}.
\end{equation}
Substituting (\ref{eq: or current derivative of W}) into (\ref{eq:Lagrange_Maxwell})
, the first equation of set (\ref{eq:Lagrange_Maxwell})  becomes
\begin{equation}\label{eq: the first equations of set}
    \begin{array}{l} {\displaystyle{\frac{d}{dt} \left(\frac{\partial L}{\partial i_{k} } \right)+\frac{\partial \Psi }{\partial i_{k} } =\sum_{r=1}^{6}\frac{\partial L_{kk}^{o}}{\partial q_r}\frac{dq_{r}}{dt}i_{k}
    +L_{kk}^{o}\frac{di_{k}}{dt}}} \\
{\displaystyle{+\sum_{s=1,\; s\neq k}^{m}\left(\sum_{r=1}^{6}\frac{\partial L_{ks}^{o}}{\partial q_r}\frac{dq_{r}}{dt}i_{s}
    +L_{ks}^{o}\frac{di_{s}}{dt}\right)}}\\
{\displaystyle{+\sum_{j=1}^n\left(\sum_{r=1}^{6}\frac{\partial M_{kj}}{\partial q_r}\frac{dq_{r}}{dt}i_{cj}+M_{kj}\frac{di_{cj}}{dt}\right)+R_ki_k \pm \sum_{s=1,\; s\neq k}^{m}R_{ks}i_{s} =0.}}
\end{array}
\end{equation}
Accounting for (\ref{eq:coil current}), the $k-$ eddy current can be
represented as
\begin{equation}\label{eq:eddy current}
    i_k=I_ke^{\jmath\omega t},
\end{equation}
where $I_k$ is the amplitude. Hence, Eq.\
(\ref{eq: the first equations of set}) can be rewritten in term of the current amplitudes as follows
\begin{equation}\label{eq: the first equations of set amplitude}
    \begin{array}{l} {\displaystyle{\sum_{r=1}^{6}\frac{\partial L_{kk}^{o}}{\partial q_r}\frac{dq_{r}}{dt}I_{k}
    +L_{kk}^{o}\jmath\omega I_k}} \\
{\displaystyle{+\sum_{s=1,\; s\neq k}^{m}\left(\sum_{r=1}^{6}\frac{\partial L_{ks}^{o}}{\partial q_r}\frac{dq_{r}}{dt}I_{s}
    +L_{ks}^{o}\jmath\omega I_s\right)}}\\
{\displaystyle{+\sum_{j=1}^n\left(\sum_{r=1}^{6}\frac{\partial M_{kj}}{\partial q_r}\frac{dq_{r}}{dt}I_{cj}+M_{kj}\jmath\omega I_{cj}\right)+R_kI_k \pm \sum_{s=1,\; s\neq k}^{m}R_{ks}I_{s}=0.}}
\end{array}
\end{equation}
Equation (\ref{eq: the first equations of set amplitude}) is nonlinear due to the velocities of generalized coordinates, $dq_r/dt$. In fact, the analysis of the existing suspension prototypes shows that the velocity $dq_r/dt$ can be assumed to be small. Also the frequency $\omega$ is usually larger than 1 MHz, which corresponds to $\sim10^7$ rad/s. Hence, Eq.\ (\ref{eq: the first equations of set amplitude}) can be rewritten \cite{Poletkin2013} as follows
\begin{equation}\label{eq: the first equations of set amplitude linear}
   {\displaystyle{\left(L_{kk}^{o}+R_k/(\jmath\omega )\right) I_k+\sum_{s=1,\; s\neq k}^{m}
    (L_{ks}^{o}\pm R_{ks}/(\jmath\omega  ))I_s=-\sum_{j=1}^nM_{kj}I_{cj}.}}
\end{equation}


It is important to note that, for higher values of the generalized velocities $dq_r/dt$, when the quasi-static approximation does not hold, Eq.\ (\ref{eq: the first equations of set amplitude}) must be used. In order to define the eddy currents $I_k$, a set of linear equations can be compiled from
(\ref{eq: the first equations of set amplitude linear}) in a matrix
form as follows:
\begin{equation}\label{eq: linear eqns eddy current}
\begin{footnotesize}\left(\begin{array}{cccccc}
  {\scriptstyle L_{11}^{o}+\frac{R_1}{\jmath\omega}} & {\scriptstyle L_{12}^{o}\pm\frac{R_{12}}{\jmath\omega}} &\ldots & {\scriptstyle L_{1k}^{o}\pm\frac{R_{1k}}{\jmath\omega}}& \ldots &{\scriptstyle L_{1m}^{o}\pm\frac{R_{1m}}{\jmath\omega}} \\
  {\scriptstyle L_{21}^{o}\pm\frac{R_{21}}{\jmath\omega}}& {\scriptstyle L_{22}^{o}+\frac{R_2}{\jmath\omega }}  & \ldots &  {\scriptstyle L_{2k}^{o}\pm\frac{R_{2k}}{\jmath\omega}}&  \ldots&  {\scriptstyle L_{2m}^{o}\pm\frac{R_{2m}}{\jmath\omega} }\\
  \vdots & \vdots & \ddots & \vdots & \ddots&\vdots \\
  {\scriptstyle L_{k1}^{o}\pm\frac{R_{k1}}{\jmath\omega}} & {\scriptstyle L_{k2}^{o}\pm\frac{R_{k2}}{\jmath\omega}} & \ldots & {\scriptstyle L_{kk}^{o}+\frac{R_k}{\jmath\omega}} & \ldots & {\scriptstyle L_{km}^{o} \pm\frac{R_{km}}{\jmath\omega}}\\
  \vdots & \vdots &\ddots & \vdots & \ddots & \vdots\\
  {\scriptstyle L_{m1}^{o}\pm\frac{R_m1}{\jmath\omega}} &{\scriptstyle L_{m2}^{o}\pm\frac{R_{m2}}{\jmath\omega} }& \ldots & {\scriptstyle L_{mk}^{o}\pm\frac{R_{mk}}{\jmath\omega}} & \ldots& {\scriptstyle L_{mm}^{o}+\frac{R_m}{\jmath\omega
  }}
\end{array}\right)
\left(\begin{array}{c}
  I_1 \\
  I_2 \\
  \vdots\\
  I_k \\
  \vdots\\
  I_m
\end{array}\right)=
\left(\begin{array}{c}
 -\sum_{j=1}^nM_{1j}I_{cj} \\
  -\sum_{j=1}^nM_{2j}I_{cj} \\
  \vdots\\
 -\sum_{j=1}^nM_{kj}I_{cj}\\
  \vdots\\
 -\sum_{j=1}^nM_{mj}I_{cj}
\end{array}\right)
\end{footnotesize},
\end{equation}
\normalsize where $L_{ks}^{o}=L_{sk}^{o}$. The solution of (\ref{eq: linear eqns eddy current}) for $I_k$ can be found by using Cramer's rule and is written as follows
\begin{equation}\label{eq:solution for I}
    I_k=\frac{\Delta_k}{\Delta},
\end{equation}
where 
\begin{equation}\label{eq: main determinant} \Delta=
\begin{footnotesize}\left|\begin{array}{cccccc}
  {\scriptstyle L_{11}^{o}+\frac{R_1}{\jmath\omega}} & {\scriptstyle L_{12}^{o}\pm\frac{R_{12}}{\jmath\omega}} &\ldots & {\scriptstyle L_{1k}^{o}\pm\frac{R_{1k}}{\jmath\omega}}& \ldots &{\scriptstyle L_{1m}^{o}\pm\frac{R_{1m}}{\jmath\omega}} \\
  {\scriptstyle L_{21}^{o}\pm\frac{R_{21}}{\jmath\omega}}& {\scriptstyle L_{22}^{o}+\frac{R_2}{\jmath\omega }}  & \ldots &  {\scriptstyle L_{2k}^{o}\pm\frac{R_{2k}}{\jmath\omega}}&  \ldots&  {\scriptstyle L_{2m}^{o}\pm\frac{R_{2m}}{\jmath\omega} }\\
  \vdots & \vdots & \ddots & \vdots & \ddots&\vdots \\
  {\scriptstyle L_{k1}^{o}\pm\frac{R_{k1}}{\jmath\omega}} & {\scriptstyle L_{k2}^{o}\pm\frac{R_{k2}}{\jmath\omega}} & \ldots & {\scriptstyle L_{kk}^{o}+\frac{R_k}{\jmath\omega}} & \ldots & {\scriptstyle L_{km}^{o} \pm\frac{R_{km}}{\jmath\omega}}\\
  \vdots & \vdots &\ddots & \vdots & \ddots & \vdots\\
  {\scriptstyle L_{m1}^{o}\pm\frac{R_{m1}}{\jmath\omega}} &{\scriptstyle L_{m2}^{o}\pm\frac{R_{m2}}{\jmath\omega} }& \ldots & {\scriptstyle L_{mk}^{o}\pm\frac{R_{mk}}{\jmath\omega}} & \ldots& {\scriptstyle L_{mm}^{o}+\frac{R_m}{\jmath\omega
  }}
\end{array}\right|
\end{footnotesize},
\end{equation}
\begin{equation}\label{eq:sub  determinant} \Delta_k=
\begin{footnotesize}\left|\begin{array}{cccccc}
  {\scriptstyle L_{11}^{o}+\frac{R_1}{\jmath\omega}} & {\scriptstyle L_{12}^{o}\pm\frac{R_{12}}{\jmath\omega}}&\ldots & -\sum_{j=1}^nM_{1j}I_{cj}& \ldots &{\scriptstyle L_{1m}^{o}\pm\frac{R_{1m}}{\jmath\omega}} \\
  {\scriptstyle L_{21}^{o}\pm\frac{R_{21}}{\jmath\omega}}& {\scriptstyle L_{22}^{o}+\frac{R_2}{\jmath\omega }}  & \ldots &  -\sum_{j=1}^nM_{2j}I_{cj} & \ldots &  {\scriptstyle L_{2m}^{o}\pm\frac{R_{2m}}{\jmath\omega} } \\
  \vdots & \vdots & \ddots & \vdots & \ddots&\vdots \\
  {\scriptstyle L_{k1}^{o}\pm\frac{R_{k1}}{\jmath\omega}} & {\scriptstyle L_{k2}^{o}\pm\frac{R_{k2}}{\jmath\omega}} & \ldots &  -\sum_{j=1}^nM_{kj}I_{cj} & \ldots & {\scriptstyle L_{km}^{o} \pm\frac{R_{km}}{\jmath\omega}}\\
  \vdots & \vdots &\ddots & \vdots & \ddots & \vdots\\
 {\scriptstyle L_{m1}^{o}\pm\frac{R_{m1}}{\jmath\omega}} &{\scriptstyle L_{m2}^{o}\pm\frac{R_{m2}}{\jmath\omega} } & \ldots & -\sum_{j=1}^nM_{mj}I_{cj} & \ldots& {\scriptstyle L_{mm}^{o}+\frac{R_m}{\jmath\omega
  }}
\end{array}\right|
\end{footnotesize}.
\end{equation}

Rewriting determinant (\ref{eq:sub  determinant}) as follows
\begin{equation}\label{eq:sub  determinant transponation} \Delta_k=
\begin{footnotesize}\left|\begin{array}{cccccc}
 {\scriptstyle L_{11}^{o}+\frac{R_1}{\jmath\omega}}& {\scriptstyle L_{21}^{o}\pm\frac{R_{21}}{\jmath\omega}} &\ldots &  {\scriptstyle L_{k1}^{o}\pm\frac{R_{k1}}{\jmath\omega}} &  \ldots &{\scriptstyle L_{m1}^{o}\pm\frac{R_{m1}}{\jmath\omega}} \\
 {\scriptstyle L_{12}^{o}\pm\frac{R_{12}}{\jmath\omega}} & {\scriptstyle L_{22}^{o}+\frac{R_2}{\jmath\omega }}  & \ldots &{\scriptstyle L_{k2}^{o}\pm\frac{R_{k2}}{\jmath\omega}}  &  \ldots& {\scriptstyle L_{m2}^{o}\pm\frac{R_{m2}}{\jmath\omega} }\\
  \vdots & \vdots & \ddots & \vdots & \ddots&\vdots \\
  -\sum_{j=1}^nM_{1j}I_{cj}& -\sum_{j=1}^nM_{2j}I_{cj} & \ldots &  -\sum_{j=1}^nM_{kj}I_{cj} & \ldots &-\sum_{j=1}^nM_{mj}I_{cj}  \\
  \vdots & \vdots &\ddots & \vdots & \ddots & \vdots\\
  {\scriptstyle L_{1m}^{o}\pm\frac{R_{1m}}{\jmath\omega}} & {\scriptstyle L_{2m}^{o}\pm\frac{R_{2m}}{\jmath\omega} } & \ldots & {\scriptstyle L_{km}^{o} \pm\frac{R_{km}}{\jmath\omega}}& \ldots& {\scriptstyle L_{mm}^{o}+\frac{R_m}{\jmath\omega
  }}
\end{array}\right|
\end{footnotesize},
\end{equation}
and accounting for the determinant properties, (\ref{eq:sub  determinant}) can be represented as the sum
\begin{equation}\label{eq:sum of sub  determinant transponation}
    \Delta_k=-\sum_{j=1}^n\Delta_{kj}I_{cj},
\end{equation}
where
\begin{equation}\label{eq:sub  determinant transponation}
\Delta_{kj}=-
\begin{footnotesize}\left|\begin{array}{cccccc}
   {\scriptstyle L_{11}^{o}+\frac{R_1}{\jmath\omega}}& {\scriptstyle L_{21}^{o}\pm\frac{R_{21}}{\jmath\omega}} &\ldots &  {\scriptstyle L_{k1}^{o}\pm\frac{R_{k1}}{\jmath\omega}} &  \ldots &{\scriptstyle L_{m1}^{o}\pm\frac{R_{m1}}{\jmath\omega}} \\
  {\scriptstyle L_{12}^{o}\pm\frac{R_{12}}{\jmath\omega}} & {\scriptstyle L_{22}^{o}+\frac{R_2}{\jmath\omega }}  & \ldots &{\scriptstyle L_{k2}^{o}\pm\frac{R_{k2}}{\jmath\omega}}  &  \ldots& {\scriptstyle L_{m2}^{o}\pm\frac{R_{m2}}{\jmath\omega} }\\
  \vdots & \vdots & \ddots & \vdots & \ddots&\vdots \\
  M_{1j}&M_{2j} & \ldots & M_{kj} & \ldots & M_{mj}  \\
  \vdots & \vdots &\ddots & \vdots & \ddots & \vdots\\
  {\scriptstyle L_{1m}^{o}\pm\frac{R_{1m}}{\jmath\omega}} & {\scriptstyle L_{2m}^{o}\pm\frac{R_{2m}}{\jmath\omega} } & \ldots & {\scriptstyle L_{km}^{o} \pm\frac{R_{km}}{\jmath\omega}}& \ldots& {\scriptstyle L_{mm}^{o}+\frac{R_m}{\jmath\omega
  }}
\end{array}\right|
\end{footnotesize}.
\end{equation}
Taking the later equation into account, the current corresponding to the k-th eddy current circuit can be directly written in terms  of the coils currents. Hence, Eq.\ (\ref{eq:solution for I}) becomes 
\begin{equation}\label{eq:solution for I in term of coil current}
    I_k=\frac{-\sum_{j=1}^n\Delta_{kj}I_{cj}}{\Delta}.
\end{equation}

Thus, instead of $m+6$ equations, set (\ref{eq:Lagrange_Maxwell}) can be reduced to six equations. Hence, the behavior of the suspension is defined only by the generalized coordinates of its mechanical part. Moreover, the number of generalized coordinates of the mechanical part can be further reduced, depending on a particular design of the suspension, as will be shown below.

Accounting for Eq. (\ref{eq:solution for I in term of coil current}), the energy stored within the electromagnetic field describing by Eq. (\ref{eq:field}) can be written via current amplitudes as follows
\begin{equation} \label{eq:field amplitude}
\begin{array}{l}{\displaystyle{W_{m} =\frac{1}{2}\sum_{j=1}^n\sum_{s=1}^{n}L_{js}^{c}I_{cj}I_{cs}+
\frac{1}{2}\frac{1}{\Delta^2}\sum_{k=1}^m\sum_{s=1}^{m}\left(L_{ks}^{o}\sum_{j=1}^n\sum_{i=1}^n\Delta_{kj}\Delta_{si}I_{cj}I_{ci}\right)}}\\
{\displaystyle{
-\frac{1}{2}\frac{1}{\Delta}\sum_{k=1}^m\sum_{j=1}^n\left(M_{kj}\sum_{s=1}^n\Delta_{ks}I_{cs}I_{cj}\right)}}.
\end{array}
\end{equation}

The set becomes
\begin{equation} \label{eq:Lagrange_Maxwell modified}
\left\{\begin{array}{l} {\displaystyle{M\ddot{q_l}+\mu_l\dot{q_l}-\frac{\partial W_m}{\partial q_l}  =F_l; \; l=1,2};} \\
{\displaystyle{M\ddot{q_3}+\mu_3\dot{q_3}+mg-\frac{\partial W_m}{\partial q_3}  =F_3;}}\\
{\displaystyle{ J_l\ddot{q_l}+\mu_l\dot{q_l}-\frac{\partial W_m}{\partial q_l}=T_{l} ; \;l=4,5,6,}} \\
 \end{array}\right.  \end{equation}
where the derivative of $W_m$ with respect to a generalized
coordinate has the following general form
\begin{equation} \label{eq:deriviation of field }
\begin{array}{l}{\displaystyle{  \frac{\partial W_m}{\partial q_r} =
\frac{1}{2}\frac{1}{\Delta^2}\sum_{k=1}^m\sum_{s=1}^{m}\left(\frac{\partial L_{ks}^{o}}{\partial q_r}\sum_{j=1}^n\sum_{i=1}^n\Delta_{kj}\Delta_{si}I_{cj}I_{ci}\right.}}\\
{\displaystyle{+\sum_{k=1}^m\sum_{s=1}^{m}\left.L_{ks}^{o}\sum_{j=1}^n\sum_{i=1}^n
\left[\frac{\partial \Delta_{kj}}{\partial q_r}\Delta_{si}+\Delta_{kj}\frac{\partial \Delta_{si}}{\partial q_r}\right]I_{cj}I_{ci}\right)}}\\
{\displaystyle{-\frac{1}{\Delta^3}\frac{\partial \Delta}{\partial q_r}\sum_{k=1}^m\sum_{s=1}^{m}\left(L_{ks}^{o}\sum_{j=1}^n\sum_{i=1}^n\Delta_{kj}\Delta_{si}I_{cj}I_{ci}\right)}}\\
{\displaystyle{
-\frac{1}{2}\frac{1}{\Delta}\sum_{k=1}^m\sum_{j=1}^n\left(\frac{\partial M_{kj}}{\partial q_r}\sum_{s=1}^n\Delta_{ks}I_{cs}I_{cj}+ M_{kj}\sum_{s=1}^n\frac{\partial \Delta_{ks}}{\partial q_r}I_{cs}I_{cj}\right)}}\\
{\displaystyle{ +\frac{1}{2}\frac{1}{\Delta^2}\frac{\partial
\Delta}{\partial
q_r}\sum_{k=1}^m\sum_{j=1}^n\left(M_{kj}\sum_{s=1}^n\Delta_{ks}I_{cs}I_{cj}\right);
\; r=1,\ldots,6.}}
\end{array}
\end{equation}

\subsection{Linearizing}

The amplitudes of the eddy currents are several orders of magnitude less than the amplitudes of the coil currents. As a result, the stored energy, which is defined by the second term in Eq.\ (\ref{eq:field amplitude}), is negligible compared to the third term. Hence, Eq.\ (\ref{eq:deriviation of field }) can be simplified as follows
\begin{equation} \label{eq:deriviation of field simplified}
\begin{array}{l}
{\displaystyle{
\frac{\partial W_m}{\partial q_r} =\overbrace{-\frac{1}{2}\frac{1}{\Delta}\sum_{k=1}^m\sum_{j=1}^n\left(\frac{\partial M_{kj}}{\partial q_r}\sum_{s=1}^n\Delta_{ks}I_{cs}I_{cj}+ M_{kj}\sum_{s=1}^n\frac{\partial \Delta_{ks}}{\partial q_r}I_{cs}I_{cj}\right)}^{\mbox{I term}}}}\\
{\displaystyle{
+\underbrace{\frac{1}{2}\frac{1}{\Delta^2}\frac{\partial
\Delta}{\partial
q_r}\sum_{k=1}^m\sum_{j=1}^n\left(M_{kj}\sum_{s=1}^n\Delta_{ks}I_{cs}I_{cj}\right)}_{\mbox{II
term}}; \; r=1,\ldots,6.}}
\end{array}
\end{equation}
The analysis of (\ref{eq:deriviation of field simplified}) shows that two sources of ponderomotive forces can be identified: those due to changing the positions of the eddy currents with respect to the coils (the first term in (\ref{eq:deriviation of field simplified})), and those due to changing the positions of the eddy currents with respect to each other within the micro-object (the second one in (\ref{eq:deriviation of field simplified})).

For further analysis, the derivative of stored magnetic energy, $W_m$, with respect to the generalized coordinates, $q_r$, is expanded into the Taylor series. Due to the above mentioned assumption of small displacements of the micro-object relative to the equilibrium position, the following functions taken from (\ref{eq:deriviation of field simplified}) can be expanded into the Taylor series, keeping only second order terms, as follows
\begin{equation}\label{eq:Taylor series}
\begin{array}{l}
 {\displaystyle{M_{kj}=m_0^{kj}+\sum_{l=1}^6m_l^{kj}q_l+\frac{1}{2}\sum_{r=1}^6\sum_{l=1}^6m_{rl}^{kj}q_rq_l}}; \\
 {\displaystyle{L_{ks}^{o}=g_0^{ks}+\sum_{l=1}^6g_l^{ks}q_l+\frac{1}{2}\sum_{r=1}^6\sum_{l=1}^6g_{rl}^{ks}q_rq_l}}; \\
 {\displaystyle{\Delta_{ks}=\overline{\Delta}_0^{ks}+\sum_{l=1}^6\overline{\Delta}_l^{ks}q_l+\frac{1}{2}\sum_{r=1}^6\sum_{l=1}^6\overline{\Delta}_{rl}^{ks}q_rq_l}}; \\
  {\displaystyle{\Delta=\overline{\Delta}_0+\sum_{l=1}^6\overline{\Delta}_lq_l+\frac{1}{2}\sum_{r=1}^6\sum_{l=1}^6\overline{\Delta}_{rl}q_rq_l}},
\end{array}
\end{equation}
where the overbar denotes a complex quantity.
The coefficients of determinants {$\Delta$} and {$\Delta_{ks}$} are complex values due to their definitions (\ref{eq: main determinant}) and (\ref{eq:sub  determinant transponation}), respectively,  and assumed to be expressed in
terms of the inductances $L_{ks}^{o}$, $M_{kj}$ and resistances $R_k$ and $R_{ks}$.

Taking into account:
\begin{equation}\label{eq:Taylor series derivatives}
\begin{array}{l}
 {\displaystyle{
 \frac{\partial M_{kj}}{\partial q_r}=m_r^{kj}+\sum_{l=1}^6m_{rl}^{kj}q_l}};
 \;
 {\displaystyle{\frac{\partial L_{ks}^{o}}{\partial q_r}=g_r^{ks}+\sum_{l=1}^6g_{rl}^{ks}q_l}}; \\
 {\displaystyle{
 \frac{\partial \Delta_{ks}}{\partial q_r}=\overline{\Delta}_r^{ks}+\sum_{l=1}^6\overline{\Delta}_{rl}^{ks}q_l}};
 \;
  {\displaystyle{
  \frac{\partial \Delta}{\partial q_r}=\overline{\Delta}_r+\sum_{l=1}^6\overline{\Delta}_{rl}q_l}},
\end{array}
\end{equation}
and (\ref{eq:Taylor series}), equation (\ref{eq:deriviation of field simplified}) can be linearized as follows:
\begin{equation} \label{eq:deriviation of field linearizing}
\begin{array}{l}{\displaystyle{
\frac{\partial W_m}{\partial q_r} =
-\frac{1}{2}\frac{1}{\overline{\Delta}_0}\sum_{k=1}^m\sum_{j=1}^{n}\left(
m_r^{kj}I_{cj}\sum_{s=1}^n\overline{\Delta}_0^{ks}I_{cs}
+m_0^{kj}I_{cj}\sum_{s=1}^n\overline{\Delta}_r^{ks}I_{cs}\right)
}}\\
{\displaystyle{
+\frac{1}{2}\frac{\overline{\Delta}_r}{\overline{\Delta}_0^2}\sum_{k=1}^m\sum_{j=1}^{n}\left(
m_0^{kj}I_{cj}\sum_{s=1}^n\overline{\Delta}_0^{ks}I_{cs}
\right)
}}\\
{\displaystyle{ -\frac{1}{2}\frac{1}{\overline{\Delta}_0}\sum_l^6\left[\sum_{k=1}^m\sum_{j=1}^{n}\left(
m_{rl}^{kj}I_{cj}\sum_{s=1}^n\overline{\Delta}_0^{ks}I_{cs}
+m_r^{kj}I_{cj}\sum_{s=1}^n\overline{\Delta}_l^{ks}I_{cs}\right.\right.
}}\\
{\displaystyle{
\left.\left.+m_{l}^{kj}I_{cj}\sum_{s=1}^n\overline{\Delta}_r^{ks}I_{cs}
+m_{0}^{kj}I_{cj}\sum_{s=1}^n\overline{\Delta}_{rl}^{ks}I_{cs}\right)\right]\cdot
q_l
}}\\
{\displaystyle{ +\frac{1}{2}\frac{1}{\overline{\Delta}_0^2}\sum_l^6\left[ \overline{\Delta}_{rl}\sum_{k=1}^m\sum_{j=1}^{n}\left(
m_{0}^{kj}I_{cj}\sum_{s=1}^n\overline{\Delta}_0^{ks}I_{cs}\right)
\right]\cdot q_l; \; r=1,\ldots,6.}}
\end{array}
\end{equation}
Accounting for (\ref{eq:deriviation of field linearizing}),
 set (\ref{eq:Lagrange_Maxwell modified}) can be rewritten as
\begin{equation} \label{eq: model}
\left\{\begin{array}{l} {\displaystyle{
M\ddot{q_l}+\mu_l\dot{q_l}+\overline{c}_{l0}+\sum_r^6\overline{c}_{lr}q_r
=F_l; \; l=1,2;
}} \\
{\displaystyle{
M\ddot{q_3}+\mu_3\dot{q_3}+{M}g+\overline{c}_{30}+\sum_r^6\overline{c}_{3r}q_r
=F_3;
}}\\
{\displaystyle{ J_l\ddot{q_l}+\mu_l\dot{q_l}+\overline{c}_{l0}+\sum_r^6\overline{c}_{lr}q_r=T_{l} ; \;l=4,5,6}}, \\
 \end{array}\right.  \end{equation}
 where $\overline{c}_{l0}$ and $\overline{c}_{lr}$ $(l,r=1,\ldots,6)$ are complex coefficients, defined by (\ref{eq:deriviation of field linearizing}). At the equilibrium point, the following  coefficients must hold:
\begin{equation} \label{eq: free terms}
\overline{c}_{30}=-Mg; \; \overline{c}_{l0}=0; \; l=1,2,4,5,6.
\end{equation}
Hence, the final linearized model describing dynamics of micro-machined inductive contactless suspension becomes
\begin{equation} \label{eq: model final}
\left\{\begin{array}{l} {\displaystyle{
M\ddot{q_l}+\mu_l\dot{q_l}+\sum_r^6\overline{c}_{lr}q_r =F_l; \;
l=1,2,3;
}} \\
{\displaystyle{ J_l\ddot{q_l}+\mu_l\dot{q_l}+\sum_r^6\overline{c}_{lr}q_r=T_{l} ; \;l=4,5,6}}. \\
 \end{array}\right.  \end{equation}
Generalized linear model (\ref{eq: model final}) 
developed here, assuming small displacements of the levitated micro-object and its quasi-static behavior, can now be applied to study the dynamics and stability of the micromachined inductive contactless suspension.

\section{Stability of Micromachined Inductive Contactless Suspensions}
Let us represent linear model (\ref{eq: model final})  in matrix form as
\begin{equation}\label{eq: model matrix}
    \mathbf{A}\ddot{\mathbf{\overline{q}}}+\mathbf{B}\dot{\mathbf{\overline{q}}}+\left(\mathbf{R}+\jmath\mathbf{P}\right)\mathbf{\overline{q}}=\mathbf{f},
\end{equation}
where $\mathbf{\overline{q}}=(\overline{q}_1,\ldots,\overline{q}_6)^{T}$ is the column-vector of generalized coordinates, which are complex variables due to (\ref{eq: model final}); $\mathbf{f}=(F_1,F_2,F_3,T_4,T_5,T_6)^{T}$ is the column-vector of generalized forces and torques applied to the micro-object; $\mathbf{A}=\mathrm{diag}(M,M,M,J_4,J_5,J_6)$ is the diagonal matrix of the micro-object mass and its moments of inertia; $\mathbf{B}=\mathrm{diag}(\mu_1,\ldots,\mu_6)$ is the diagonal matrix of damping coefficients; $\mathbf{R}=\left(\mathrm{Re}\{\overline{c}_{lr}\}\right)$ and $\mathbf{P}=\left(\mathrm{Im}\{\overline{c}_{lr}\}\right)$. 


According to Eq.\ (\ref{eq:deriviation of field linearizing}), the complex coefficients can be defined as
\begin{equation} \label{eq:complex coefficients}
\begin{array}{l}
{\displaystyle{\overline{c}_{lr}= -\frac{1}{2}\frac{1}{\overline{\Delta}_0}\sum_{k=1}^m\sum_{j=1}^{n}\left(
m_{rl}^{kj}I_{cj}\sum_{s=1}^n\overline{\Delta}_0^{ks}I_{cs}
+m_r^{kj}I_{cj}\sum_{s=1}^n\overline{\Delta}_l^{ks}I_{cs}\right.
}}\\
{\displaystyle{
\left.+m_{l}^{kj}I_{cj}\sum_{s=1}^n\overline{\Delta}_r^{ks}I_{cs}
+m_{0}^{kj}I_{cj}\sum_{s=1}^n\overline{\Delta}_{rl}^{ks}I_{cs}\right)
}}
{\displaystyle{ +\frac{1}{2}\frac{\overline{\Delta}_{rl}}{\overline{\Delta}_0^2}\sum_{k=1}^m\sum_{j=1}^{n}\left(
m_{0}^{kj}I_{cj}\sum_{s=1}^n\overline{\Delta}_0^{ks}I_{cs}\right).}}
\end{array}
\end{equation}

The physical meanings of matrices $\mathbf{A}$,
$\mathbf{B}$ and $\mathbf{R}$ are obvious. Matrix $\mathbf{P}$ presents the coefficients of the nonconservative positional forces due to the dissipation of eddy currents. Eq.\ (\ref{eq: model matrix})  can be rewritten using only real values, and at the equilibrium point the linear model is equivalent to
\begin{equation}\label{eq: model matrix 12}
\left(
  \begin{array}{c|c}
    \mathbf{A}  & 0 \\
    \hline
    0 & \mathbf{A} \\
  \end{array}
\right)
\left(
    \begin{array}{l}
      \ddot{\mathbf{q}} \\
      \hline
      \ddot{\mathbf{q}}*
    \end{array}
\right)+ \left(
  \begin{array}{c|c}
    \mathbf{B}  & 0 \\
    \hline
    0 & \mathbf{B} \\
  \end{array}
\right) \left(
    \begin{array}{l}
      \dot{\mathbf{q}} \\
      \hline
      \dot{\mathbf{q}}*
    \end{array}
\right) + \left(
  \begin{array}{c|c}
    \mathbf{R}  & -\mathbf{P} \\
    \hline
   \mathbf{P} & \mathbf{R} \\
  \end{array}
\right) \left(
    \begin{array}{l}
      \mathbf{q} \\
      \hline
      \mathbf{q}*
    \end{array}
\right)=0
    ,
\end{equation}
where $(\mathbf{q}|\mathbf{q}*)^{T}$ is the block column-vector of twelve variables; $\mathbf{q}=\Re\{\overline{q}\}$ is the real part of $\overline{q}$; $\mathbf{q}*=\Im\{\overline{q}\}$ is the imaginary part of $\overline{q}$, and all block matrices have 12$\times$12 elements. It is obvious that
\begin{equation}\label{eq: matrix}
    \left(
  \begin{array}{c|c}
    \mathbf{R}  & -\mathbf{P} \\
    \hline
   \mathbf{P} & \mathbf{R} \\
  \end{array}
\right)=
 \left(
  \begin{array}{c|c}
    \mathbf{R}  & 0 \\
    \hline
   0 & \mathbf{R} \\
  \end{array}
\right)+
 \left(
  \begin{array}{c|c}
    0  & -\mathbf{P} \\
    \hline
   \mathbf{P} & 0 \\
  \end{array}
\right),
\end{equation}
and
\begin{equation}\label{eq: matrix skew-sym}
    \left(
  \begin{array}{c|c}
    0  & -\mathbf{P} \\
    \hline
   \mathbf{P} & 0 \\
  \end{array}
\right)=
  -\left(
  \begin{array}{c|c}
    0  & -\mathbf{P} \\
    \hline
   \mathbf{P} & 0 \\
  \end{array}
\right)^{T}
\end{equation}
is a skew-symmetric matrix which corresponds to the positional nonconservative forces.

Analysis of  model (\ref{eq: model matrix}) reveals the following general issues related to  stability of MIS, which are in particular formulated in terms of three theorems, proofs of which are provided  in \ref{app:Th}.

\begin{thm}[Unstable levitation I]\label{thm: instability}
If a micromachined inductive suspension is subjected to only electromagnetic forces defined by (\ref{eq:deriviation of field }) (without dissipation forces, so that $\mathbf{B}=0$), then  stable levitation in this suspension is impossible.
\end{thm}
This fact can be referred to  the main feature of inductive contactless suspension. However, if the levitating micro-object is a   perfect conductor, then $\mathbf{P}=0$. Hence, when matrix $\mathbf{R}$ is  positive definite, stable levitation without dissipative forces becomes possible. Also another obvious conclusion can be formulated in the following corollary.
\begin{clr}\label{clr: instability}
If a micromachined inductive suspension is subjected to only electromagnetic forces, and  the potential part of the electromagnetic forces is absent ($\mathbf{R}=0$), then stable levitation in the suspension is impossible.
\end{clr}
Even if the dissipative force{s} are added to  such {a} system without potential forces, the stable levitation in MIS is still impossible, this fact can be formulated in the second theorem below.
\begin{thm}[Unstable levitation II]\label{thm: instability II}
If a micromachined inductive suspension is subjected to electromagnetic forces having only positional $\mathbf{P}\neq0$ ($\mathbf{R}=0$) and dissipative forces ($\mathbf{B}>0$), then stable levitation is impossible.
\end{thm}

The stable levitation in MIS can be only achieved  by adding the dissipative force. Upon holding the following necessary and sufficient conditions given in the theorem below{,} the suspension can be asymptotically stable.
\begin{thm}[Asymptotically stable levitation]\label{thm: asymptotical stability}
By adding dissipative forces ($\mathbf{B}>0$) to a micromachined inductive suspension subjected to electromagnetic forces defined by (\ref{eq:deriviation of field }) and having a positive definite matrix of potential forces ($\mathbf{R}>0$), the suspension can be asymptotically stable.
\end{thm}
 For the asymptotically stable levitation in a MIS the necessary condition is that  matrix $ \mathbf{A}$, $ \mathbf{B}$ and $\mathbf{R}$ should be positive definite  according to Metelitsyn's inequality \cite[page 32]{SeyranianMailybaev2003,Metelitsyn1952}. The sufficient practical condition for asymptotically stable levitation is
 \begin{equation}\label{eq:sufficient condition}
   \mu_\textrm{min}>p_\textrm{max}\sqrt{a_\textrm{max}/r_\textrm{min}}, 
 \end{equation}
 where $ \mu_\textrm{min}$, and $r_\textrm{min}$ are the respective minimum values of  $\mathbf{B}$ and  $\mathbf{R}$; $p_\textrm{max}$ and $a_\textrm{max}$ are the respective maximum values of  $ \mathbf{P}$ and $ \mathbf{A}$ (please see Theorem \ref{thm: asymptotical stability}  in \ref{app:Th}).

  Operating MIS in  air, inequality (\ref{eq:sufficient condition}) automatically holds due to the fact that damping forces dominate in the micro-world. Note that inequality (\ref{eq:sufficient condition}) should be separately verified upon using the MIS in a vacuum environment.



\begin{figure*}[!t]
    \centering
    \subfigure[]
    {
        \includegraphics[width=1.6in]{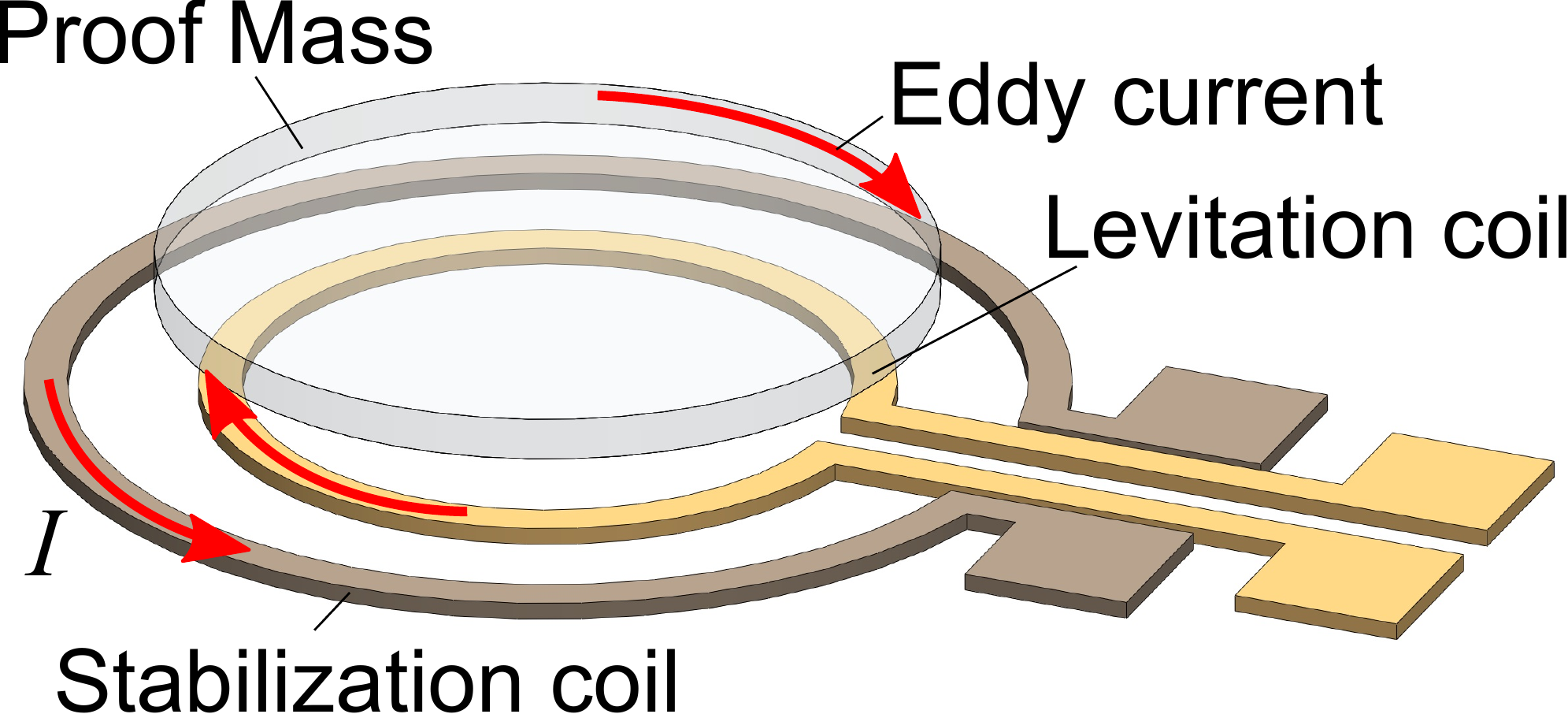}
    \label{fig:Design0}
    }
    \subfigure[]
    {
        \includegraphics[width=1.3in]{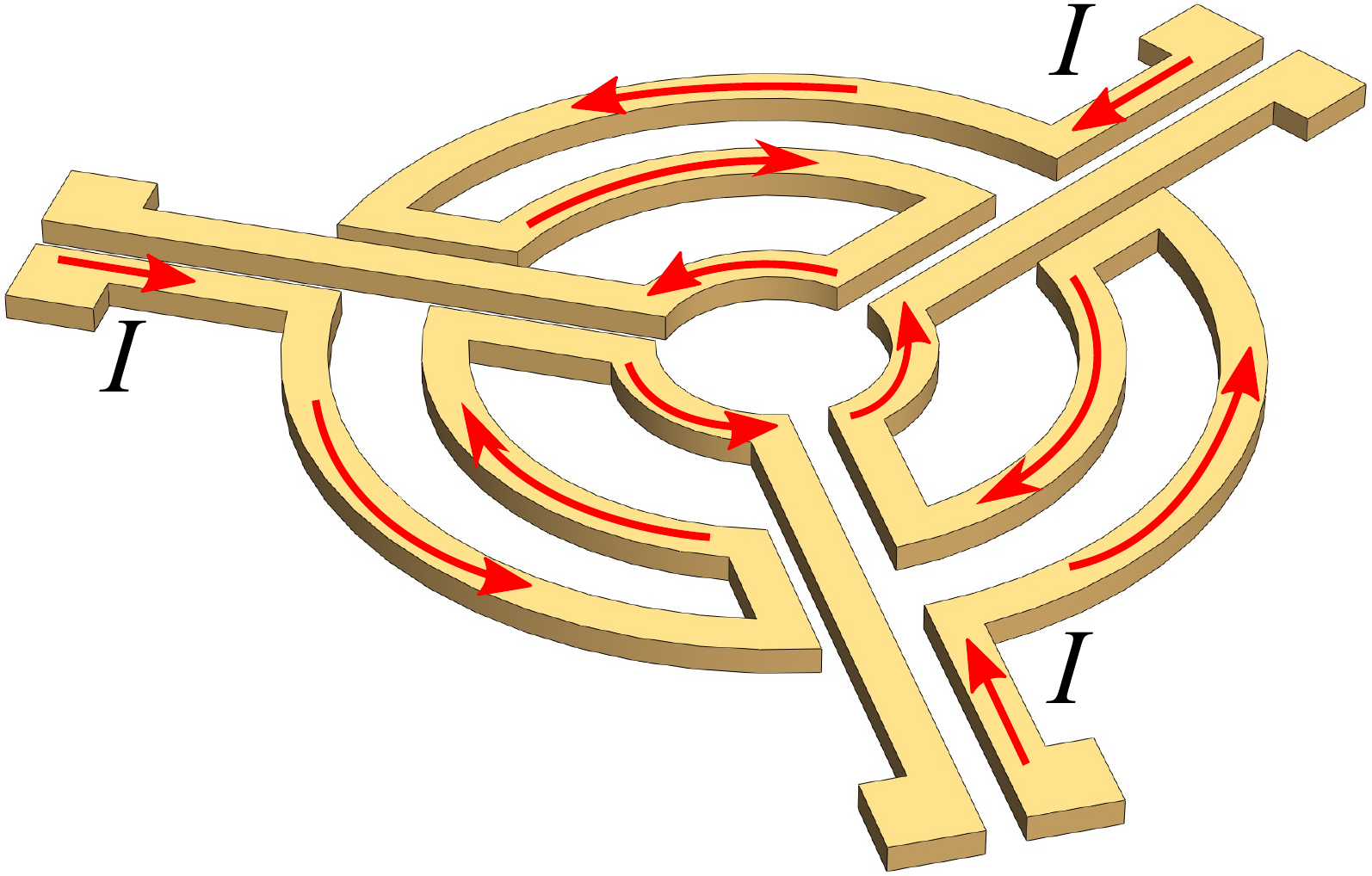}
    \label{fig:Design2}
    }
    \subfigure[]
    {
        \includegraphics[width=1.3in]{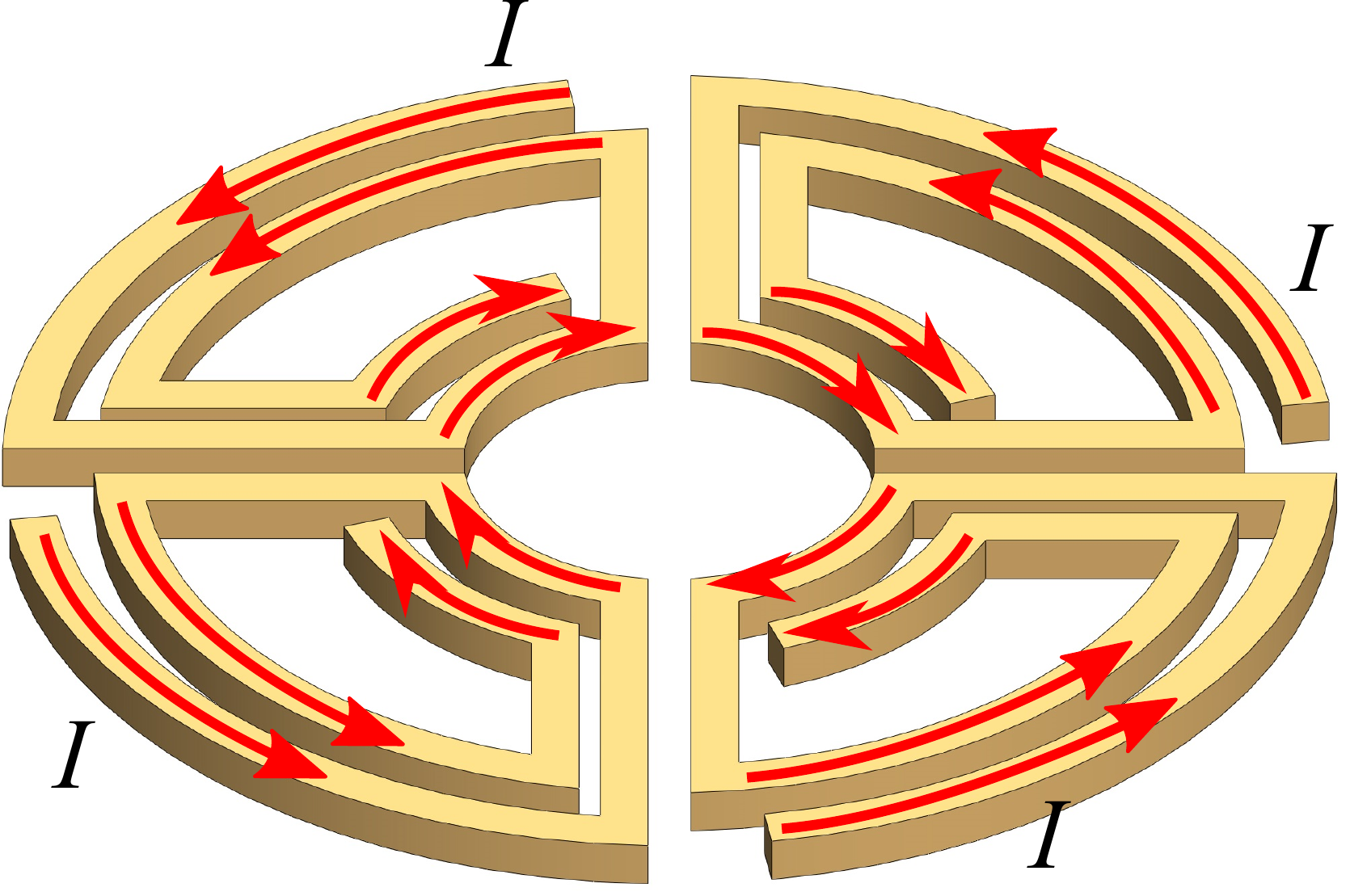}
    \label{fig:Design3}
    }
    \subfigure[]
    {
        \includegraphics[width=1.4in]{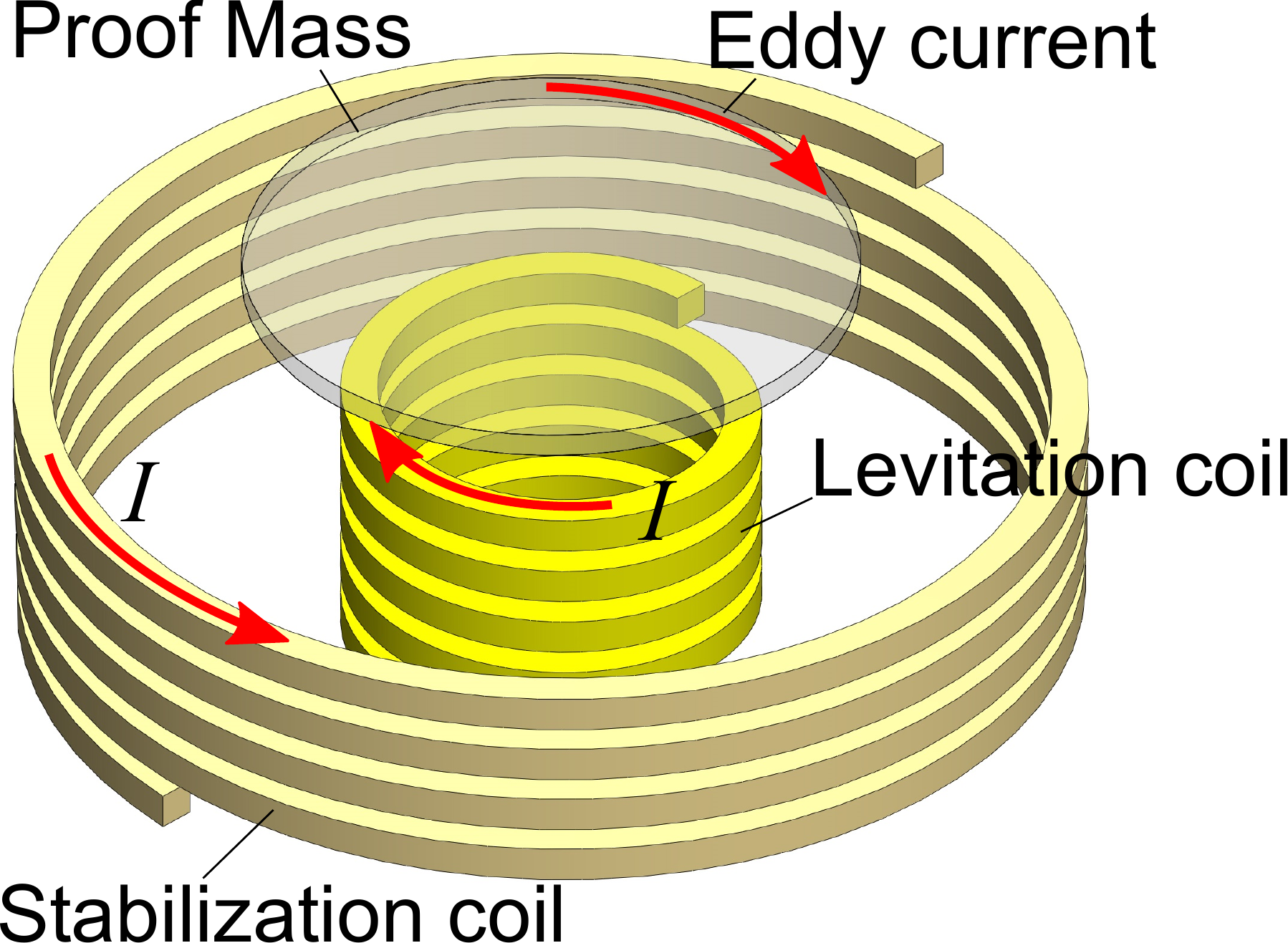}
    \label{fig:Design4}
    }
    \subfigure[]
    {
        \includegraphics[width=1.4in]{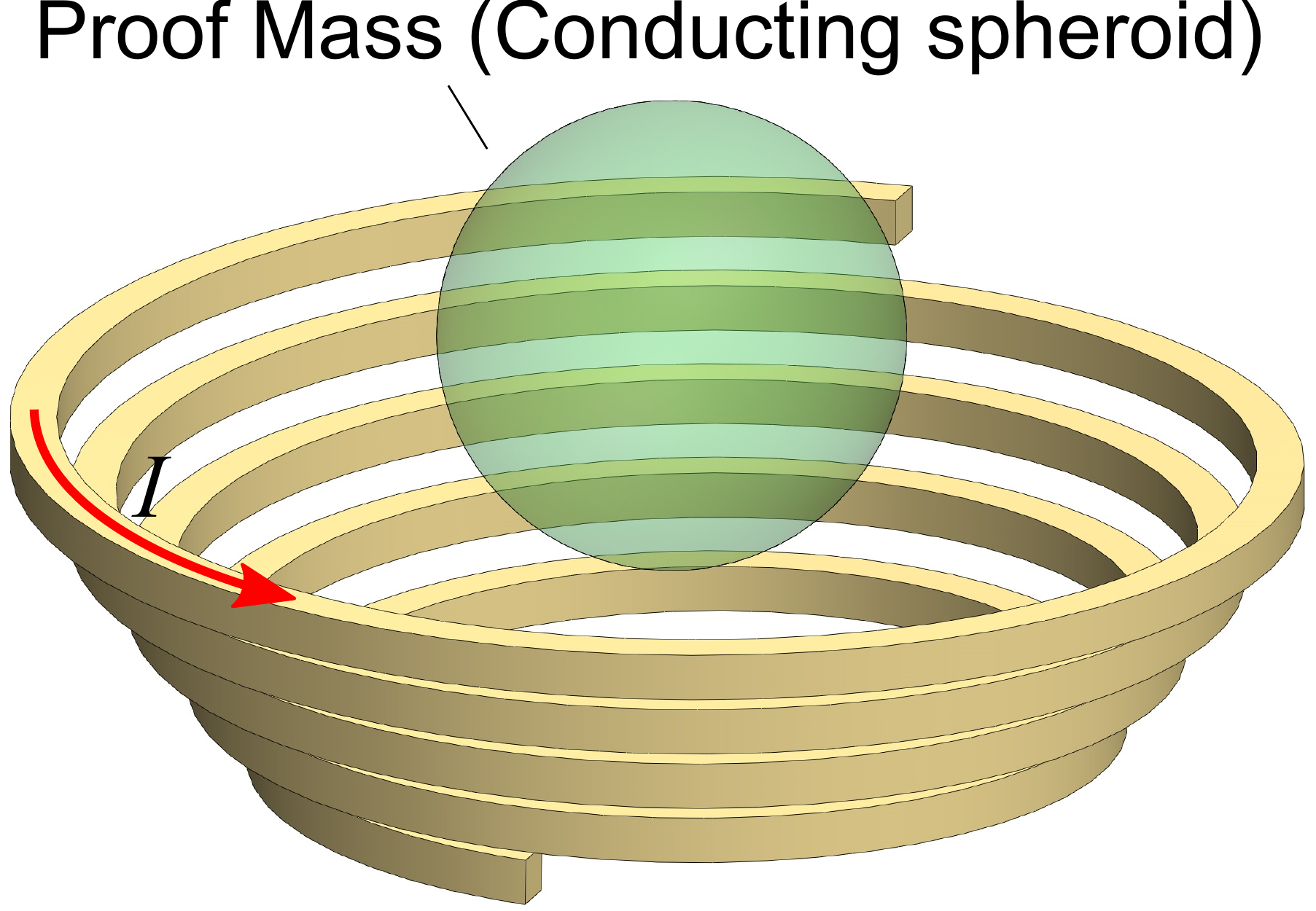}
    \label{fig:Design5}
    }
    \caption{ Axially symmetric designs based on planar {and 3D} micro-coils: $I$ is the electric current. }
    \label{fig:Design planar}
\end{figure*}

\section{Various Designs of Inductive Contactless Suspensions}

In this section, we apply the qualitative approach developed above to analyze the dynamics and stability of several symmetric and axially symmetric designs of micromachined inductive suspensions.

A variety of axially symmetric designs of inductive suspensions based on planar and 3D micro-coils are shown in Fig.~\ref{fig:Design planar}. In particular, the MIS design shown in Fig.~\ref{fig:Design planar}(a) was utilized in the suspension prototype reported in \cite{Williams1996} and proposed for its potential application as a gyroscope. The designs shown in Fig.~\ref{fig:Design planar}(b) and (c) were employed in micro-gyroscope prototypes reported in \cite{WilliamsShearwoodMellorEtAl1997} and \cite{Shearwood2000,Zhang2006} in which the rotation of a disk-shaped rotor was demonstrated. The design of a MIS based on 3D micro-coils shown in Fig.~\ref{fig:Design
planar}(d) was realized in the prototype reported in \cite{Badilita2009}. Fig.~\ref{fig:Design
planar}(e) shows the possible design of a MIS based on spiral shaped 3D micro-coils in order to levitate, for instance, a conducting micro-sphere.

\begin{figure*}[!t]
    \centering
      \subfigure[]
    {
        \includegraphics[width=2.0in]{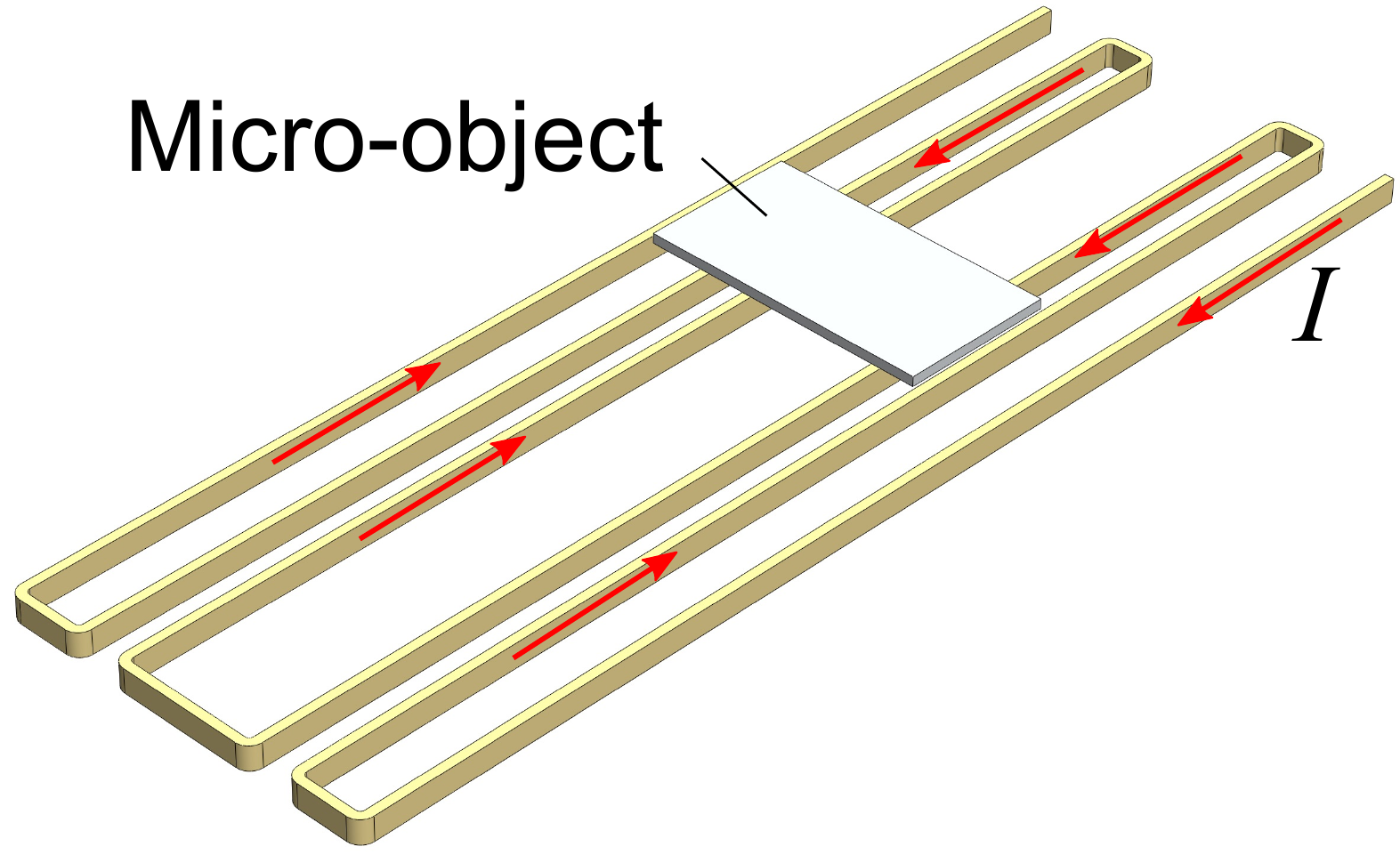}
    \label{fig:Design4}
    }
     \subfigure[]
    {
        \includegraphics[width=1.8in]{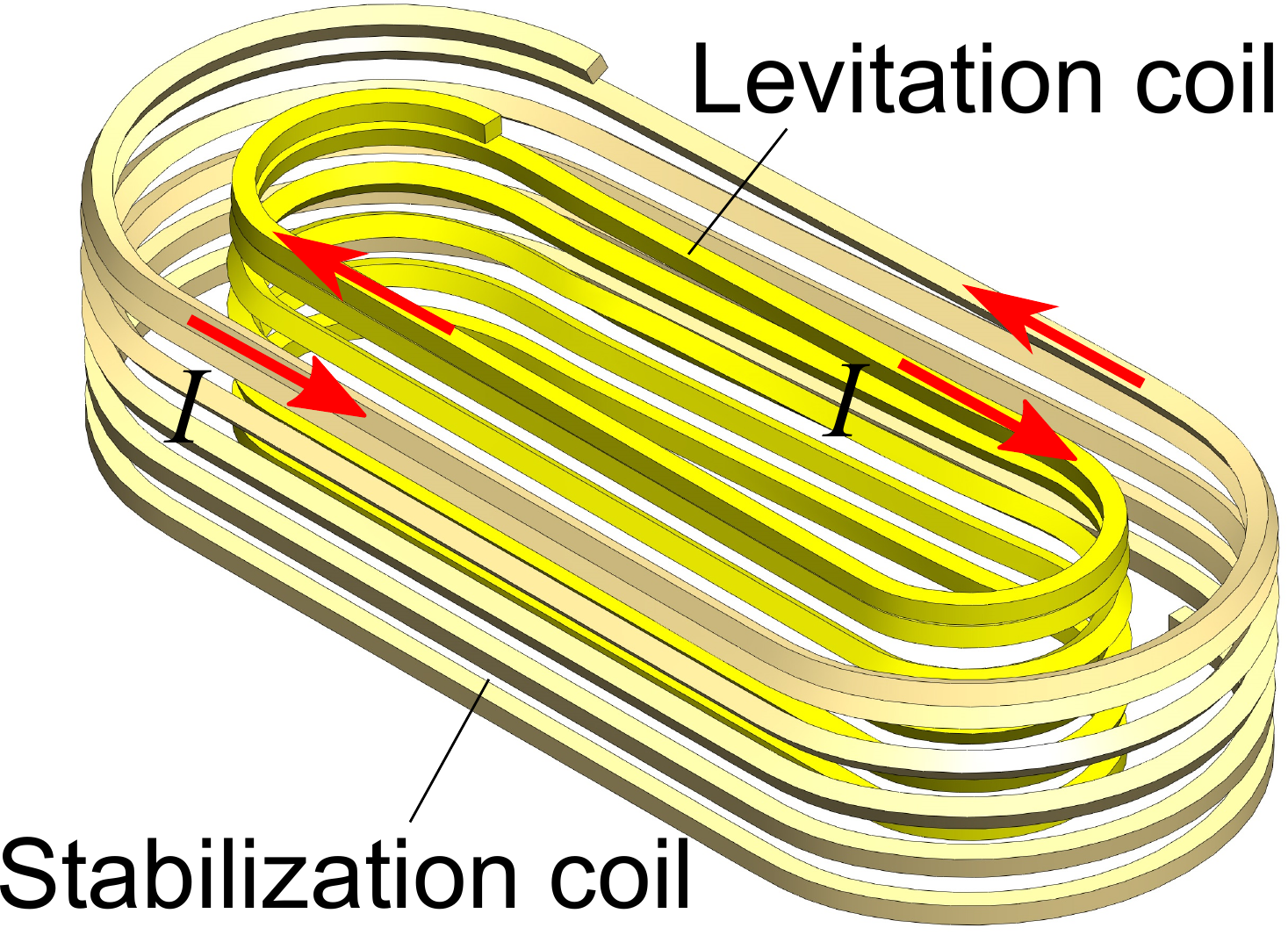}
    \label{fig:Design3}
    }
   \caption{ Symmetric designs  based on planar and 3D micro-coils{: $I$ is the electric current}. }
    \label{fig:Design sym}
\end{figure*}

Examples of MIS symmetric designs are shown in Fig.~\ref{fig:Design sym}. 
The design shown in Fig.~\ref{fig:Design sym}(a) was recently utilized in a prototype of accelerator for sorting micro-objects \cite{Sari2014}. Fig.~\ref{fig:Design sym}(b) presents the design based on 3D micro-coils, which can be employed as a linear-transporter of micro-objects. The prototype based on this design will be demonstrated below and its stability will be studied theoretically and experimentally.

\begin{figure*}[!b]
    \centering
     \subfigure[]
    {
        \includegraphics[width=1.2in]{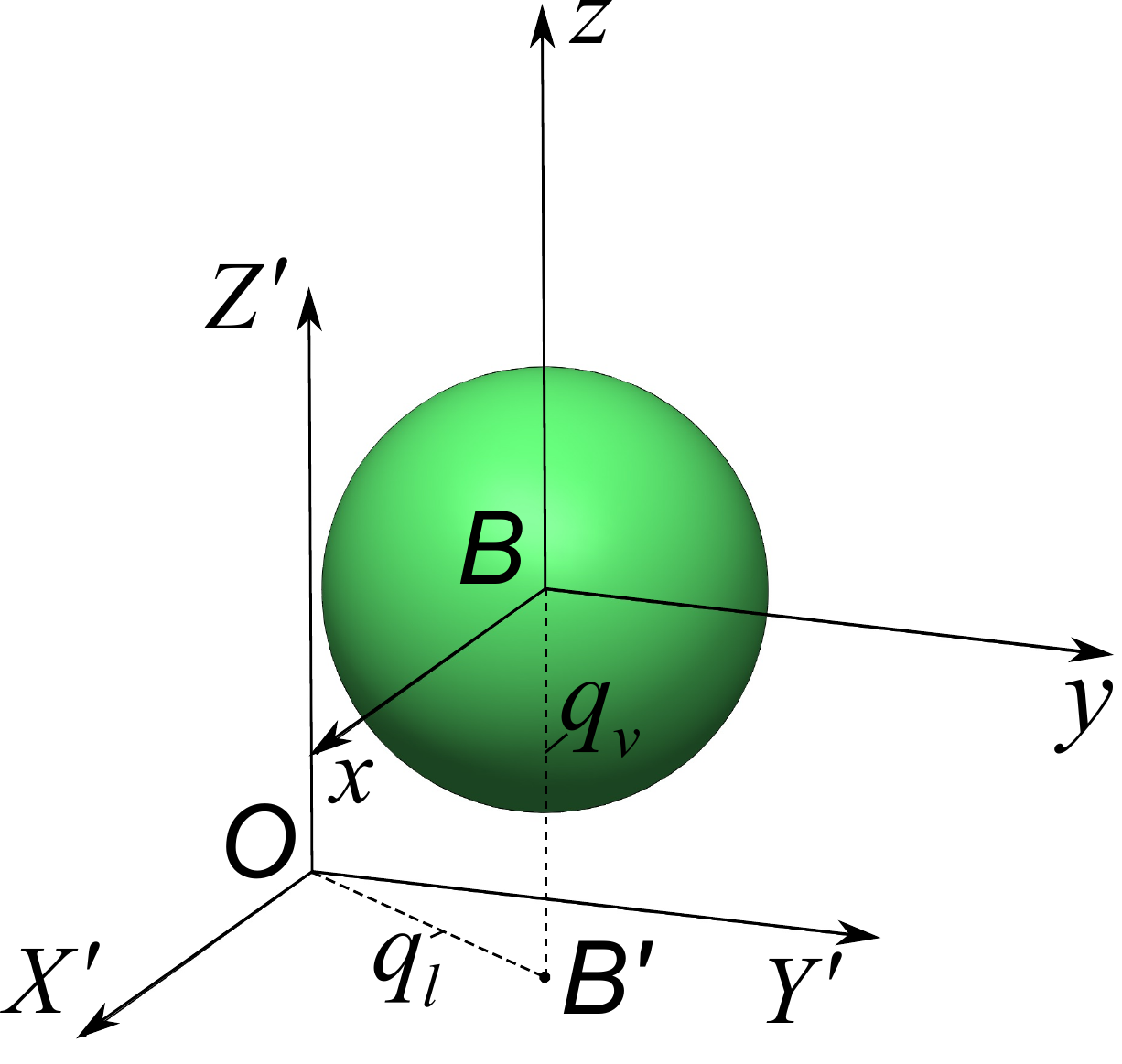}
    \label{fig:CF sphere}
    }
     \subfigure[]
    {
        \includegraphics[width=1.2in]{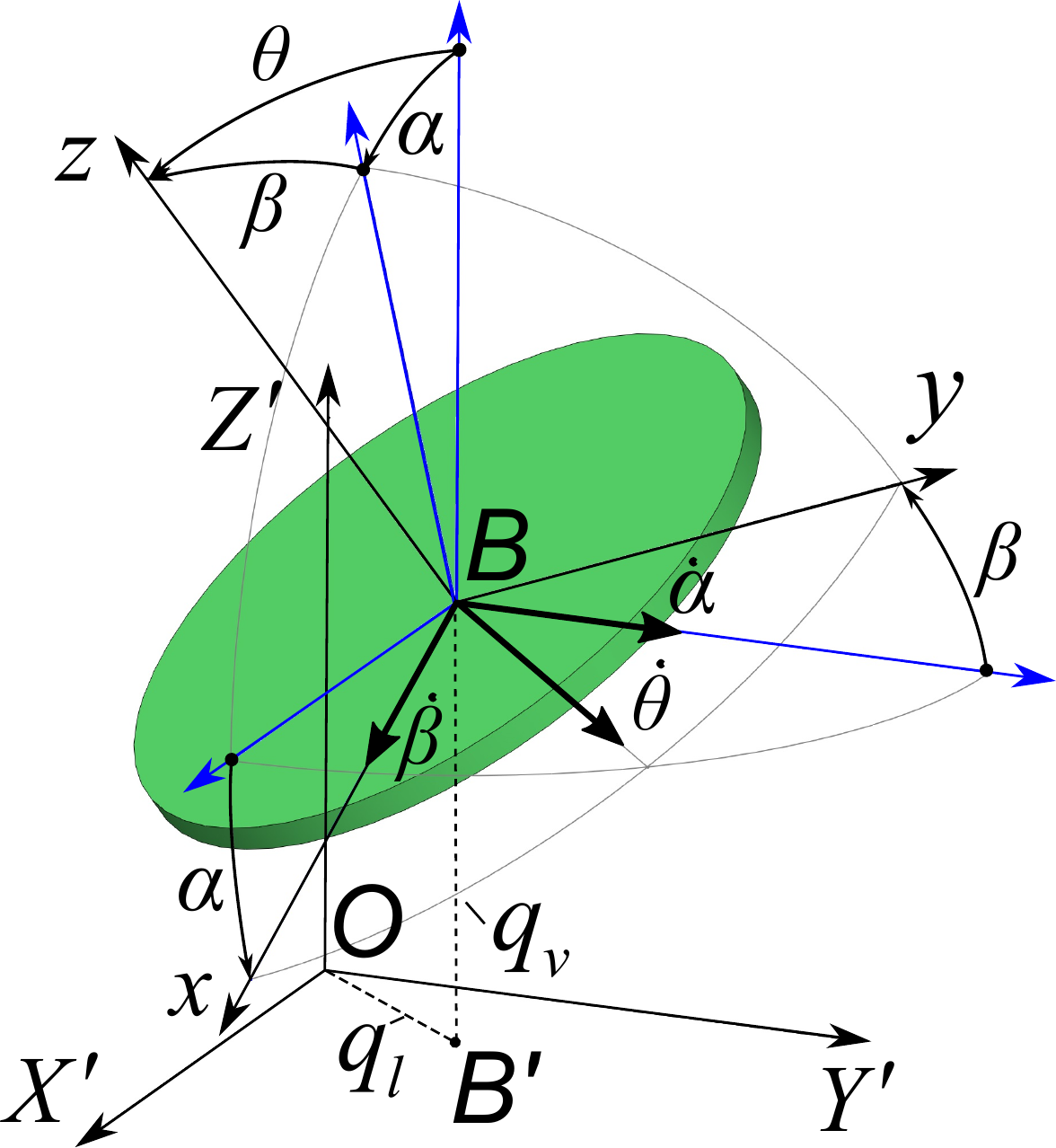}
    \label{fig:CF disk}
    }
    \subfigure[]
    {
        \includegraphics[width=1.2in]{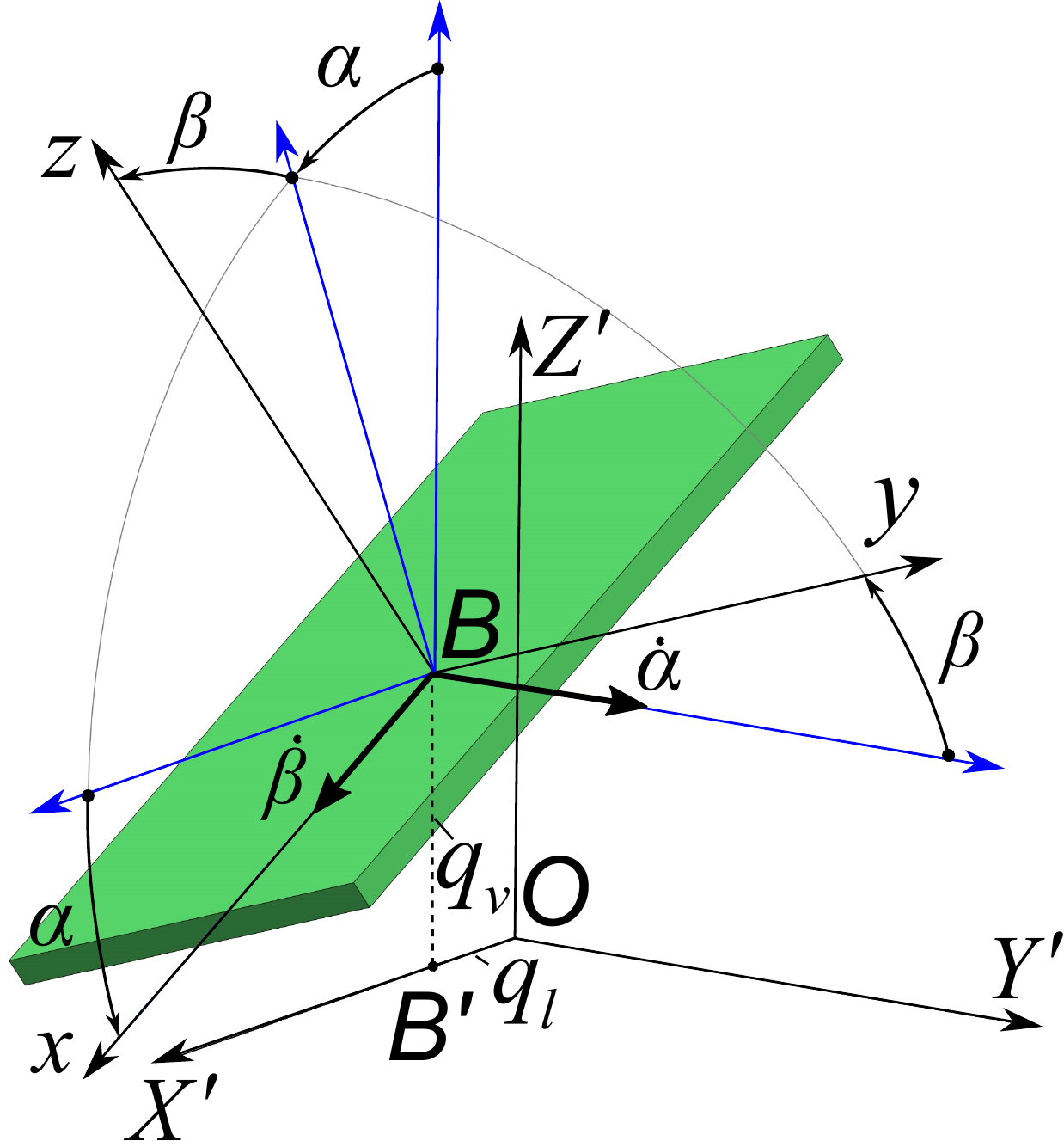}
    \label{fig:CF rec}
    }
    \caption{ Coordinate frames and generalized coordinates to define the position of spherical, disk and rectangle shaped proof masses for axial{ly symmetric} and symmetric designs. }
    \label{fig: coordinates}
\end{figure*}
\begin{table}[!b]
  \caption{The  structures of the analytical model of the suspension  as a function of design.}\label{tab:model structure v}
  \centering
  \begin{tabular}{lcl}
    \toprule
   Design & Levitating micro-object & Model Structure\\
  \midrule
 \multirow{2}{*}{Axially symmetric} &\includegraphics[width=0.3in]{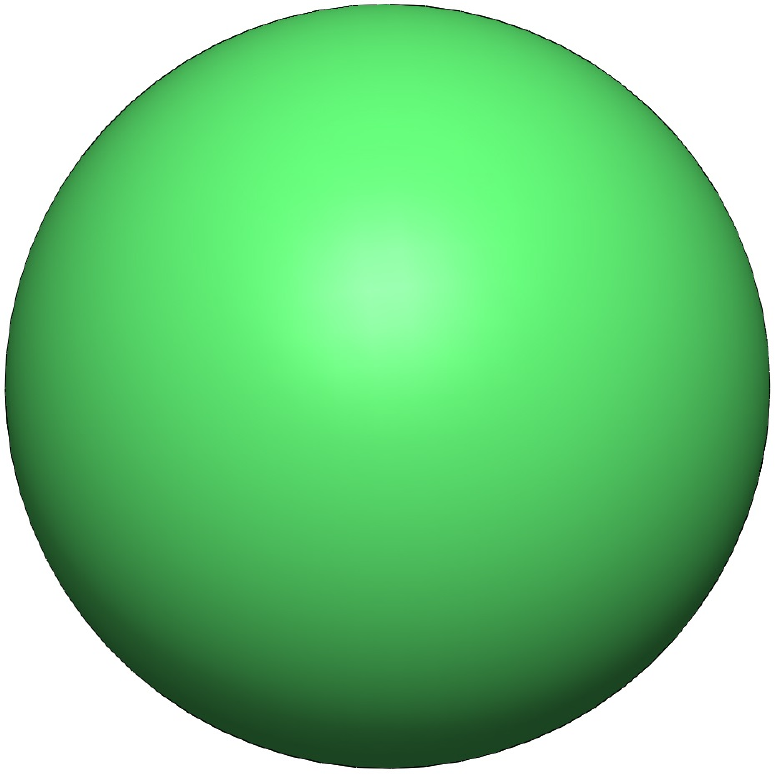}& $ \begin{footnotesize}
        \left\{\begin{array}{l} {\displaystyle{
        {\scriptstyle M\ddot{q_v}+\mu_v\dot{q_v}+\overline{c}_{vv}q_v =F_v;
         }}} \\
        {\displaystyle{
        {\scriptstyle M\ddot{q_l}+\mu_l\dot{q_l}+\overline{c}_{ll}q_l=F_{l} .}}} \\
         \end{array}\right.  \end{footnotesize}$ \\
        \cmidrule(r){2-3}
   &  \includegraphics[width=0.5in]{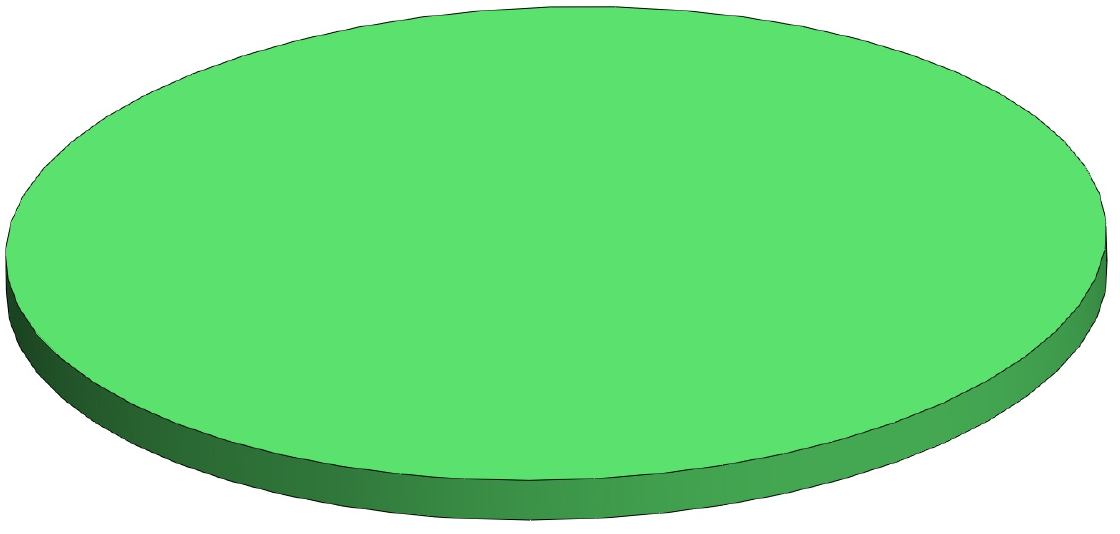}  &
  $ \begin{footnotesize}
    \left\{\begin{array}{l} {\displaystyle{
{\scriptstyle M\ddot{q_v}+\mu_v\dot{q_v}+\overline{c}_{vv}q_v =F_v;}}} \\
{\displaystyle{
{\scriptstyle M\ddot{q_l}+\mu_l\dot{q_l}+\overline{c}_{ll}q_l+\overline{c}_{l\theta}\theta=F_{l} ;}}}\\
{\displaystyle{
{\scriptstyle J_{\theta}\ddot{\theta}+\mu_{\theta}\dot{\theta}+\overline{c}_{\theta l}q_l+\overline{c}_{\theta \theta}\theta=T_{\theta} .}}}\\
 \end{array}\right.
 \end{footnotesize} $ \\   \midrule
 Symmetric    &
 \includegraphics[width=0.5in]{fig//Disk.pdf} $\;$ $\;$ \includegraphics[width=0.6in]{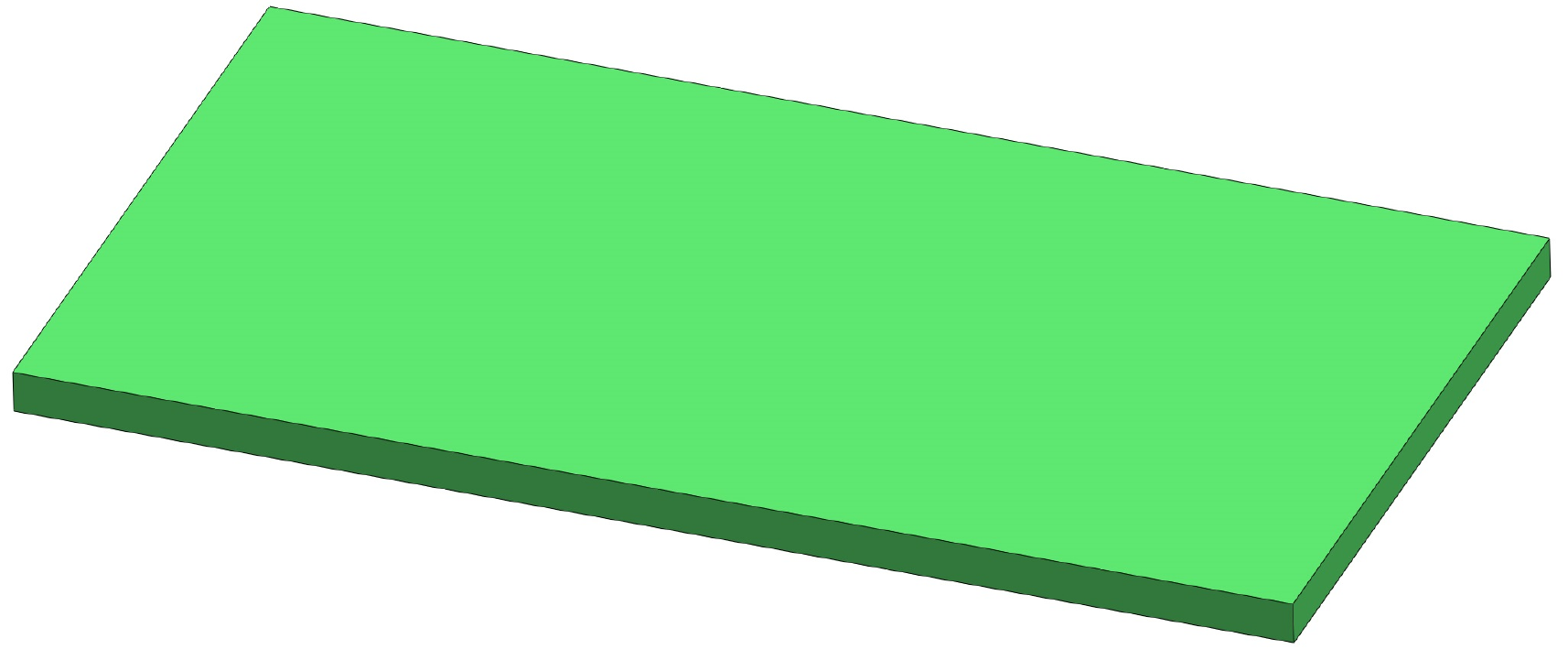}
 & $ \begin{footnotesize}\left\{\begin{array}{l} {\displaystyle{
{\scriptstyle M\ddot{q_v}+\mu_v\dot{q_v}+\overline{c}_{vv}q_v =F_v;}}} \\
{\displaystyle{
{\scriptstyle M\ddot{q_l}+\mu_l\dot{q_l}+\overline{c}_{ll}q_l+\overline{c}_{l\alpha}\alpha=F_{l} ;
}}}
\\
{\displaystyle{
{\scriptstyle J_{\alpha}\ddot{\alpha}+\mu_{\alpha}\dot{\alpha}+\overline{c}_{\alpha l}q_l+\overline{c}_{\alpha \alpha}\alpha=T_{\alpha} ;
}}}
\\
{\displaystyle{
{\scriptstyle J_{\beta}\ddot{\beta}+\mu_{\beta}\dot{\beta}+\overline{c}_{\beta \beta}\beta=T_{\beta} .
}}}
 \end{array}\right.\end{footnotesize} $  \\
    \bottomrule
  \end{tabular}
\end{table}
Due to the symmetry in the considered designs, the number of equations in  set (\ref{eq: model final}) can be reduced. For the case of axially symmetric designs and a spherical proof mass, the position of the levitated sphere is described by two generalized coordinates, namely $q_v$  and $q_l$ representing the vertical and lateral linear displacements, as shown in Fig.~\ref{fig: coordinates}(a). Let us assign the origin of the coordinate frame $X'Y'Z'$  to the equilibrium point, $O$, in such a way that the $Z'$ axis is parallel to the $Z$ axis. The coordinate frame $xyz$ is assigned to the mass center of the proof mass. Then the generalized coordinate $q_v$ characterizes the linear displacement of sphere's centre-of-mass, parallel to the $Z'$ axis from the $X'Y'$ surface. The generalized coordinate $q_l$ characterizes the linear displacement of the sphere's centre-of-mass on the $X'Y'$ surface from the $O$ point. Hence, the model is reduced to a set of two equations. The behaviour of the disk-shaped proof mass without rotation can be described by three generalized coordinates \cite{Poletkin2013}. In addition to the two linear coordinates, $q_l$ and $q_v$, the angular generalized coordinate, $\theta$ is used as shown in Fig.~\ref{fig: coordinates}(b). For the symmetric designs shown in Fig.~\ref{fig:Design sym}, it can be assumed that the levitated micro-object is in a neutral equilibrium state along the transportation line. Directing the $Y'$ axis parallel to this line of transportation and locating {the point, $O$,} on the symmetry axis of the design, the generalized coordinates can be introduced as shown in Fig.~\ref{fig: coordinates}(c). The generalized coordinates $q_l$ and $q_v$ characterize the linear displacement of the micro-object along the $X'$ axis and the vertical one parallel to the   $Z'$ axis, respectively{, while}  two generalized coordinates $\alpha$ and $\beta$  characterize its angular position.

 Thus, depending on the design and the shape of the levitating micro-object, the model structures describing the behaviour of the MIS, in particular the number of equations and elements of the complex matrix $\mathbf{\overline{C}}$ are already known from the defined generalized coordinates above and summarized in Table \ref{tab:model structure v}.

We suggest the following procedure for {designing MIS}. Assuming that a micromachined inductive suspension is intended for using in air, the application of our approach is reduced to the analysis of the coefficients of matrix $\mathbf{R}=\left(\mathrm{Re}\{\overline{c}_{lr}\}\right)$, whose elements are defined in (\ref{eq:complex coefficients}) as
functions of the design parameters. A result of this analysis would be to find the domains of these design parameters where the matrix $\mathbf{R}>0$ is positive definite, or to demonstrate that such domains do not exist ($\mathbf{R}<0$ is everywhere negative definite). Additionally, for a vacuum environment, it becomes necessary to define the coefficients of the matrix $\mathbf{P}=\left(\mathrm{Im}\{\overline{c}_{lr}\}\right)$, which give the required values of the damping coefficients, $\mu_r$, in order to fulfill the condition for stable
levitation as defined in Theorem \ref{thm: asymptotical stability}.

\subsection{Axially Symmetric Design}
\begin{figure*}[!b]
    \centering
     \subfigure[]
    {
        \includegraphics[width=1.8in]{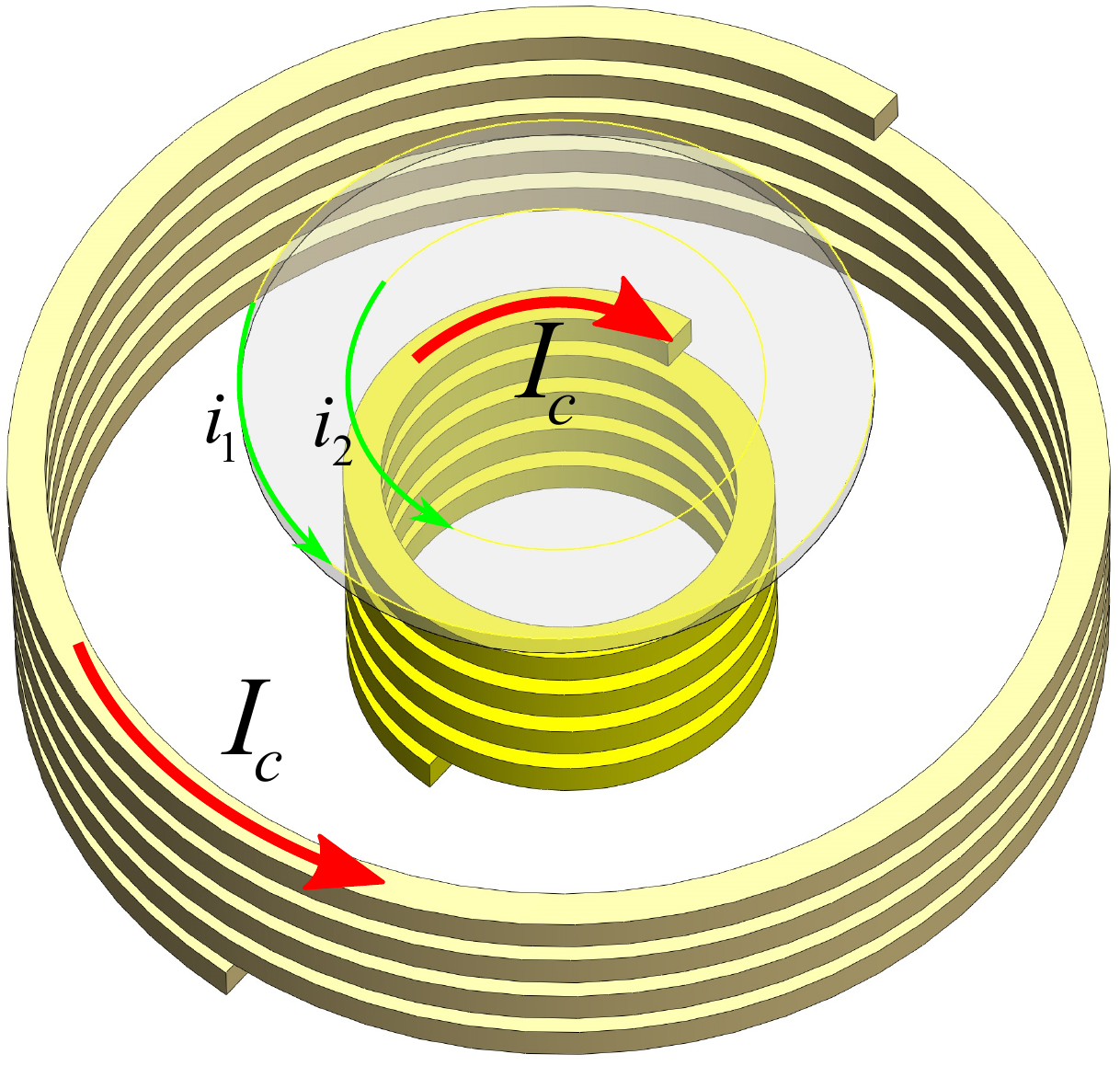}
    \label{fig:two eddy currents}
    }
     \subfigure[]
    {
        \includegraphics[width=1.8in]{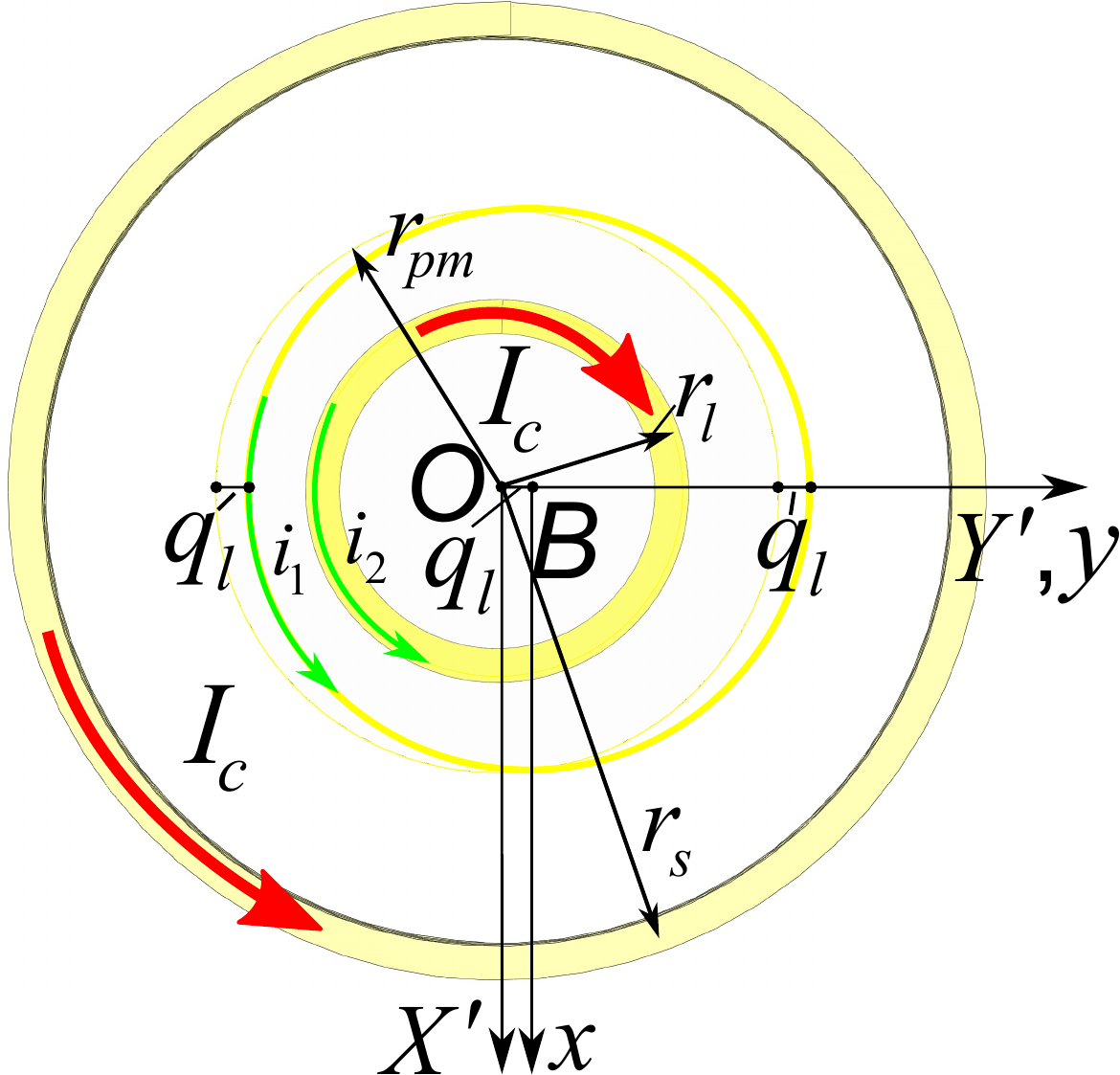}
    \label{fig:behaviour of center eddy currnet circuit}
    }
    \caption{Schematic of 3D micromachined inductive suspension with two representative circuits for the induced eddy current.}
    \label{fig: two eddy current scheme}
\end{figure*}

In this section, the micromachined inductive suspension design based on 3D micro-coils shown in Fig.~\ref{fig:Design planar}(d) is analyzed. Using the theoretical model presented above, two eddy current circuits are taken into account instead of one as was done in {our}  previous study \cite{Lu2014}. As a result, when evaluating its dynamics and stability, an improvement in accuracy will be demonstrated, without introducing any coefficients of similarity. As it was shown in \cite{Lu2012}, the induced eddy currents are distributed along the levitated proof mass in such a way that two circuits having maximum values of eddy current {density} can be identified. Hence, the eddy current circuit can be represented as shown in Fig.~\ref{fig:two eddy currents}. The eddy current circuit $i_1$ is the same as was given in \cite{Poletkin2013} and this current flows along the edge of the proof mass. At the same time, the circuit for eddy current $i_2$ is defined by the levitation coil and has a circular path with radius equal to the radius of the levitation coil. Unlike the $i_1$ circuit, the position in space of the $i_2$ circuit is dependent only on the two generalized coordinates $\theta$ and $q_v$, and independent on the lateral displacement $q_l$ of the proof mass, as shown in Fig.~\ref{fig:behaviour of center eddy currnet circuit}. This figure presents the case for which the lateral displacement of the proof mass takes place along the $Y'$ axis. Due to the fact that the position of the $i_1$ circuit with respect to the $i_2$ circuit within the micro-object is variable, both sources of ponderomotive forces defined in (\ref{eq:deriviation of field simplified}) are acting on the proof mass.

\begin{figure*}[!t]
    \centering
     \subfigure[]
    {
        \includegraphics[width=1.4in]{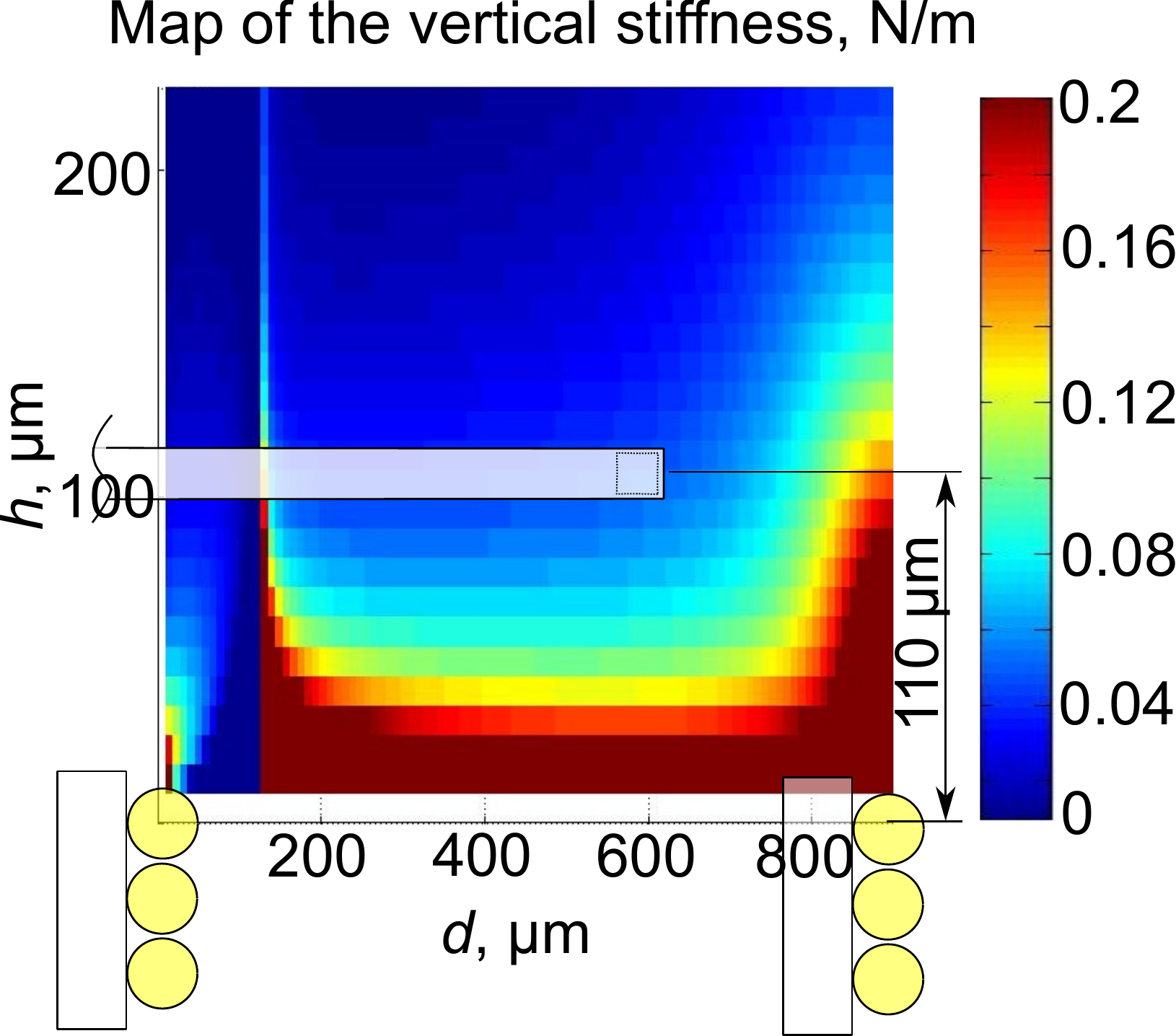}
    \label{fig:st ver axsym}
    }
     \subfigure[]
    {
        \includegraphics[width=1.4in]{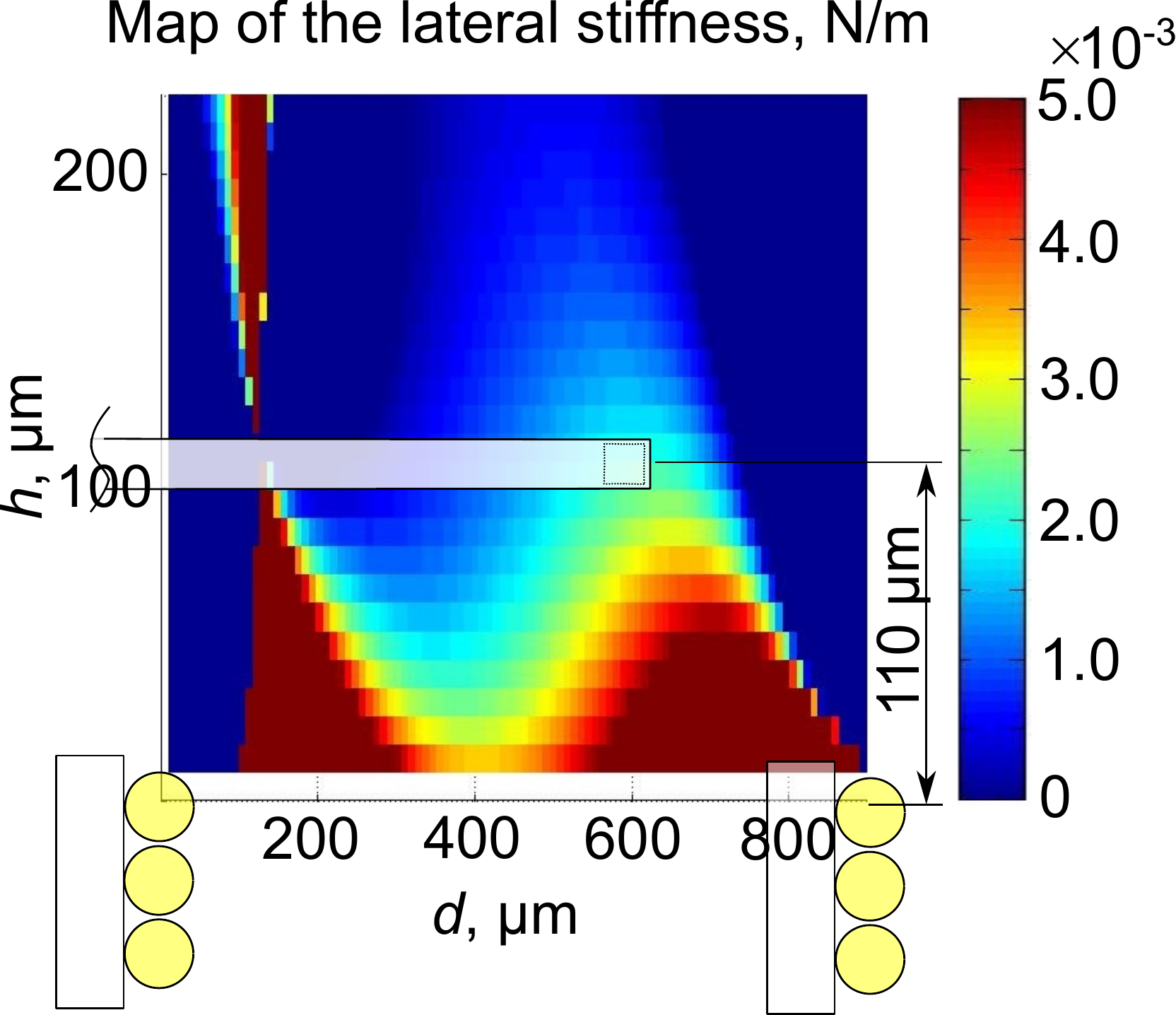}
    \label{fig:st lat axsym}
    }
    \subfigure[]
    {
        \includegraphics[width=1.4in]{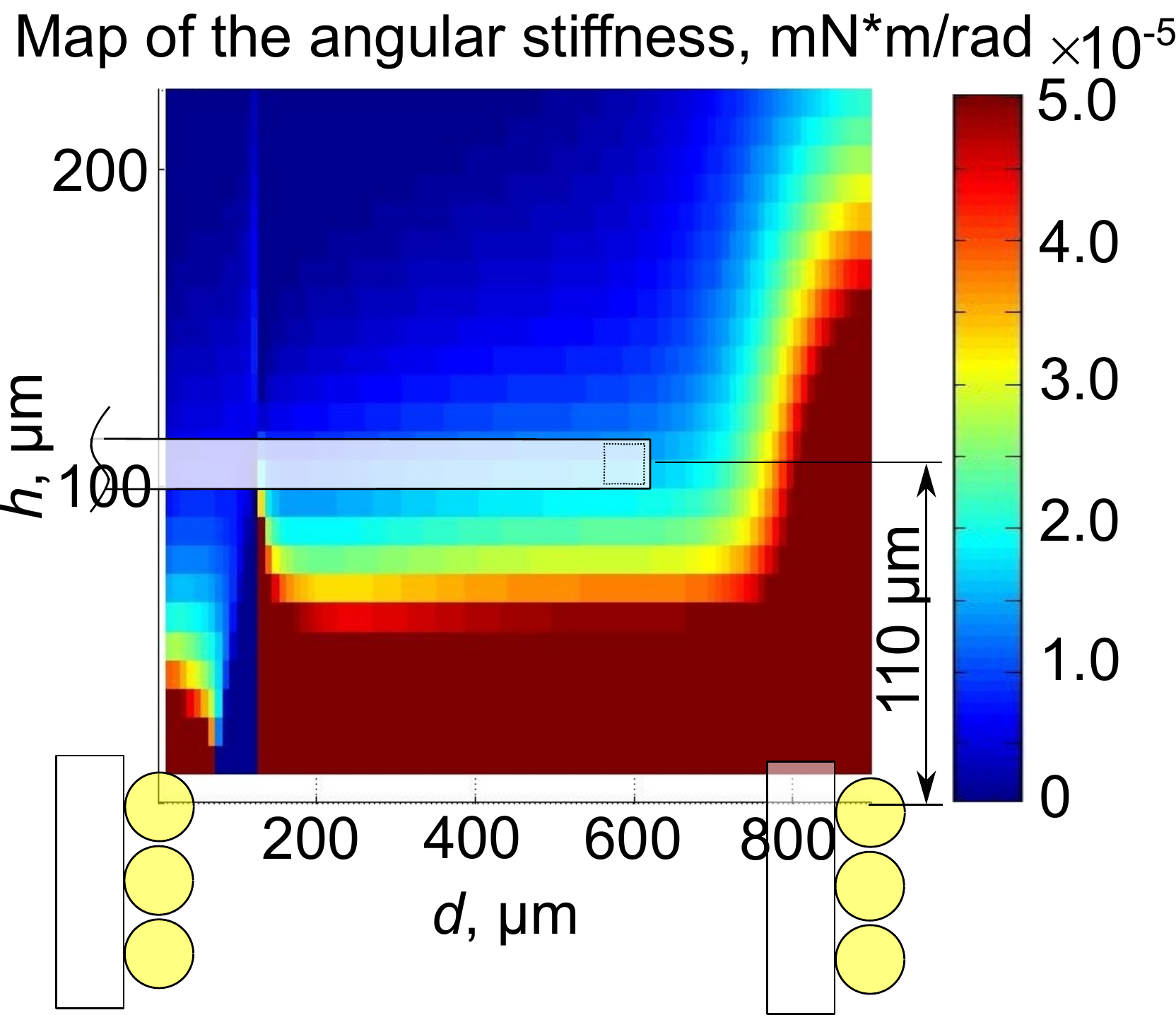}
    \label{fig:st ang axsym}
    }
    \caption{Maps of vertical, lateral and angular stiffness coefficients of the suspension; the square with dashed lines is the area of interest for calculation. }
    \label{fig: map of stiffness of axsym}
\end{figure*}

\begin{table}[!b]
\caption{Comparison of suspension stiffness from modelling and experiment results. }
\centering
\begin{tabular}{lccc}
\toprule
Stiffness  & Measured values &  Modelling& New model\\
component &      reported in \cite{Lu2014}            &   reported in \cite{Lu2014}  &   this work              \\
\midrule
Lateral, [N$\cdot$m$^{-1}$] &3.0$\times$10$^{-3}$  & 3.0$\times$10$^{-3}$&2.0$\times$10$^{-3}$\\
Vertical, [N$\cdot$m$^{-1}$] & 4.5$\times$10$^{-2}$& 4.2$\times$10$^{-2}$& 4.5$\times$10$^{-2}$\\
Angular, [m$\cdot$N$\cdot$rad$^{-1}$] & 1.5$\times$10$^{-8}$ & 0.8$\times$10$^{-8}$&1.4$\times$10$^{-8}$\\
\bottomrule
\end{tabular}
\label{tb:mechanical}
\end{table}

Now let us assume that the suspension is operated in air. Hence, the elements of the matrix $\mathbf{R}=\left(\mathrm{Re}\{\overline{c}_{lr}\}\right)$ will be defined as functions of the design parameters, which are the same as previously proposed  in \cite{Poletkin2014a,Lu2014}. Thus,  $d=r_{pm}-r_l$ is the difference between the radius of the disk-shaped proof mass and the radius of the levitation coil (see, Fig.~\ref{fig:behaviour of center eddy currnet circuit}), and $h$ is the levitation height. In order to calculate the stiffness elements as described by (\ref{eq:complex coefficients}),  equations (\ref{eq:Taylor series}) are compiled as shown in \ref{app:A}. We are using the same design of 3D MIS and the experimental parameters  given in Ref. \cite{Lu2014}: the radii of the stabilization and levitation coils are 1.9 and 1.0 mm, respectively; the pitch of coil winding is 25 $\mu$m; the number of windings for the stabilization and levitation coils are 12 and 20, respectively; the radius of the proof mass is 1.6 mm. For an excitation current of 109 mA in both coils, the maps of the suspension stiffness coefficients are shown in Fig.~\ref{fig: map of stiffness of axsym}. The results of the calculation are shown in Table~\ref{tb:mechanical} together with experimental and modelling results published in \cite{Lu2014} in order to enable the direct comparison. The analysis of Table~\ref{tb:mechanical} shows that the developed technique allows us to evaluate the stiffness with a good enough accuracy without  similarity coefficients using in \cite{Lu2014}.

\subsection{Symmetric Design. Micro-linear Transporter Based on 3D Micro-coils}
\begin{figure*}[!t]
    \centering
         {
        \includegraphics[width=4.6in]{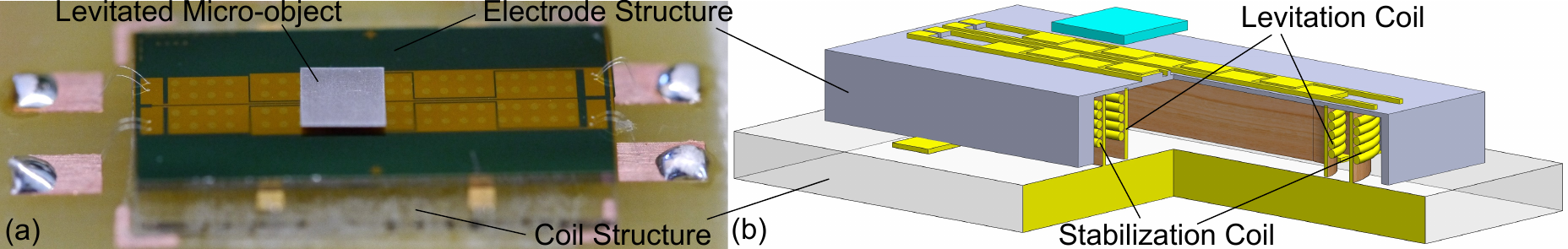}
       }
    \caption{ Micro-transporter: a)  the prototype of micro-transporter glued onto a PCB
under experimental testing; b) the schematic of the micro-transporter design.  }
    \label{fig: Symmetric design}
\end{figure*}

\begin{figure*}[!b]
    \centering
         {
        \includegraphics[width=2.3in]{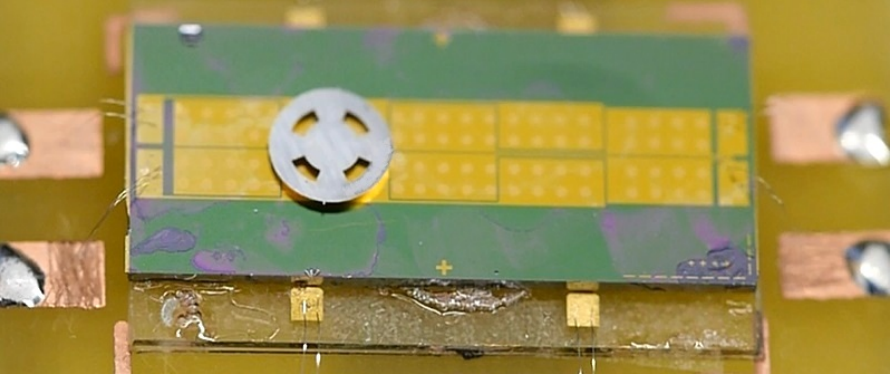}
       }
    \caption{ Successful levitation of disk shaped micro-object.  }
    \label{fig: Symmetric design disk shape}
\end{figure*}

In this section, we analyze the symmetric design based on 3D micro-coils for its potential application as a micro-transporter, shown in Fig.~\ref{fig:Design3}. Fig.~\ref{fig: Symmetric design} shows the prototype of such a micro-transporter, together with a schematic cut-away drawing. The micro-transporter consists of two structures fabricated independently, namely the coil structure and the electrode structure, which are aligned and assembled by flip-chip bonding into one device with dimensions 20 mm $\times$ 7.0~mm $\times$ 1~mm as shown in Fig.~\ref{fig: Symmetric design}(a).  Fig.~\ref{fig: Symmetric design}(b) illustrates the interiour of the micro-transporter design in a sectional view which presents the position of the 3D-coils inside the electrode structure.

The coil structure consists of two racetrack shaped solenoidal 3D wire-bonded microcoils to be used as Maglev rails, namely stabilization and levitation coils, fabricated on a Pyrex substrate using SU-8 2150 and UV photolithography. For electrostatic propelling of the micro-object, an array of electrodes is fabricated on a 510 $\mu$m thick silicon substrate having an oxide layer for passivation. Electrodes are patterned on top of the oxide layer by UV lithography on evaporated Cr/Au layers (20/150nm). The prototype provides stable levitation for a rectangular- as well as a disk-shaped micro-object as shown in Fig.~\ref{fig: Symmetric design}(a) and Fig.~\ref{fig: Symmetric design disk shape}.
\begin{figure*}[!b]
    \centering
     \subfigure[]
    {
        \includegraphics[width=2.0in]{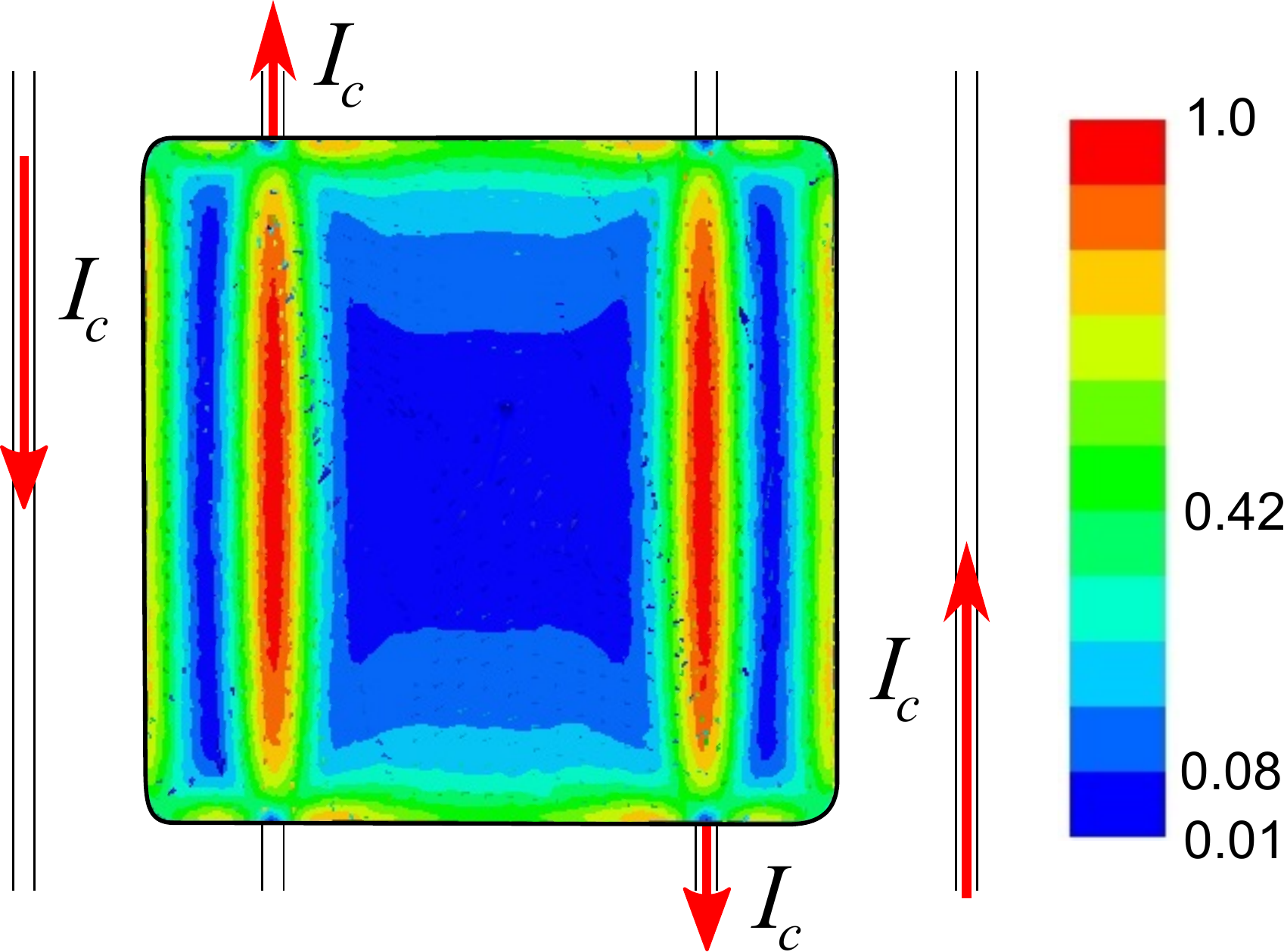}
    \label{fig: eddy currents density}
    }
     \subfigure[]
    {
        \includegraphics[width=1.5in]{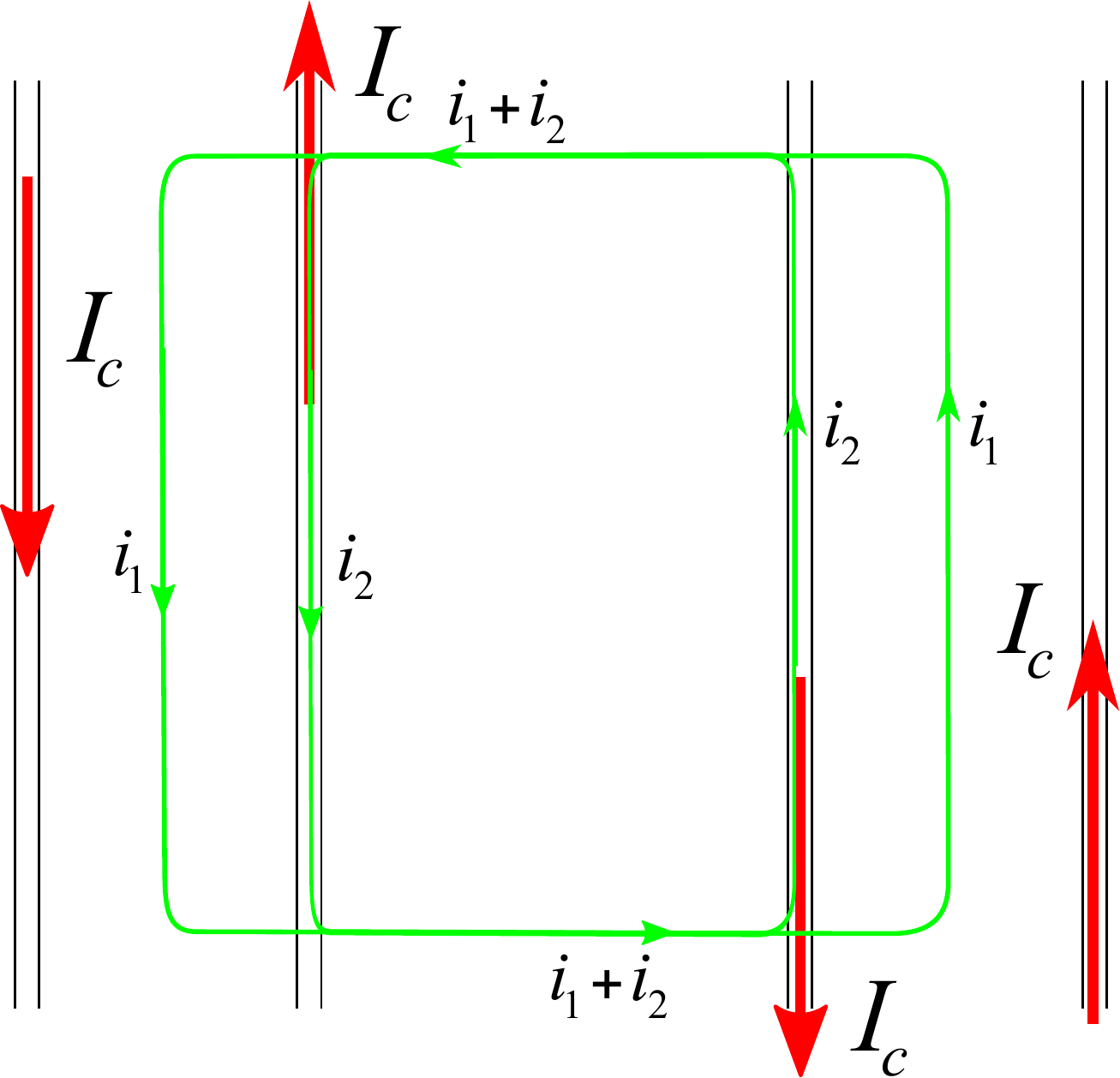}
    \label{fig:Represetative circuit}
    }
    \caption{ Eddy current induced in the rectangular micro-object: a) distribution of eddy current density; b) representative circuit.}
    \label{fig: Symmetric design represetative eddy circuit}
\end{figure*}

We study the stability of the prototype operated in air environment for the case of a rectangular-shaped levitated micro-object. According to the procedure proposed above, first a representative eddy current circuit is defined. Then the equations for coefficients of  Taylor series (\ref{eq:Taylor series}) are computed. The distribution of eddy currents generated by the micro-coils can be studied using a similar design of the prototype consisting of four straight wires and a rectangular shaped micro-object. Taking into account that the levitation height of the micro-object is significantly smaller than its lateral dimensions, the eddy current  distribution can be represented as shown in Fig.~\ref{fig: Symmetric design represetative eddy circuit}. The simulation was performed for a levitation height of 100 $\mu$m and coil currents of 100 mA. The distribution in Fig.~\ref{fig: eddy currents density} is presented in dimensionless relative values, i.e., the ratio of the current density to its maximum value. The analysis of Fig.~\ref{fig: eddy currents density} shows that the representative eddy current circuit can consist of two circuits as shown in Fig.\ref{fig:Represetative circuit} covering a particular eddy current density range between 0.42 and 1.0. It is important to note that the behaviour of the eddy current circuit, $i_2$, is similar to the one in the axially symmetric design and its position in space does not depend on the lateral displacement of the micro-object characterized by the generalized coordinate, $q_l$. Accounting for this fact, the equations to calculate the coefficients of (\ref{eq:complex coefficients}) can be compiled as shown in \ref{app:B}.

\begin{table}[!t]
\caption{Parameters of the prototype of micro-transporter.  }
\centering
\begin{tabular}{lc}
\toprule
The levitation coil width, $c_l$\footnotemark[1] & 1500 $\mathrm{\mu m}$\\
The stabilization coil width, $c_s$\footnotemark[1] & 2900 $\mathrm{\mu m}$\\
The coils pitch of winding, $p$ & 25 $\mathrm{\mu m}$\\
Number of windings for stabilization coil, $N$ & 10\\
Number of windings for levitation coil, $M$ & 14\\
Length of the track, $l_c$ & 7000 $\mathrm{\mu m}$\\
Length of micro-object, $b$ & 2400 $\mathrm{\mu m}$\\
\bottomrule
\end{tabular}
\label{tb:param}
\end{table}
\footnotetext[1]{Parameter is defined in  \ref{app:B}, Fig.~\ref{fig: scheme for transporter}.}

The structure of the model is given in Table~\ref{tab:model structure v}; a condition for the stable levitation in air becomes as follows:
\begin{subequations}
\label{eq:condition}
\begin{empheq}[left={}\empheqlbrace]{align}
  &c_{vv}>0; \; c_{ll}>0; \; c_{\alpha\alpha}>0; \; c_{\beta\beta}>0; \label{eq:a}\\
  & c_{ll}\cdot c_{\alpha\alpha}>c_{l\alpha}^2. \label{eq:b}
\end{empheq}
\end{subequations}

\begin{figure*}[!t]
    \centering
     \subfigure[]
    {
        \includegraphics[width=2.2in]{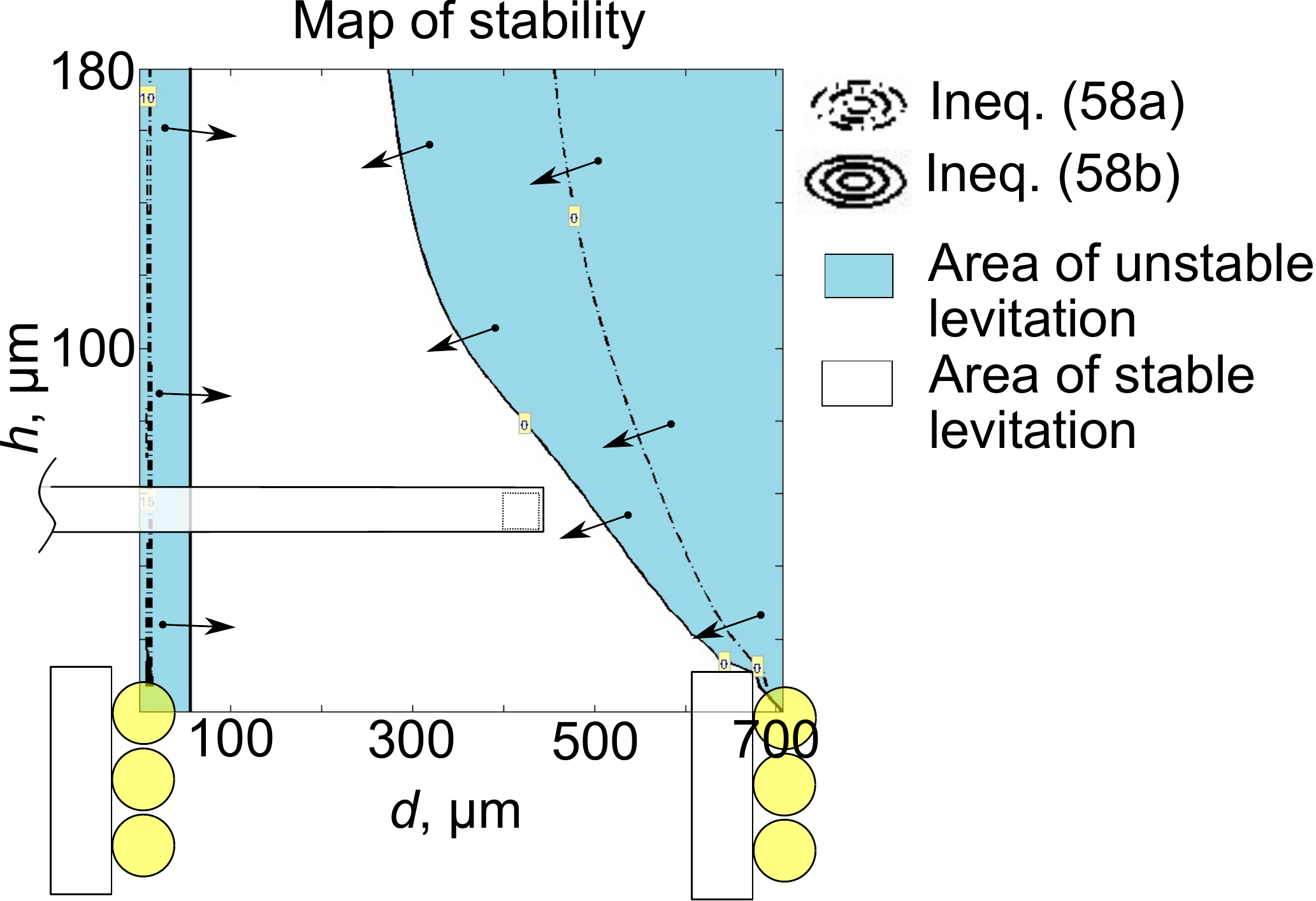}
    \label{fig:stability map}
    }
     \subfigure[]
    {
        \includegraphics[width=2.2in]{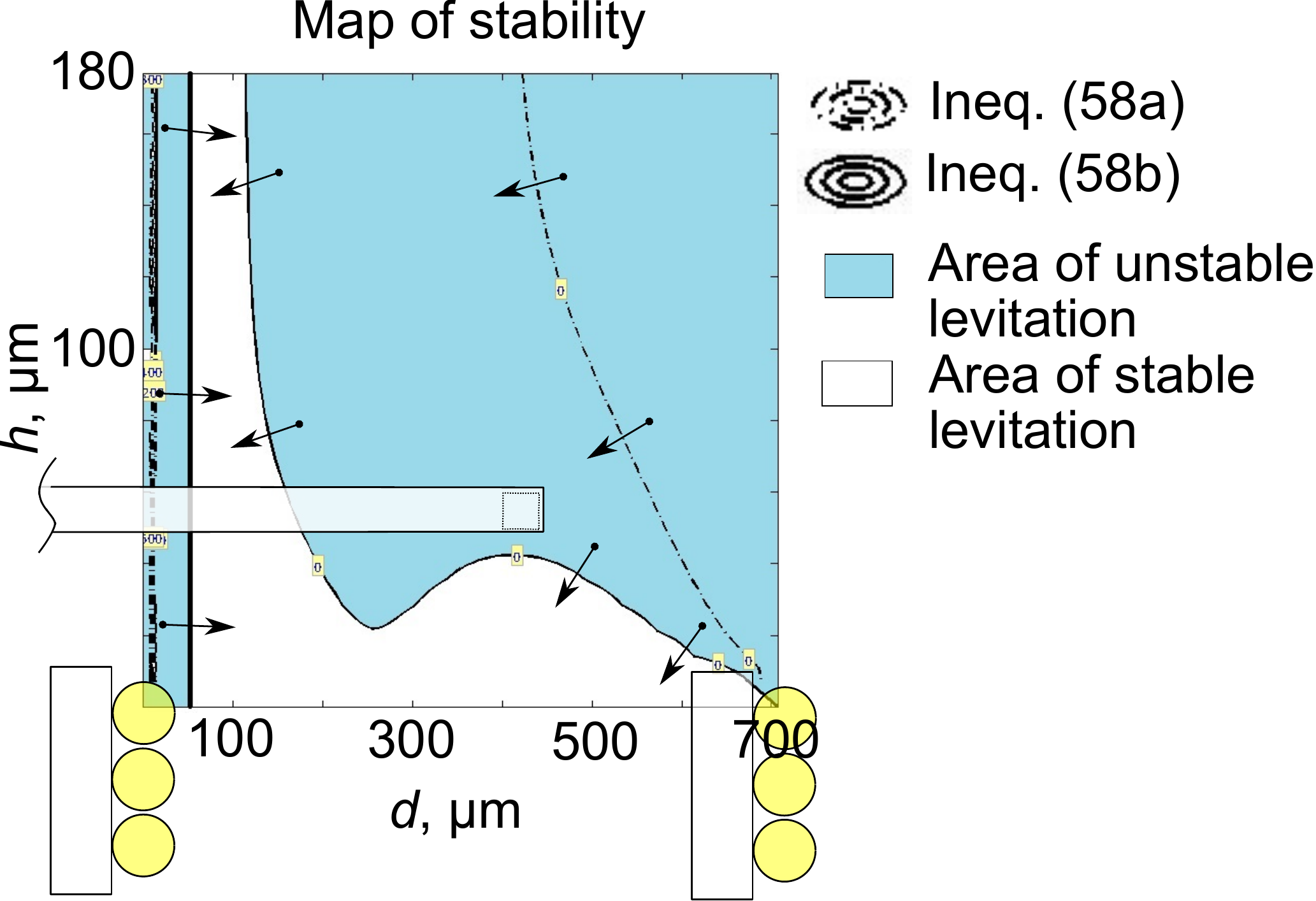}
    \label{fig:unstability map}
    }
    \caption{ Map of stable levitation of the prototype 3D micro-transporter: a) the stability map for the length of the micro-object: $b=2.8 mm$; b) for the length $b=0.5 mm$.   }
    \label{fig:map}
\end{figure*}

Geometrical parameters of the transporter prototype are defined by the schematic shown in \ref{app:B}, Fig.~\ref{fig: scheme for transporter}.
 The parameters of this particular prototype are presented in Table~\ref{tb:param}. Considering a current of 120 mA in each coil and a phase shift of 180$^{\circ}$, the map of stability in terms of levitation height $h$ and width $d=(a-c_s)/2$ is shown in Fig.~\ref{fig:map}. The figure shows two cases, namely, when the length of the micro-object is $b=2.8$ mm and $b=0.5$ mm. In general, the analysis of the map indicates that stable levitation in this prototype is possible for a rectangular shaped micro-object with a length of 2.8 mm, when the width is within the range from 1.7 to 2.8 mm. The experimental study for the square-shaped micro-object having a size of 1.5 mm proves the fact that, for a micro-object with a width less than 1.7 mm, stable levitation is not possible. Another important feature which is reflected by this approach is that decreasing the length of a micro-object leads to a decrease in the area of stable levitation, and for a particular value of the length (in this case: $b=0.5$ mm) stable levitation for any width is impossible. This fact was verified experimentally and agrees well with the theory as shown in Fig.~\ref{fig:unstability map}.

For a micro-object with a width of 2.8 mm, the qualitative approach developed here predicted the top limit of the levitation height to be around 40 $\mu$m. The experimental study demonstrates that the levitation height can be larger than 120 $\mu$m. We see a disagreement between experimental results and the prediction of the model, however, the accuracy can be improved by adding more eddy current circuits during the calculation, as was mentioned above.

\section{Discussion and conclusion}


In this work we have developed the qualitative approach to study the dynamics and stability of micromachined inductive contactless suspensions, taking into account three types of forces: potential, dissipative and positional (nonconservative).
The generalized linear model of MIS has been obtained based on this approach, the analysis of which provides the general properties of the suspension as a dynamic system. In particular, Theorem \ref{thm: instability} of unstable levitation was formulated, where we proved that the stable levitation in MIS subjected to only electromagnetic forces without dissipative forces is impossible. It is worth noting that 
the issue  of destabilising effect
of induced eddy current on a levitron was previously discussed in work in \cite{KrechetnikovMarsden2006}, but it was not mathematically formulated. Also Theorem \ref{thm: instability} extends the classical theorem for the case of a stable potential system having equal natural frequencies subjected to nonconservative positional forces  \cite[page 202, Theorem 6.12]{Merkin2012}.

From Theorem \ref{thm: asymptotical stability} we have an opportunity to estimate the order of magnitude of a component of positional forces having the maximum value, for instance, for 3D MIS  by using the result of experimental measurements given in Table \ref{tb:mechanical}.  In this particular case{,} we have $a_{max}$=\SI{2.4e-7}{\kilo\gram}, $r_{min}$=\SI{1.5e-8}{\newton \meter \per \radian} and the minimum damping coefficient corresponding to angular displacement   is around $\mu_{min}$=\SI{1.0e-8}{\newton \meter \second \per \radian}, the maximum component of stiffness of positional forces must be less than \SI{2.5e-9}{\newton \per \meter}$>p_{max}$. For the one dipole approximation the component of stiffness potential and   positional forces  has a simple linear dependence like $\mathbf{P}=k\mathbf{R}$, where  $k$ must be less than \num{1e-7} for 3D MIS operating in air. The interesting point is that the positional forces are even smaller.
This fact was confirmed by employing the MIS in vacuum \cite{Zhang2006}. Another issue which can be pointed out here is that the stability of MIS can be adapted to a particular vacuum environment by increasing its appropriate component of stiffness.

The qualitative approach developed herewith allows us to propose the general procedure for {designing MIS}. In a particular case, for the symmetric MIS designs levitating rectangular- and disk-shaped micro-objects and axially symmetric designs levitating sphere and disk-shaped proof mass, the general structures of the analytical models describing their behaviour were obtained. Then, this approach was applied to study the dynamics of the prototype of axially symmetric MIS levitating the disk-shaped proof mass and the stability of the prototype of symmetric MIS levitating the rectangular-shaped proof mass, both based on 3D micro-coil technology. In the first prototype, its stiffness components were calculated, which agree well with experimental measurements without using similarity coefficients. In the second prototype, which has been proposed for potential application as a linear transporter, its stability was studied. Theoretical analysis of its stability map as a function of  width of rectangular-shaped proof mass and the levitation height, showed that the stable levitation is possible when the width of proof mass is larger than the width of the levitation coil, $c_l$, and less than the width of stabilization coils, $c_s$. In the case of equality between widths of proof mass and the stabilization coil the stable levitation is impossible. This fact agrees with the experimental study. Also theoretically we show the effect of the proof mass length on the stability. Reducing the  length of the proof mass, decreases the area of stability. In particular, for a width of the proof mass of $2.8$ mm, when the length of the proof mass is less than $0.5$ mm, the stable levitation in the presented design is not possible. This observation was also verified experimentally.


\section*{Acknowledgment}
KP acknowledges with thanks the support of the Alexander von Humboldt Foundation. JGK acknowledges support from the European Research Council (ERC) under grant no. 290586 NMCEL.

\section*{References}

\bibliography{DRAFT}

\appendix


\section{Theorems}
\label{app:Th}

Three theorems with proofs are provided below, which establish the general stability issues of a micromachined inductive contactless suspension.\\ \\
\textbf{Theorem 1 (Unstable levitation I).}
\textit{If a micromachined inductive suspension is subjected to only electromagnetic forces defined by} (\ref{eq:deriviation of field }) \textit{(without dissipation forces, so that }$\mathbf{B}=0$\textit{)}, \textit{then  stable levitation in this suspension is impossible.}\\
\textbf{Proof}. According to the statement of the theorem, model
(\ref{eq: model matrix}) is rewritten as
\begin{equation}\label{eq: model matrix B=0}
    \mathbf{A}\ddot{\mathbf{\overline{q}}}+\left(\mathbf{R}+\jmath\mathbf{P}\right)\mathbf{\overline{q}}=0.
\end{equation}
Let us consider two cases. The first case is when matrix $\mathbf{R}$ is negative definite ($\mathbf{R}<0$), and the second case is when matrix $\mathbf{R}$ is positive definite ($\mathbf{R}>0$).

The case of $\mathbf{R}<0$ is the trivial one, since system (\ref{eq: model matrix B=0}) becomes unstable. 
Due to the fact that adding the positional forces to such a system cannot provide stable levitation \cite[page 203, Theorem 6.13]{Merkin2012}. For $\mathbf{R}>0$,  system (\ref{eq: model matrix B=0})  can be transformed. Introducing a new  complex vector $\mathbf{\overline{u}}$ such that
\begin{equation}\label{eq: new variable}
    \mathbf{\overline{q}}=\Lambda\mathbf{\overline{u}},
\end{equation}
where $\Lambda$ is the orthogonal matrix of transformation, matrices $\mathbf{A}$ and $\mathbf{R}$ can be represented as
\begin{equation}\label{eq: diagonal C and A}
     \Lambda^T\mathbf{A}\Lambda=\mathbf{I},\;
    \Lambda^T\mathbf{R}\Lambda=\mathbf{R}_0,
\end{equation}
where $\mathbf{R}_0=\mathrm{diag}(r_1,\ldots,r_6)$ and $\mathbf{I}$ is  the identity matrix. 
Accounting for (\ref{eq: new variable}) and (\ref{eq: diagonal C and
A}),  {model} (\ref{eq: model matrix B=0}) becomes
\begin{equation}\label{eq: model matrix B=0 new variable}
    \mathbf{I}\ddot{\mathbf{\overline{u}}}+\left(\mathbf{R}_0+\jmath\mathbf{\hat{P}}\right)\mathbf{\overline{u}}=0,
\end{equation}
where $\mathbf{\hat{P}}=\Lambda^T\mathbf{P}\Lambda$.  Eq.\ (\ref{eq: model matrix B=0 new variable}) can be rewritten in real values as
\begin{equation}\label{eq: model matrix 12 B=0}
\left(
  \begin{array}{c|c}
    \mathbf{I}  & 0 \\
    \hline
    0 & \mathbf{I} \\
  \end{array}
\right) \left(
    \begin{array}{l}
      \ddot{\mathbf{u}} \\
      \hline
      \ddot{\mathbf{u}}*
    \end{array}
\right) + \left(
  \begin{array}{c|c}
    \mathbf{R}_0  & -\mathbf{\hat{P}} \\
    \hline
   \mathbf{\hat{P}} & \mathbf{R}_0 \\
  \end{array}
\right) \left(
    \begin{array}{l}
      \mathbf{u} \\
      \hline
      \mathbf{u}*
    \end{array}
\right)=0
    ,
\end{equation}
for which the characteristic equation is
\begin{equation}\label{eq: model matrix 12 B=0 chr}
\mathrm{det}\left(
  \begin{array}{c|c}
   \mathbf{I}\lambda^2+ \mathbf{R}_0  & -\mathbf{\hat{P}} \\
    \hline
   \mathbf{\hat{P}} & \mathbf{I}\lambda^2+\mathbf{R}_0 \\
  \end{array}
\right)=0,
\end{equation}
or
\begin{equation}\label{eq: model matrix 12 B=0 chr modified}
\mathrm{det}\left(
   \left(\mathbf{I}\lambda^2+ \mathbf{R}_0\right)^2  + \mathbf{\hat{P}}^2
  \right)=0.
\end{equation}
Due to the fact that matrix $\mathbf{\hat{P}}^2>0$ is positive
definite, the following characteristic equation $\mathrm{det}\left(
   \tilde{\lambda}^2  + \mathbf{\hat{P}}^2
  \right)=0$ has twelve imaginary roots, $\tilde{\lambda}_i=\pm \jmath
  a_i$, $a_i>0$,
  $i=(1,\ldots,6)$. Hence, accounting for $\tilde{\lambda}_i=\lambda^2+r_i$, {where $r_i>0$, $i=(1,\ldots,6)$,} the roots of (\ref{eq: model matrix 12 B=0 chr
  modified}) become
\begin{equation}\label{eq: roots}
     \begin{array}{c}
        \lambda_i=\pm\jmath\sqrt{(r_i-\jmath a_i)},\; i=(1,\ldots,6),\\
       \lambda_j=\pm\jmath\sqrt{(r_j+\jmath a_j)},\; j=(7,\ldots,12),
     \end{array}
\end{equation}
Finally, we have
\begin{equation}\label{eq: roots extand}
    \begin{array}{c}
      \lambda_i=\pm\sqrt{\frac{\sqrt{r_i^2+a_i^2}-r_i}{2}}\pm\jmath\sqrt{\frac{\sqrt{r_i^2+a_i^2}+r_i}{2}}, \;
    i=(1,\ldots,6)\\
      \lambda_j=\mp\sqrt{\frac{\sqrt{r_j^2+a_j^2}-r_j}{2}}\pm\jmath\sqrt{\frac{\sqrt{r_j^2+a_j^2}+r_j}{2}}, \;
    j=(7,\ldots,12)
    \end{array}
\end{equation}
From (\ref{eq: roots extand}), it is seen that the real part of the roots have positive values. This fact proves the theorem.\\
\textbf{Corollary 1.1.}
\textit{If a micromachined inductive suspension is subjected to only electromagnetic forces, and  the potential part of the electromagnetic forces is absent }($\mathbf{R}=0$)\textit{, then stable levitation in the suspension is impossible.}\\
This fact follows directly from (\ref{eq: roots extand}). Substituting $r_i=0$ into (\ref{eq: roots extand}), the roots still have a positive real part. Also, the corollary agrees with  theorem \cite[page 197, Theorem 6.10]{Merkin2012} about the equilibrium of a system subjected only to positional forces.\\ \\
\textbf{Theorem 2 (Unstable levitation II).}
\textit{If a micromachined inductive suspension is subjected to electromagnetic forces having only positional} $\mathbf{P}\neq0$ ($\mathbf{R}=0$)\textit{ and dissipative forces} ($\mathbf{B}>0$)\textit{, then stable levitation is impossible.}\\
\textbf{Proof}. 
We consider the following equation
\begin{equation}\label{eq: model matrix C=0}
    \mathbf{A}\ddot{\mathbf{\overline{q}}}+\mathbf{B}\dot{\mathbf{\overline{q}}}+\jmath\mathbf{R}\mathbf{\overline{q}}=0.
\end{equation}
As it was done above, the complex vector $\mathbf{\overline{u}}$ given in (\ref{eq: new variable}) is used, hence matrices $\mathbf{A}$ and $\mathbf{B}$ can be represented as:
\begin{equation}\label{eq: diagonal A and B}
     \Lambda^T\mathbf{A}\Lambda=\mathbf{I},\;
    \Lambda^T\mathbf{B}\Lambda=\mathbf{B}_0,
\end{equation}
where $\mathbf{B}_0=\mathrm{diag}(\hat{\mu}_1,\ldots,\hat{\mu}_6)$. Taking {later equations} into account,  Eq.\ (\ref{eq: model matrix C=0}) is rewritten as
\begin{equation}\label{eq: model matrix C=0 new variable}
    \mathbf{I}\ddot{\mathbf{\overline{u}}}+\mathbf{B}_0\dot{\mathbf{\overline{u}}}+\jmath\mathbf{\hat{P}}\mathbf{\overline{u}}=0,
\end{equation}
where $\mathbf{\hat{P}}=\Lambda^T\mathbf{P}\Lambda$. Eq.\ (\ref{eq: model matrix C=0 new variable}) can be rewritten in real values as
\begin{equation}\label{eq: model matrix 12 C=0}
\left(
  \begin{array}{c|c}
    \mathbf{I}  & 0 \\
    \hline
    0 & \mathbf{I} \\
  \end{array}
\right) \left(
    \begin{array}{l}
      \ddot{\mathbf{u}} \\
      \hline
      \ddot{\mathbf{u}}*
    \end{array}
\right) + \left(
  \begin{array}{c|c}
    \mathbf{B}_0  & 0 \\
    \hline
    0 & \mathbf{B}_0 \\
  \end{array}
\right) \left(
    \begin{array}{l}
      \dot{\mathbf{u}} \\
      \hline
      \dot{\mathbf{u}}*
    \end{array}
\right) + \left(
  \begin{array}{c|c}
    0  & -\mathbf{\hat{P}} \\
    \hline
   \mathbf{\hat{P}} & 0\\
  \end{array}
\right) \left(
    \begin{array}{l}
      \mathbf{u} \\
      \hline
      \mathbf{u}*
    \end{array}
\right)=0
    .
\end{equation}
The characteristic equation is:
\begin{equation}\label{eq: model matrix 12 C=0 chr}
\mathrm{det}\left(
  \begin{array}{c|c}
   \mathbf{I}\lambda^2+ \mathbf{B}_0\lambda  & -\mathbf{\hat{P}} \\
    \hline
   \mathbf{\hat{P}} & \mathbf{I}\lambda^2+\mathbf{B}_0\lambda \\
  \end{array}
\right)=0,
\end{equation}
or
\begin{equation}\label{eq: model matrix 12 C=0 chr modified}
\mathrm{det}\left(
   \left(\mathbf{I}\lambda^2+ \mathbf{B}_0\lambda\right)^2  + \mathbf{\hat{P}}^2
  \right)=0.
\end{equation}
Using the same reasoning as for Theorem \ref{thm: instability}, the roots are:
\begin{equation}\label{eq: roots for UnstableII}
\begin{array}{c}
      \lambda_i=\frac{-\hat{\mu}_i+\sqrt{\hat{\mu}_i^2\mp \jmath 4a_i}}{2}, \;
    i=(1,\ldots,6)\\
      \lambda_j=\frac{-\hat{\mu}_i-\sqrt{\hat{\mu}_i^2\mp \jmath 4a_i}}{2}, \;
    j=(7,\ldots,12)
    \end{array}
\end{equation}
Here we need to prove that the real part of $\Re(\lambda_i)>0$ is positive. Accounting for
\begin{equation}\label{eq: sqrt}
 {\displaystyle
  {\sqrt{\hat{\mu}_i^2\mp \jmath 4a_i}=\sqrt{\frac{\sqrt{\hat{\mu}_i^4+16a_i^2}+\hat{\mu}_i^2}{2}}
  \mp\jmath\sqrt{\frac{\sqrt{\hat{\mu}_i^4+16a_i^2}-\hat{\mu}_i^2}{2}},}
  }
\end{equation}
the real part of $\lambda_i$ is
\begin{equation}\label{eq: real part lamba}
 {\displaystyle
  {\Re(\lambda_i)=\frac{1}{2}\left(-\hat{\mu}_i+\sqrt{\frac{\sqrt{\hat{\mu}_i^4+16a_i^2}+\hat{\mu}_i^2}{2}}\right).}
  }
\end{equation}
 We can write
\begin{equation}\label{eq: real part lamba positive}
 {\displaystyle
  {-\hat{\mu}_i+\sqrt{\frac{\sqrt{\hat{\mu}_i^4+16a_i^2}+\hat{\mu}_i^2}{2}}>0.}
  }
\end{equation}
Inequality (\ref{eq: real part lamba positive}) is rewritten as
\begin{equation}\label{eq: real part lamba positive rewritten}
 {\displaystyle
  {\sqrt{\hat{\mu}_i^4+16a_i^2}>\hat{\mu}_i^2,}
  }
\end{equation}
which yields
\begin{equation}\label{eq: real part lamba positive final}
 {\displaystyle
  {16a_i^2>0.}
  }
\end{equation}
This fact shows that the real part of $\lambda_i$ is positive. Hence
the theorem is proved.

It is important to note that Theorem \ref{thm: instability II} corresponds to Theorem \cite[page 198, Theorem 6.11]{Merkin2012}, which claims that the equilibrium of a system subjected to arbitrary nonconservative positional forces and linear dissipative forces is always unstable.\\ \\
\textbf{Theorem 3 (Asymptotically stable levitation).} 
\textit{By adding dissipative forces} ($\mathbf{B}>0$) \textit{to a micromachined inductive suspension subjected to electromagnetic forces defined by} (\ref{eq:deriviation of field }) \textit{and having a positive definite matrix of potential forces} ($\mathbf{R}>0$)\textit{, the suspension can be asymptotically stable.}\\
\textbf{Proof}. In order to prove the theorem, Metelitsyn's inequality \cite{Metelitsyn1952} is used \cite[page 32]{SeyranianMailybaev2003}. The necessary condition is that matrix $ \mathbf{A}$, $ \mathbf{B}$ and $\mathbf{R}$ should be positive definite. The condition follows from the statement of the theorem. According to \cite[page 1099]{Seyranian2003}, a sufficient practical condition for asymptotically stable levitation for the present case becomes as follows
 \begin{equation}
   \mu_\textrm{min}>p_\textrm{max}\sqrt{a_\textrm{max}/r_\textrm{min}}, \nonumber
 \end{equation}
 where $ \mu_\textrm{min}$, and $r_\textrm{min}$ are the respective minimum values of  $\mathbf{B}$ and  $\mathbf{R}$; $p_\textrm{max}$ and $a_\textrm{max}$ are the respective maximum values of  $ \mathbf{P}$ and $ \mathbf{A}$. This fact proves that the real part of  eigenvalues is negative when the inequality (\ref{eq:sufficient condition}) holds true. Thus, the system is asymptotically stable.

\section{Equation compilation for axially symmetric MIS}
\label{app:A}
In order to find stiffness components for the model of 3D axially symmetric MIS, the structure of which has defined in Table~\ref{tab:model structure v},  terms of  Taylor series in (\ref{eq:Taylor series}) are calculated. Using the notation defined in Sec. \ref{sec:Qualitative Technique} and Fig.~\ref{fig:behaviour of center eddy currnet circuit}, we can write for $m_0^{kj}$ \cite{Lu2014}:
\begin{equation}\label{eq:m_0}
\begin{array}{l}
{\displaystyle
    m_0^{11}=\sum_{\iota=0}^{N-1} \mu_0\cdot r_s\left[\ln\frac{8r_s}{\sqrt{(h+\iota\cdot p)^2+(d-c)^2}}-1.92\right];} \\
{\displaystyle
    m_0^{22}=\sum_{\iota=0}^{M-1} \mu_0\cdot r_l\left[\ln\frac{8r_l}{h+\iota\cdot p}-1.92\right];} \\
{\displaystyle
    m_0^{12}=\sum_{\iota=0}^{M-1} \mu_0\cdot (r_{l}+d)\left[\ln\frac{8(r_{l}+d)}{\sqrt{(h+\iota\cdot p)^2+d^2}}-1.92\right];} \\
{\displaystyle
    m_0^{21}=\sum_{\iota=0}^{N-1} \mu_0\cdot r_s\left[\ln\frac{8r_s}{\sqrt{(h+\iota\cdot p)^2+c^2}}-1.92\right]}, \\
\end{array}
\end{equation}
where $\mu_0=4\pi\times10^{-7}$ H/m  is the magnetic permeability of vacuum,  $p$ is the winding pitch of the coils , $c=r_s-r_l$,
$N$ and $M$ are numbers of winding  for stabilization and levitation coils, respectively. According to \cite{Poletkin2014a} for the set of terms $m_l^{kj}$ we have:
\begin{equation}\label{eq:m_l}
\begin{array}{l}
{\displaystyle
    m_{\nu}^{11}=-\sum_{\iota=0}^{N-1} \mu_0\cdot r_s\frac{ h+\iota\cdot p}{(h+\iota\cdot p)^2+(c-d)^2};} \\
{\displaystyle
    m_{\nu}^{22}=-\sum_{\iota=0}^{M-1} \mu_0\cdot r_l\frac{ 1}{h+\iota\cdot p};} \\
{\displaystyle
    m_{\nu}^{12}=-\sum_{\iota=0}^{M-1} \mu_0\cdot (r_{l}+d)\frac{ h+\iota\cdot p}{(h+\iota\cdot p)^2+d^2};} \\
{\displaystyle
    m_{\nu}^{21}=-\sum_{\iota=0}^{N-1} \mu_0\cdot r_s\frac{ h+\iota\cdot p}{(h+\iota\cdot p)^2+c^2}}; \\
{\displaystyle
    m_{l}^{11}= m_{l}^{22}= m_{l}^{12}= m_{l}^{21}=0};\\
{\displaystyle
    m_{\theta}^{11}= m_{\theta}^{22}= m_{\theta}^{12}= m_{\theta}^{21}=0}.\\
\end{array}
\end{equation}
{Terms of the second derivatives}, $m_{ll}^{kj}$, are defined as follows. For $m_{\nu\nu}^{kj}$, we have:
\begin{equation}\label{eq:m_nunu}
\begin{array}{l}
{\displaystyle
    m_{\nu\nu}^{11}=\sum_{\iota=0}^{N-1} \mu_0\cdot r_s\frac{ (h+\iota\cdot p)^2-(c-d)^2}{\left[(h+\iota\cdot p)^2+(c-d)^2\right]^2};} \\
{\displaystyle
    m_{\nu\nu}^{22}=\sum_{\iota=0}^{N-1} \mu_0\cdot r_l\frac{ 1}{h+\iota\cdot p};} \\
{\displaystyle
    m_{\nu\nu}^{12}=\sum_{\iota=0}^{N-1} \mu_0\cdot (r_l+d)\frac{ (h+\iota\cdot p)^2-d^2}{\left[(h+\iota\cdot p)^2+d^2\right]^2};} \\
{\displaystyle
    m_{\nu\nu}^{21}=\sum_{\iota=0}^{N-1} \mu_0\cdot r_s\frac{ (h+\iota\cdot p)^2-c^2}{\left[(h+\iota\cdot p)^2+c^2\right]^2};} \\
\end{array}
\end{equation}
Taking into account the behaviour of the second circuit of eddy current shown in Fig.~\ref{fig:behaviour of center eddy currnet circuit}, for $m_{ll}^{kj}$ we can write:
\begin{equation}\label{eq:m_ll}
\begin{array}{l}
{\displaystyle
    m_{ll}^{11}=\sum_{\iota=0}^{N-1} \mu_0\cdot \frac{r_s}{2}\frac{(d-c)^2r_s-(h+\iota\cdot p)^2(r_s-2c+2d) }{(r_s-(c-d))\left[(h+\iota\cdot p)^2+(d-c)^2\right]^2};} \\
{\displaystyle
    m_{ll}^{12}=\sum_{\iota=0}^{N-1} \mu_0\cdot \frac{r_l}{2}\frac{d^2r_l-(h+\iota\cdot p)^2(r_l+2d) }{(r_l+d)\left[(h+\iota\cdot p)^2+d^2\right]^2};} \\
{\displaystyle
    m_{ll}^{21}= m_{ll}^{22}=0.} \\
\end{array}
\end{equation}
For $m_{\theta\theta}^{kj}$ we can write:
\begin{equation}\label{eq:m_ff}
\begin{array}{l}
{\displaystyle
    m_{\theta\theta}^{11}=(r_l+d)^2\sum_{\iota=0}^{N-1} \mu_0\cdot \frac{r_s}{2}\frac{(h+\iota\cdot p)^2 -(d-c)^2 }{\left[(h+\iota\cdot p)^2+(d-c)^2\right]^2};} \\
{\displaystyle
    m_{\theta\theta}^{12}=r_l^2\sum_{\iota=0}^{N-1} \mu_0\cdot \frac{r_s}{2}\frac{(h+\iota\cdot p)^2 -d^2 }{\left[(h+\iota\cdot p)^2+d^2\right]^2};} \\
{\displaystyle
    m_{\theta\theta}^{21}=(r_l+d)^2\sum_{\iota=0}^{M-1} \mu_0\cdot \frac{r_l}{2}\frac{(h+\iota\cdot p)^2 -d^2 }{\left[(h+\iota\cdot p)^2+d^2\right]^2};} \\
    {\displaystyle
    m_{\theta\theta}^{22}=r_l^2\sum_{\iota=0}^{M-1} \mu_0\cdot \frac{r_l}{2}\frac{1 }{(h+\iota\cdot p)^2.}} \\
\end{array}
\end{equation}
Terms of Taylor series for  self and mutual  inductances of eddy current circuits like $g_0^{ks}$, $g_l^{ks}$ and $g_{ll}^{ks}$ can be written as follows.
Terms $g_0^{ks}$ are:
\begin{equation}\label{eq:g_0}
\begin{array}{l}
{\displaystyle
    g_0^{11}=\mu_0\cdot (r_l+d)\left[\ln\frac{8(r_l+d)}{\chi}-1.92\right];} \\
{\displaystyle
    g_0^{22}=\mu_0\cdot r_l\left[\ln\frac{8r_l}{\chi}-1.92\right];} \\
{\displaystyle
    g_0^{12}=g_0^{12}=\mu_0\cdot (r_l+d)\left[\ln\frac{8(r_l+d)}{d}-1.92\right],} \\
\end{array}
\end{equation}
where $\chi$ is the characteristic length for eddy current circuit.
Although,
\begin{equation}\label{eq:g_l}
\begin{array}{l}
{\displaystyle
    g_l^{12}=g_{\theta}^{12}}=g_{\nu}^{12}=0 \\
\end{array}
\end{equation}
and
\begin{equation}\label{eq:g_ss}
\begin{array}{l}
{\displaystyle
    g_{\theta\theta}^{12}=g_{\nu\nu}^{12}}=0, \\
\end{array}
\end{equation}
however the second derivative with respect to generalized coordinate $l$ is  zero. Due to the behaviour of eddy current circuits shown in Fig.~\ref{fig:behaviour of center eddy currnet circuit} it becomes as
\begin{equation}\label{eq:g_ll}
\begin{array}{l}
{\displaystyle
    g_{ll}^{12}=g_{ll}^{21}= \mu_0\cdot \frac{r_l}{2}\frac{r_l }{(r_l+d)d^2}.} \\
\end{array}
\end{equation}
Using the equations above we can define determinants (\ref{eq: main determinant}) and (\ref{eq:sub  determinant transponation}) as follows. Determinants $\Delta_0^{ks}$ are:
\begin{equation}\label{eq:delat 0}
\begin{array}{cc}
 \Delta_0^{11}=\left|
    \begin{array}{cc}
       m_0^{11} &  m_0^{21} \\
      g_0^{21} & g_0^{22} \\
    \end{array}
  \right|; &
  \Delta_0^{12}= \left|
    \begin{array}{cc}
       m_0^{12} &  m_0^{22} \\
      g_0^{21} & g_0^{22} \\
    \end{array}
  \right|; \\ \\
  \Delta_0^{21}=\left|
    \begin{array}{cc}
       g_0^{11} &  g_0^{12} \\
      m_0^{11} & m_0^{21} \\
    \end{array}
  \right|; &
  \Delta_0^{22}= \left|
    \begin{array}{cc}
       g_0^{11} &  g_0^{12} \\
      m_0^{12} & m_0^{22} \\
    \end{array}
  \right|. \\
 \end{array}
\end{equation}
Determinant $\Delta_0$ is
\begin{equation}\label{eq:delat}
  \Delta_0=\left|
    \begin{array}{cc}
       g_0^{11} &  g_0^{12} \\
      g_0^{21} & g_0^{22} \\
    \end{array}
  \right|.
\end{equation}
Accounting for (\ref{eq:m_l}) and (\ref{eq:g_l}),  $\Delta_l^{ks}$ are:
\begin{equation}\label{eq:delat l}
\begin{array}{cc}
 \Delta_{\nu}^{11}=\left|
    \begin{array}{cc}
       m_{\nu}^{11} &  m_{\nu}^{21} \\
      g_0^{21} & g_0^{22} \\
    \end{array}
  \right|; &
  \Delta_{\nu}^{12}= \left|
    \begin{array}{cc}
       m_{\nu}^{12} &  m_{\nu}^{22} \\
      g_0^{21} & g_0^{22} \\
    \end{array}
  \right|; \\ \\
  \Delta_{\nu}^{21}=\left|
    \begin{array}{cc}
       g_0^{11} &  g_0^{12} \\
      m_{\nu}^{11} & m_{\nu}^{21} \\
    \end{array}
  \right|; &
  \Delta_{\nu}^{22}= \left|
    \begin{array}{cc}
       g_0^{11} &  g_0^{12} \\
      m_{\nu}^{12} & m_{\nu}^{22} \\
    \end{array}
  \right|, \\
 \end{array}
\end{equation}
and others $\Delta_l^{ks}$ and $\Delta_{\theta}^{ks}$ are  zero. The first derivative of (\ref{eq:delat}) is also zero.
Accounting for (\ref{eq:m_nunu}), {(\ref{eq:m_ll})},  {(\ref{eq:m_ff})} and (\ref{eq:g_ll}), the second derivatives of the determinants are written as follows. With respect to $\nu$:
\begin{equation}\label{eq:delat nunu}
\begin{array}{cc}
 \Delta_{\nu\nu}^{11}=\left|
    \begin{array}{cc}
       m_{\nu\nu}^{11} &  m_{\nu\nu}^{21} \\
      g_0^{21} & g_0^{22} \\
    \end{array}
  \right|; &
  \Delta_{\nu\nu}^{12}= \left|
    \begin{array}{cc}
       m_{\nu\nu}^{12} &  m_{\nu\nu}^{22} \\
      g_0^{21} & g_0^{22} \\
    \end{array}
  \right|; \\ \\
  \Delta_{\nu\nu}^{21}=\left|
    \begin{array}{cc}
       g_0^{11} &  g_0^{12} \\
      m_{\nu\nu}^{11} & m_{\nu\nu}^{21} \\
    \end{array}
  \right|; &
  \Delta_{\nu\nu}^{22}= \left|
    \begin{array}{cc}
       g_0^{11} &  g_0^{12} \\
      m_{\nu\nu}^{12} & m_{\nu\nu}^{22} \\
    \end{array}
  \right|; \\
 \end{array}
\end{equation}
with respect to $\theta$:
\begin{equation}\label{eq:delat ff}
\begin{array}{cc}
 \Delta_{\theta\theta}^{11}=\left|
    \begin{array}{cc}
       m_{\theta\theta}^{11} &  m_{\theta\theta}^{21} \\
      g_0^{21} & g_0^{22} \\
    \end{array}
  \right|; &
  \Delta_{\theta\theta}^{12}= \left|
    \begin{array}{cc}
       m_{\theta\theta}^{12} &  m_{\theta\theta}^{22} \\
      g_0^{21} & g_0^{22} \\
    \end{array}
  \right|; \\ \\
  \Delta_{\theta\theta}^{21}=\left|
    \begin{array}{cc}
       g_0^{11} &  g_0^{12} \\
      m_{\theta\theta}^{11} & m_{\theta\theta}^{21} \\
    \end{array}
  \right|; &
  \Delta_{\theta\theta}^{22}= \left|
    \begin{array}{cc}
       g_0^{11} &  g_0^{12} \\
      m_{\theta\theta}^{12} & m_{\theta\theta}^{22} \\
    \end{array}
  \right|; \\
 \end{array}
\end{equation}
and with respect to $l$ we have
\begin{equation}\label{eq:delat ll}
\begin{array}{c}
 \Delta_{ll}^{11}=\left|
    \begin{array}{cc}
       m_{ll}^{11} &  0 \\
      g_0^{21} & g_0^{22} \\
    \end{array}
  \right|+ \left|
    \begin{array}{cc}
       m_{0}^{11} &  m_{0}^{21} \\
      g_{ll}^{21} &0 \\
    \end{array}
  \right|; \\ \\
  \Delta_{ll}^{12}= \left|
    \begin{array}{cc}
       m_{ll}^{12} &  0\\
      g_0^{21} & g_0^{22} \\
    \end{array}
  \right|+
  \left|
    \begin{array}{cc}
       m_{0}^{12} & m_{0}^{22}\\
      g_{ll}^{21} & 0\\
    \end{array}
  \right|; \\ \\
  \Delta_{ll}^{21}=\left|
    \begin{array}{cc}
      0 &  g_{ll}^{12} \\
      m_{0}^{11} & m_{0}^{21} \\
    \end{array}
  \right|+
  \left|
    \begin{array}{cc}
     g_{0}^{11} &  g_{0}^{12} \\
      m_{ll}^{11} &0 \\
    \end{array}
  \right|; \\ \\
  \Delta_{ll}^{22}= \left|
    \begin{array}{cc}
       0 &  g_{ll}^{12} \\
      m_{0}^{12} & m_{0}^{22} \\
    \end{array}
  \right|+
  \left|
    \begin{array}{cc}
       g_{0}^{11} &  g_{0}^{12} \\
      m_{ll}^{12} & 0 \\
    \end{array}
  \right|. \\
 \end{array}
\end{equation}
For determinant (\ref{eq:delat}) only the second derivative with respect to $l$  exists and becomes as
\begin{equation}\label{eq:ddelta ll}
  \Delta_{ll}=\left|
    \begin{array}{cc}
       0 &  g_{ll}^{12} \\
      g_0^{21} & g_0^{22} \\
    \end{array}
  \right|+
  \left|
    \begin{array}{cc}
       g_0^{11} &  g_{0}^{12} \\
      g_{ll}^{21} & 0 \\
    \end{array}
  \right|.
\end{equation}

\section{Equation compilation for 3D linear transporter}
\label{app:B}
\begin{figure*}[!t]
    \centering
     \subfigure[]
     {
        \includegraphics[width=2.0in]{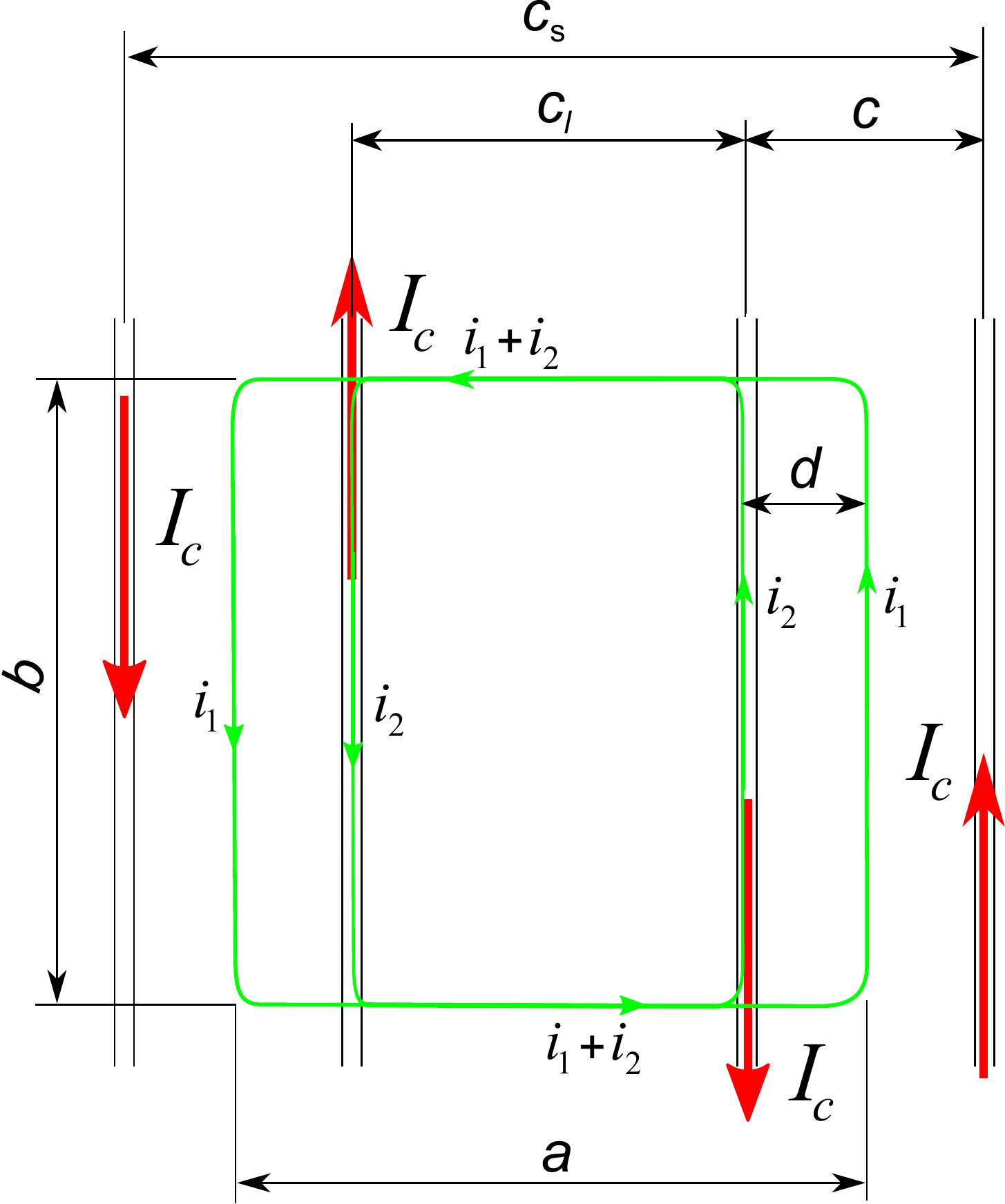}
      \label{fig: scheme for transporter}
      }
       \subfigure[]
     {
        \includegraphics[width=2.2in]{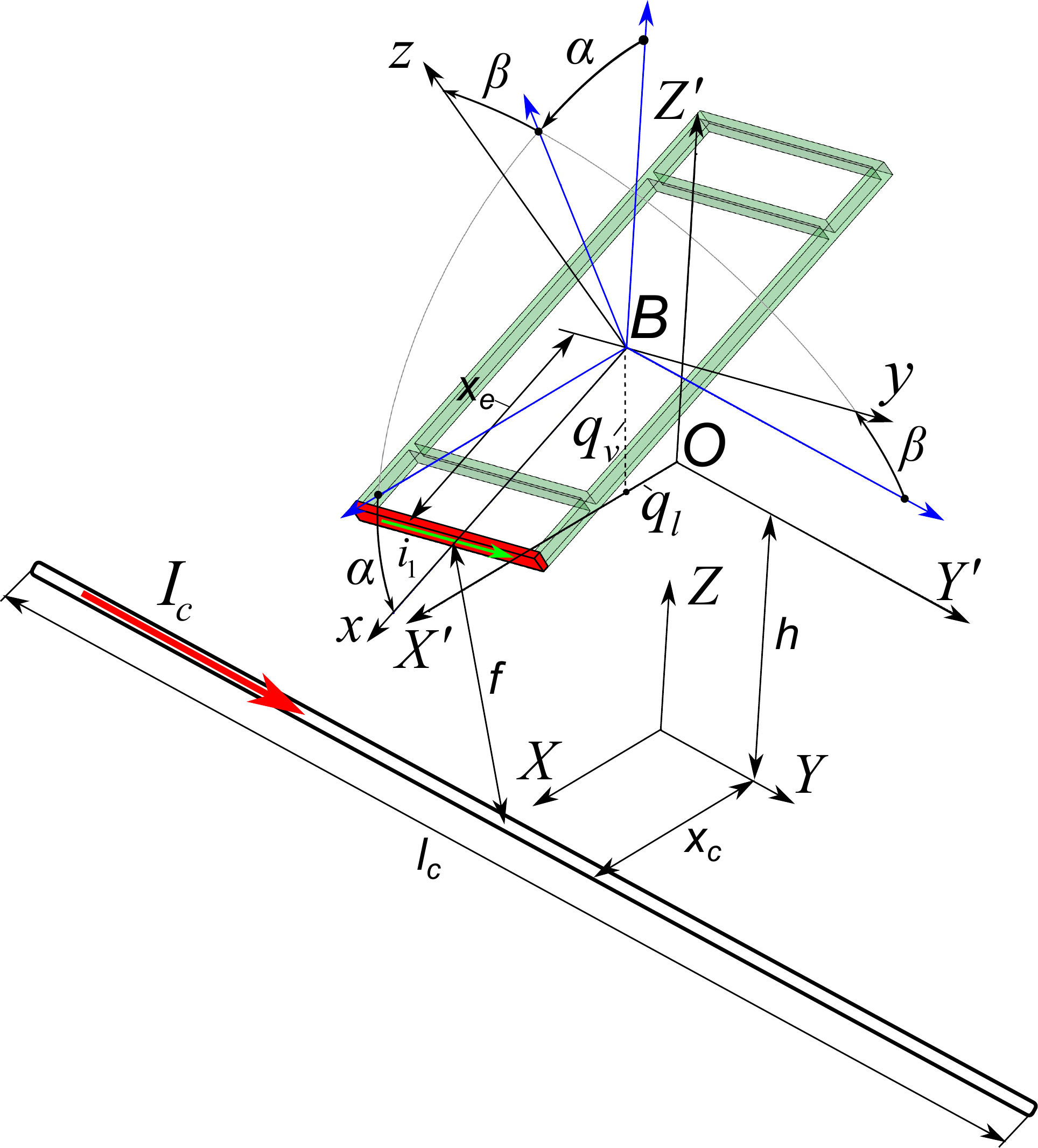}
      \label{fig: scheme for mutual inductance}
      }
    \caption{ Scheme of transporter for calculation.  }
\label{fig: scheme for transporter general}
\end{figure*}
The scheme for the calculation of the stiffness and stability of the transporter is shown in Fig.~\ref{fig: scheme for transporter}. Using equations for self-inductances of a rectangle and straight wire having a square cross-section  \cite[pages 320 and 315]{Rosa1908}, respectively,  and mutual inductance of two parallel wires \cite[page 306]{Rosa1908}, terms $g_0^{ks}$ of (\ref{eq:Taylor series}) can be defined as:
\begin{equation}\label{eq:g_0 rec}
\begin{array}{l}
{\displaystyle
    g_0^{11}=\frac{\mu_0}{\pi}\cdot(c_l+2d+b)\left[\ln\frac{2(c_l+2d)b}{\chi}-\frac{(c_l+2d)\ln(c_l+2d+b)+b\ln(c_l+2d+b)}{c_l+2d+b}\right.} \\
{\displaystyle \left.+\frac{\sqrt{(c_l+2d)^2+b^2}}{c_l+2d+b}-\frac{1}{2}+0.477\frac{\chi}{c_l+2d+b} \right];}\\
{\displaystyle
    g_0^{22}=\frac{\mu_0}{\pi}\cdot(c_l+b)\left[\ln\frac{2c_lb}{\chi}-\frac{c_l\ln(c_l+b)+b\ln(c_l+b)}{c_l+b}+\frac{\sqrt{c_l^2+b^2}}{c_l+b}
    -\frac{1}{2}+0.477\frac{\chi}{c_l+b}\right];} \\
{\displaystyle
    g_0^{12}=g_0^{21}=\frac{\mu_0c_l}{\pi}\left[\ln\frac{c_l}{\chi}+\frac{1}{2}\right] +\frac{\mu_0b}{\pi}\left[\ln\frac{b+\sqrt{b^2+d^2}}{d}-\frac{\sqrt{b^2+d^2}}{b}+\frac{d}{b}\right]} \\
    {\displaystyle
   -\frac{\mu_0b}{\pi}\left[\ln\frac{b+\sqrt{b^2+(d+c_l)^2}}{d+c_l}-\frac{\sqrt{b^2+(d+c_l)^2}}{b}+\frac{d+c_l}{b}\right],} \\
\end{array}
\end{equation}
where $d=(a-c_l)/2$. As it follows from the analysis of scheme shown in Fig.~\ref{fig: scheme for transporter},  the mutual inductances between the coils' wires and the levitated micro-object are reduced to the analysis of the mutual inductance of the system of the parallel wires. 
In order to compile the terms of  (\ref{eq:Taylor series}) let us define the mutual inductance between coil's straight wire and an element of eddy current circuit as it is shown in Fig.~\ref{fig: scheme for mutual inductance}. The element of eddy current circuit is highlighted in red. Using equation of mutual  inductance of two parallel wires \cite[page 306]{Rosa1908}
 having the same length, $l_c$,  the following auxiliary function can be defined as
 \begin{equation}\label{eq:auxiliary function}
   {\displaystyle
   M_a(l_c,f(x_e,x_c))=\frac{\mu_0l_c}{\pi}\left[\ln\frac{1+\sqrt{1+\xi^2}}{\xi}-\sqrt{1+\xi^2}+\xi\right],}
 \end{equation}
 where $\xi=f/l_c$ is the dimensionless parameter, $l_c$ is the length of the coil wire, $f$ is the distance between two wires, which can be calculated as
 \begin{equation}\label{eq:f}
   {\displaystyle
   f(x_e,x_c)=\sqrt{(h+q_v-x_e\sin\alpha)^2+(x_c-x_e\cos\alpha+q_l)^2},}
 \end{equation}
 where $x_c$ is the coordinate of location of the coil wire along the $X$-axis and $x_e$ is the coordinate of location of the element of eddy current circuit along the $X'$-axis (equilibrium state)as shown in Fig.~\ref{fig: scheme for mutual inductance}. Assuming that the displacements are small and using auxiliary function (\ref{eq:auxiliary function}), the mutual inductance between element of the eddy current circuit and coil's wire as shown in Fig.~\ref{fig: scheme for mutual inductance} can be written as \cite[page 45]{Grover2004}:
  \begin{equation}\label{eq:mutual inductance between element and coil wire}
   {\displaystyle
   M(x_e,x_c)=\left(M_a((l_c+b\cos\beta)/2,f(x_e,x_c))-M_a((l_c-b\cos\beta)/2,f(x_e,x_c))\right).}
 \end{equation}
 Noting that the latter equation is derived for the case when the geometrical centers of the coil wire and element of eddy current circuit are aligned.  In order to take into account the number of {windings,}  Eq.\ (\ref{eq:mutual inductance between element and coil wire}) can be  modified as follows
  \begin{equation}\label{eq:mutual inductance between element and coil wire number}
   {\displaystyle
   M_{\iota}(x_e,x_c)=\left(M_a((l_c+b\cos\beta)/2,f_{\iota}(x_e,x_c))-M_a((l_c-b\cos\beta)/2,f_{\iota}(x_e,x_c))\right).}
 \end{equation}
 where
  \begin{equation}\label{eq:f number}
   {\displaystyle
   f_{\iota}(x_e,x_c)=\sqrt{(h+p\cdot\iota+q_v-x_e\sin\alpha)^2+(x_c-x_e\cos\alpha+q_l)^2}.}
 \end{equation}
Hence, considering pairwise wires of coils and accounting for (\ref{eq:mutual inductance between element and coil wire number})  terms  $m_0^{kj}$ are
\begin{equation}\label{eq:m_trans}
\begin{array}{l}
{\displaystyle
    m_0^{11}=2\sum_{\iota=0}^{N-1}\left[M_{\iota}(c_l/2+d,c_s/2)-M_{\iota}(-c_l/2-d,c_s/2)\right]  ;} \\
{\displaystyle
    m_0^{22}=2\sum_{\iota=0}^{M-1}\left[M_{\iota}(c_l/2,c_l/2)-M_{\iota}(-c_l/2,c_l/2)\right]  ;}  \\
{\displaystyle
    m_0^{12}=2\sum_{\iota=0}^{M-1}\left[M_{\iota}(c_l/2+d,c_l/2)-M_{\iota}(-c_l/2-d,c_l/2)\right];} \\
{\displaystyle
    m_0^{21}=2\sum_{\iota=0}^{N-1}\left[M_{\iota}(c_l/2,c_s/2)-M_{\iota}(-c_l/2,c_s/2)\right]  .} \\
\end{array}
\end{equation}

For deriving derivatives of (\ref{eq:mutual inductance between element and coil wire number}) with respect  generalized coordinates $q_v$, $q_l$ and $\alpha$, a general rule can be used such as
\begin{equation}\label{eq:general rule}
  \begin{array}{l}
    {\displaystyle
   \frac{\partial M_a}{\partial q}= \frac{\partial M_a}{ \partial \xi}\frac{\partial \xi}{\partial q};}\\
    {\displaystyle
   \frac{\partial^2 M_a}{\partial q^2}=\frac{\partial^2 M_a}{ \partial \xi^2}\left(\frac{\partial \xi}{\partial q}\right)^2+\frac{\partial M_a}{ \partial \xi}\frac{\partial^2 \xi}{\partial q^2}.}\\
  \end{array}
 \end{equation}
The $\xi$ - derivatives of $M_a$ are
\begin{equation}\label{eq:derivatives of auxiliary function}
   \begin{array}{l}
   {\displaystyle
   \frac{\partial M_a}{ \partial \xi}=\frac{\mu_0l_c}{\pi}\left[1-\frac{1}{\xi}-\frac{\xi}{1+\sqrt{1+\xi^2}}\right];}\\
    {\displaystyle
   \frac{\partial^2 M_a}{ \partial \xi^2}=\frac{\mu_0l_c}{\pi}\left[\frac{1}{\xi^2}-\frac{1}{(1+\sqrt{1+\xi^2})\sqrt{1+\xi^2}}\right].}
    \end{array}
 \end{equation}
 The derivatives of $\xi$ with respect to $q_v$ at the equilibrium point:
 \begin{equation}\label{eq:xi qv}
   \begin{array}{l}
   {\displaystyle
   \frac{\partial \xi}{ \partial q_v}=\frac{1}{l_c}\frac{h}{\sqrt{h^2+(x_c-x_e)^2}};}\\
    {\displaystyle
    \frac{\partial^2 \xi}{ \partial q_v^2}=\frac{1}{l_c}\frac{(x_c-x_e)^2}{\sqrt[3]{h^2+(x_c-x_e)^2}}.}
    \end{array}
 \end{equation}
 The derivatives of $\xi$ with respect to {$q_l$} at the equilibrium point:
 \begin{equation}\label{eq:xi ql}
   \begin{array}{l}
   {\displaystyle
   \frac{\partial \xi}{ \partial q_l}=\frac{1}{l_c}\frac{x_c-x_e}{\sqrt{h^2+(x_c-x_e)^2}};}\\
    {\displaystyle
    \frac{\partial^2 \xi}{ \partial q_l^2}=\frac{1}{l_c}\frac{h^2}{\sqrt[3]{h^2+(x_c-x_e)^2}}.}
    \end{array}
 \end{equation}
 The derivatives of $\xi$ with respect to $\alpha$ at the equilibrium point:
 \begin{equation}\label{eq:xi ql}
   \begin{array}{l}
   {\displaystyle
   \frac{\partial \xi}{ \partial \alpha}=\frac{1}{l_c}\frac{x_eh}{\sqrt{h^2+(x_c-x_e)^2}};}\\
    {\displaystyle
    \frac{\partial^2 \xi}{ \partial \alpha^2}=\frac{1}{l_c}\frac{x_e^2(x_c-x_e)^2}{\sqrt[3]{h^2+(x_c-x_e)^2}}.}
    \end{array}
 \end{equation}
 The cross-derivatives of $\xi$ are
  \begin{equation}\label{eq:xi cross}
   \begin{array}{l}
   {\displaystyle
   \frac{\partial^2 \xi}{ \partial q_v \partial \alpha}=\frac{1}{l_c}\frac{x_e(x_c-x_e)^2}{\sqrt[3]{h^2+(x_c-x_e)^2}};}\\
    {\displaystyle
    \frac{\partial^2 \xi}{ \partial q_l \partial \alpha }=-\frac{1}{l_c}\frac{x_eh(x_c-x_e)}{\sqrt[3]{h^2+(x_c-x_e)^2}};}\\
      {\displaystyle
    \frac{\partial^2 \xi}{ \partial q_l \partial q_v }=-\frac{1}{l_c}\frac{h(x_c-x_e)}{\sqrt[3]{h^2+(x_c-x_e)^2}}.}\\
    \end{array}
 \end{equation}
 For generalized coordinates $q_v$, $q_l$ and $\alpha$ the derivative of   $M_{\iota}(x_e,x_c)$ with respect to these coordinates has the following general form:
 \begin{equation}\label{eq:derivative of mutual inductance between element and coil wire number}
\begin{array}{l}
{\displaystyle
    \frac{\partial M_{(x_e,x_c),\iota}}{\partial q}= \frac{\partial M_a}{\partial \xi'_{\iota}}\frac{\partial \xi'_{\iota}}{\partial q}-\frac{\partial M_a}{\partial \xi''_{\iota}}\frac{\partial \xi''_{\iota}}{\partial q},}\\
\end{array}
\end{equation}
where $\xi'_{\iota}=2f_{\iota}(x_e,x_c)/(l_c+b)$ and $\xi''_{\iota}=2f_{\iota}(x_e,x_c)/(l_c-b)$.
Let us consider separately derivative of $M_{\iota}$ with respect to $\beta$.  Starting with the estimation of derivative of $\xi'_{\imath}$ with respect to $\beta$ at the equilibrium point  we have
\begin{equation}\label{eq:xi plus beta}
  \begin{array}{l}
    {\displaystyle
   \frac{\partial \xi_{\iota}'}{\partial \beta}= 0;}\\
    {\displaystyle
         \frac{\partial^2 \xi_{\iota}'}{\partial \beta^2}= 2f_{\iota}(x_e,x_c)\frac{b}{(l_c+b)^2}. }\\
    \end{array}
 \end{equation}
Accounting for the later equations, the first and the second  $\beta$ - derivative of $M_a((l_c+b\cos\beta)/2,f_{\iota}(x_e,x_c))$ for equilibrium point can be written as
\begin{equation}\label{eq:general rule beta}
  \begin{array}{l}
    {\displaystyle
   \frac{\partial M_a}{\partial \beta}= 0;}\\
    {\displaystyle
   \frac{\partial^2 M_a}{\partial \beta^2}=\frac{\partial M_a}{ \partial \xi_{\iota}'}\frac{\partial^2 \xi_{\iota}'}{\partial \beta^2}
    -\frac{\mu_0b}{2\pi}\left[\ln\frac{1+\sqrt{1+\xi_{\iota}'^2}}{\xi_{\iota}'}-\sqrt{1+\xi_{\iota}'^2}+\xi_{\iota}'\right].}\\
  \end{array}
 \end{equation}
 For terms  $m_l^{kj}$, we can write
\begin{equation}\label{eq:m_l_trans}
\begin{array}{l}
{\displaystyle
    m_v^{11}=2\sum_{\iota=0}^{N-1}\left[  \frac{\partial M_{(c_l/2+d,c_s/2),\iota}}{\partial q}- \frac{\partial M_{(-c_l/2-d,c_s/2),\iota}}{\partial q}\right]  ;} \\
{\displaystyle
    m_v^{22}=2\sum_{\iota=0}^{M-1}\left[\frac{\partial M_{(c_l/2,c_l/2),\iota}}{\partial q}- \frac{\partial M_{(-c_l/2,c_l/2),\iota}}{\partial q}
   \right]  ;}  \\
{\displaystyle
    m_v^{12}=2\sum_{\iota=0}^{M-1}\left[\frac{\partial M_{(c_l/2+d,c_l/2),\iota}}{\partial q}- \frac{\partial M_{(-c_l/2-d,c_l/2),\iota}}{\partial q}
   \right];} \\
{\displaystyle
    m_v^{21}=2\sum_{\iota=0}^{N-1}\left[\frac{\partial M_{(c_l/2,c_s/2),\iota}}{\partial q}- \frac{\partial M_{(-c_l/2,c_s/2),\iota}}{\partial q}
    \right]  ;} \\
  {\displaystyle
    m_{l}^{11}= m_{l}^{22}= m_{l}^{12}= m_{l}^{21}=0};\\
{\displaystyle
    m_{\alpha}^{11}= m_{\alpha}^{22}= m_{\alpha}^{12}= m_{\alpha}^{21}=0};\\
{\displaystyle
    m_{\beta}^{11}= m_{\beta}^{22}= m_{\beta}^{12}= m_{\beta}^{21}=0}.\\
\end{array}
\end{equation}
Terms $m_{ll}^{kj}$ can be written as follows. For generalized coordinate, $q_l$, we have
\begin{equation}\label{eq:m_ql_trans}
\begin{array}{l}
{\displaystyle
    m_{ll}^{11}=2\sum_{\iota=0}^{N-1}\left[  \frac{\partial^2 M_{(c_l/2+d,c_s/2),\iota}}{\partial q_l^2}- \frac{\partial^2 M_{(-c_l/2-d,c_s/2),\iota}}{\partial q_l^2}\right]  ;} \\
{\displaystyle
     m_{ll}^{12}=2\sum_{\iota=0}^{M-1}\left[\frac{\partial^2 M_{(c_l/2+d,c_l/2),\iota}}{\partial q_l^2}- \frac{\partial^2 M_{(-c_l/2-d,c_l/2),\iota}}{\partial q_l^2}
   \right];}  \\
{\displaystyle
    m_{ll}^{22}=0; m_{ll}^{21}=0.  } \\
\end{array}
\end{equation}
For other generalized coordinates $q_v, \alpha$ and $\beta$, the second derivatives can be found by using the general rule given below
\begin{equation}\label{eq:m_q_trans}
\begin{array}{l}
{\displaystyle
    m_{qq}^{11}=2\sum_{\iota=0}^{N-1}\left[  \frac{\partial^2 M_{(c_l/2+d,c_s/2),\iota}}{\partial q^2}- \frac{\partial^2 M_{(-c_l/2-d,c_s/2),\iota}}{\partial q^2}\right]  ;} \\
{\displaystyle
     m_{qq}^{12}=2\sum_{\iota=0}^{M-1}\left[\frac{\partial^2 M_{(c_l/2+d,c_l/2),\iota}}{\partial q^2}- \frac{\partial^2 M_{(-c_l/2-d,c_l/2),\iota}}{\partial q^2}
   \right];}  \\
{\displaystyle
    m_{qq}^{22}=2\sum_{\iota=0}^{M-1}\left[\frac{\partial^2 M_{(c_l/2,c_l/2),\iota}}{\partial q^2}- \frac{\partial^2 M_{(-c_l/2,c_l/2),\iota}}{\partial q^2}
   \right]; } \\
{\displaystyle
    m_{qq}^{21}=2\sum_{\iota=0}^{M-1}\left[\frac{\partial^2 M_{(c_l/2,c_s/2),\iota}}{\partial q^2}- \frac{\partial^2 M_{(-c_l/2,c_s/2),\iota}}{\partial q^2}
   \right]. } \\
\end{array}
\end{equation}
Similar to  axially symmetric design \ref{app:A}, terms $g_l^{ks}$ are zero for all generalized coordinates. Terms $g_{ll}^{ks}$  are zero only for coordinates $q_v, \alpha$ and $\beta$, while accounting for (\ref{eq:auxiliary function}) the second derivative with respect to $q_l$ can be written as
\begin{equation}\label{eq:g_ll trans}
\begin{array}{l}
{\displaystyle
    g_{ll}^{12}=g_{ll}^{21}= 2\left[ M_a(b,f(c_l/2+d,c_l/2))- M_a(b,f(c_l/2+d,-c_l/2))\right].} \\
\end{array}
\end{equation}
Hence, the determinants $\Delta^{ks}$, $\Delta_0^{ks}$, $\Delta_v^{ks}$ , $\Delta_{vv}^{ks}$ and $\Delta_{ll}^{ks}$  can be defined similar to (\ref{eq:delat}),  (\ref{eq:delat 0}), (\ref{eq:delat l}),  (\ref{eq:delat nunu}) and (\ref{eq:ddelta ll}), respectively, whereas  both $\Delta_{\alpha\alpha}^{ks}$ and $\Delta_{\beta\beta}^{ks}$ are similar to (\ref{eq:delat ff}).

\end{document}